\documentclass[11pt]{article}
\usepackage[utf8]{inputenc}
\usepackage{jheppub}
\usepackage{amsmath,amssymb,amsfonts,graphicx,slashed,color,amsthm,mathtools,upgreek, enumerate, tensor, scalerel, subfig}
\usepackage{arydshln}
\usepackage[dvipsnames]{xcolor}

\usepackage{bbold}
\allowdisplaybreaks

\title{Entanglement spreading and emergent locality in Brownian SYK chains
}

\author[a]{\!Jatin Narde,} 
\author[a]{\! Onkar Parrikar,}
\author[a]{\! Harshit Rajgadia,}
\author[a]{\! Sandip Trivedi}
\affiliation[\,a]{Department of Theoretical Physics, Tata Institute for Fundamental Research, 1 Homi Bhabha Road, Mumbai, Maharashtra 400005, India.}

\newcommand{\beq}{\begin{equation}}
\newcommand{\eeq}{\end{equation}}
\newcommand{\beqn}{\begin{eqnarray}}
\newcommand{\eeqn}{\end{eqnarray}}
\newcommand{\eff}{\text{eff}}
\newcommand{\pa}{\partial}

\renewcommand{\d}{\mathrm{d}} 
\newcommand{\eps}{\epsilon}

\newcommand{\re}{\text{ref}}
\newcommand{\cH}{\mathcal{H}}

\newcommand{\D}{\mathcal{D}}

\newcommand{\qt}{\tilde{q}}
\newcommand{\jt}{\tilde{J}}
\newcommand{\jeff}{J_\text{eff}}

\newcommand{\wl}{w_\lambda}

\newcommand{\bra}[1]{\langle #1 \vert}
\newcommand{\ket}[1]{\vert #1 \rangle}

\DeclareMathOperator{\Tr}{Tr}

\newcommand{\nt}{\addtocounter{equation}{1}\tag{\theequation}}

\renewcommand{\tilde}{\widetilde}
\renewcommand{\bar}{\overline}

\begin{document}

\abstract{
The Ryu-Takayanagi (RT) formula and its interpretation in terms of quantum error correction (QEC) implies an emergent locality for the spread of quantum information in holographic CFTs, where information injected at a point in the boundary theory spreads within a sharp light-cone corresponding to the butterfly velocity. This emergent locality is a necessary condition for the existence of a geometric bulk dual with an RT-like formula for entanglement entropy. In this paper, we use tools from QEC to study the spread of quantum information and the emergence of a sharp light-cone in an analytically tractable model of chaotic dynamics, namely a one-dimensional Brownian SYK chain. We start with an infinite temperature state in this model and inject a qudit at time $t=0$ at some point $p$ on the chain. We then explicitly calculate the amount of information of the qudit contained in an interval of length $2\ell$ (centered around $p$) at some later time $t=T$. We find that at strong coupling, this quantity shows a sharp transition as a function of $\ell$ from near zero to near maximal correlation. The transition occurs at $\ell \sim v_B T$, with $v_B$ being the butterfly velocity. Underlying the emergence of this sharp light-cone is a non-linear generalization of the diffusion equation called the FKPP equation, which admits sharp domain wall solutions at late times and strong coupling. These domain wall solutions can be understood on physical grounds from properties of operator growth in chaotic systems.   
}

\maketitle

\parskip=5pt

\section{Introduction}
The Ryu-Takayanagi formula \cite{Ryu:2006bv, Hubeny:2007xt, Faulkner:2013ana, Engelhardt:2014gca} and its interpretation in terms of quantum
error correction \cite{Almheiri:2014lwa, Dong:2016eik, Harlow:2016vwg} suggests a form of emergent locality in thermal states of holographic conformal field theories -- quantum information
injected at a point in such a CFT spreads uniformly within a sharp
light-cone, and is robustly protected against any
errors outside this light-cone. This is easy to see from the bulk point of view \cite{Mezei_Stanford_2017}: under suitable assumptions we can model the injected information as a probe object falling into a black hole. When the object reaches the near-horizon geometry of the black hole, we can approximate its trajectory by a null geodesic. At any sufficiently large value of boundary time, we now consider the smallest subregion in the boundary CFT whose entanglement wedge encloses the bulk probe object. Since the bulk probe is falling towards the black hole, this minimal subregion must grow with time in the boundary CFT, so that the corresponding extremal surface can reach sufficiently deep into the bulk (see figure \ref{fig:Entanglement_Wedge}). The interpretation from the boundary point of view is clear -- quantum information spreads under time evolution. Remarkably, as long as the probe is a localized object in the bulk spacetime, the gravitational picture suggests that the corresponding spread of information in the boundary CFT occurs within a \emph{sharply} defined emergent light cone controlled by a velocity $v^{\text{EW}}_B$, which one might call the entanglement wedge (EW) butterfly velocity.\footnote{This emergent velocity must obviously be smaller than the speed of light by boundary causality. From the bulk point of view, this is guaranteed by the fact that the entanglement wedge of a subregion is always bigger than its causal wedge \cite{Headrick:2014cta}.} This emergent light-cone must be sharp because the entanglement wedge is sharply defined by the extremal surface in the bulk, and a coherent probe is either contained inside this surface or outside. Thus, the emergence of a sharp light-cone for the spread of quantum information is a necessary condition for the emergence of a dual bulk spacetime geometry satisfying a Ryu-Takayanagi-like formula for entanglement entropy, and may be regarded as an information theoretic diagnostic of bulk locality (see \cite{Heemskerk:2009pn, Maldacena:2015iua, Caron-Huot:2021enk, Chandrasekaran:2021tkb, Chandrasekaran:2022qmq} for some previous work on signatures of bulk locality). A further remarkable fact is that in a large class of holographic theories, the EW butterfly velocity precisely agrees with the standard butterfly velocity $v_B$ defined as the rate at which out-of-time-order correlation (OTOC) functions propagate for generic single trace operators \cite{Mezei_Stanford_2017, Dong:2022ucb, Chua:2025vig}. This universality in operator growth (for probe operators) and the spread of quantum information in holographic CFTs is a boundary manifestation of the equivalence principle in the bulk.    

In this paper we study the spread of quantum information in an analytically tractable model of chaotic dynamics, namely a one-dimensional Brownian SYK chain. The SYK model \cite{Sachdev_1993, Kitaev2015v1, Kitaev2015v2, Maldacena:2016hyu} is a disorder-averaged theory of $N$ Majorana fermions. The model is known to have chaotic dynamics (it saturates the MSS bound \cite{Maldacena:2015waa} on the chaos exponent) and contains the Schwarzian mode \cite{Maldacena:2016upp} in its low-energy Hilbert space. Brownian SYK \cite{Saad:2018bqo, Sunderhauf:2019djv, Jian:2020krd, Stanford:2021bhl, Jian:2021hve,Balasubramanian:2023xdp, Milekhin:2023bjv, Zhang:2023vpm, Stanford:2023npy} is an even more tractable version of this model where the disorder average happens independently at each instance of time -- this makes the large-$N$ Schwinger-Dyson equations local in time, making them easier to study analytically. Since we are interested in the spread of quantum information, we will here consider a one-dimensional chain of such Brownian SYK dots, with the Hamiltonian comprising of random, nearest-neighbor hopping terms, in addition to the random $q$-fermion local/on-site terms. The strength $J$ of the on-site disorder naturally plays the role of the coupling constant in this model, with small $J$ corresponding to weak coupling and large $J$ corresponding to strong coupling. Our goal is to study how quantum information -- say, a qudit -- injected at a point (say, $x=0$) in the chain at time $t=0$ spreads under time evolution. In analogy with the gravitational discussion, we  would like to find the smallest subregion centered around $x=0$ which contains all of the information of the qudit at a later time $t$. Of course, we do not have recourse to a bulk geometry or the RT formula in this model. Instead, we use the quantum error correction interpretation of the RT formula: we interpret the Hilbert space of the injected qudit as a \emph{code subspace} and explicitly calculate the mutual information between a reference system maximally entangled with this code subspace and a subregion of size $2\ell$ centered around $x=0$ at a given time $t$. At weak coupling, the mutual information shows a gradual growth with $\ell$.\footnote{The mutual information is a monotonically growing function of $\ell$. This follows from strong subadditivity, or equivalently, from the monotonicity of relative entropy.} Remarkably, at strong coupling, we find a sharp transition in the mutual information as a function of the subregion length from near zero to a near maximal value, with the transition occurring at $\ell \sim  v_Bt$, where the velocity $v_B$ is determined in terms of the coupling constant. Thus, at strong coupling, we find a sharp light cone for the propagation of quantum information. Furthermore, by relating our entanglement entropy calculation to operator growth, we find that the velocity $v_B$ for the spread of information precisely coincides with the velocity of operator growth and OTOC propagation in this model, pointing to a gravity-like universality.

The mechanism by which this sharp light cone emerges is interesting, and can be understood on physical grounds. We find that in the Brownian SYK chain, the spread of information is controlled by a non-linear generalization of the diffusion equation, called the \emph{Fisher-Kolmogorov-Petrovsky-Piskunov (FKPP) equation} \cite{Fisher, KPP}. The FKPP equation is well-studied in the context of population growth and wave propagation, and has also appeared previously in calculations of OTOCs \cite{Xu_2019, PhysRevB.107.014201}. In our context, the FKPP equation arises from the large-$N$ Schwinger Dyson equations corresponding to the Schwinger-Keldysh path integral with multiple time folds, which in turn arises in the calculation of the mutual information via the replica trick.

 It is a well-known fact about the FKPP equation that, for the kind of boundary conditions which are  obtained here, it admits domain wall/ traveling wave solutions at sufficiently late times. In our study, the FKPP equation arises in the continuum limit with two scales: $J$, which governs the strength of the self interactions at each node, and ${\tilde J}$, which governs the strength of the quadratic, nearest-neighbour coupling.  The spatial width of the domain wall is of order $\sqrt{\frac{\jt}{J}}$.
When  $J\ll {\tilde J}$, the domain walls are broad  (this is expected given that at $J=0$ the FKPP equation reduces to the diffusion equation) and there is no sharp light-cone structure. On the other hand, when $J\gg{\tilde J}$ the domain walls become sharp and this gives rise to a sharply defined light-cone for the spread of quantum information. Let us also note that the velocity with which the domain wall propagates is given by $v\sim \sqrt {J {\tilde J}}$. The strong coupling limit referred to above, more precisely, corresponds to taking $J\rightarrow \infty, {\tilde J}\rightarrow 0$, keeping $J{\tilde J}$ fixed. In this limit, we see that the thickness of the domain wall goes to zero and it becomes very sharp, while the velocity of propagation remains finite. 

The sharpness of these domain wall solutions can be understood from a physical point of view in terms of the growth of \emph{operator size}, defined as $\phi_u = \frac{1}{N}\langle K_u\rangle$, where $\langle K_u\rangle $ is the average number of Majorana fermions constituting an operator at site $u$. For a typical operator supported inside some subregion $C$, $\phi_u =1/2$ for points in $C$ and zero outside. There is a sense in which $\phi_u=0$ (corresponding to the identity operator and the fermion parity operator) is an unstable fixed point under time evolution, while $\phi_u = 1/2$ (corresponding to the entropically favored operators of size $\frac{N}{2}$) is a stable fixed point. Under time evolution, an operator that is entirely supported in $C$ develops some small non-trivial support at the nearest neighbor sites in the complementary subregion (owing to the presence of kinetic terms in the Hamiltonian), at which point the local (i.e., on-site) dynamics takes over and ``pushes'' the operator into the entropically favored sector of size $N/2$ operators. The time scale for this to happen is, of course, controlled by the on-site coupling constant; the transfer of operator size happens efficiently at strong coupling, giving rise to a sharp light-cone structure. 

In \cite{jonay2018} (see also \cite{Zhou:2018myl, Zhou:2019pob, vardhan_moudgalya2024}), it was conjectured that a similar form of locality in entanglement dynamics emerges in general chaotic systems. These authors wrote down an effective \emph{membrane theory} for entanglement dynamics in such systems. It was later argued by Mezei \cite{Mezei_2017,Mezei_Membrane_Theory_Holography} that in holographic theories, the membrane description is a direct consequence of the Ryu-Takayanagi formula for holographic entanglement entropy. It will be interesting to understand the relation between the domain wall solutions to our Schwinger-Dyson equations and  entanglement membranes further and to develop a membrane theory for entanglement dynamics in  Brownian SYK chains. We leave this for future work.


\section{Preliminaries} \label{sec:pre}
\subsection{Information spreading in gravity}\label{sec:holography}
\begin{figure}
    \centering
\includegraphics[width=0.5\linewidth]{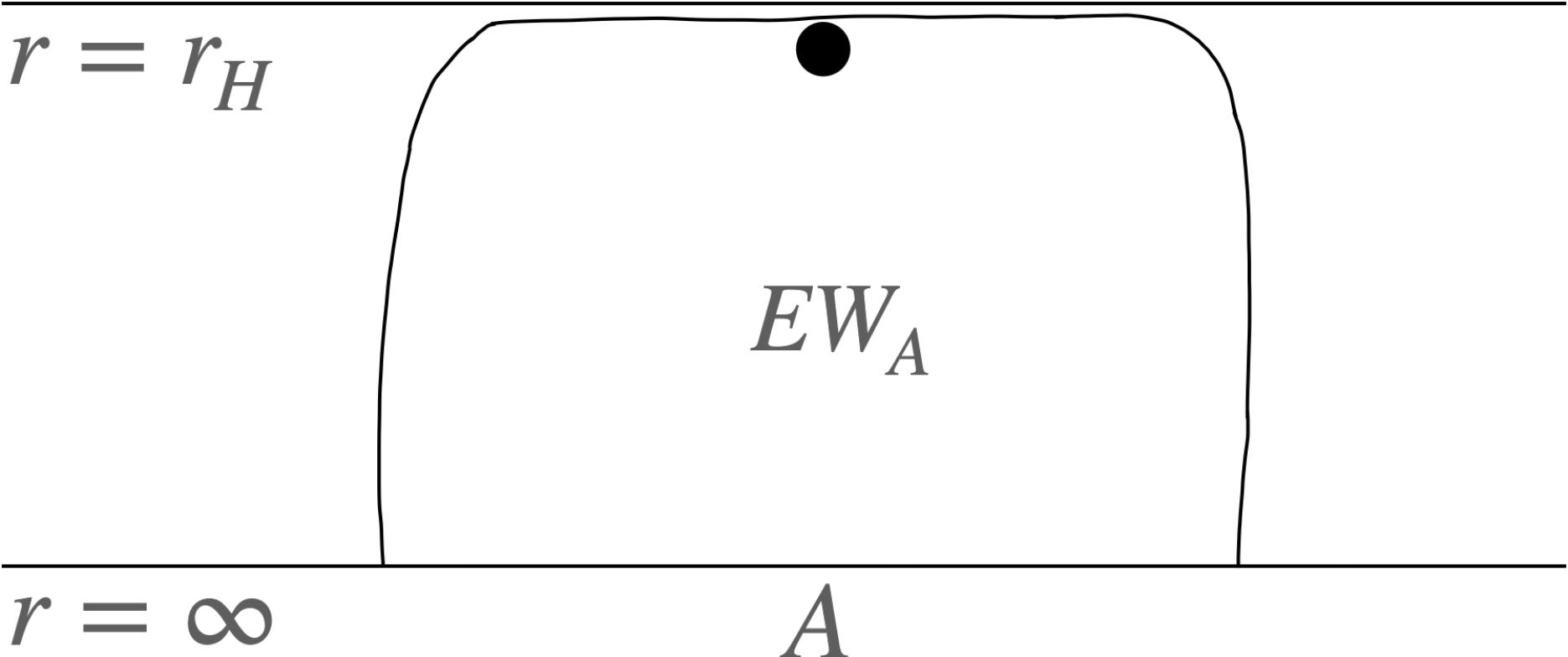}
    \caption{An illustration of the smallest boundary region $A$ whose entanglement wedge $EW_A$ contains the probe particle in the bulk.}
    \label{fig:Entanglement_Wedge}
\end{figure}

Consider the asymptotically AdS, eternal black hole geometry dual to the thermofield double state in the dual CFT. Imagine bringing in a probe particle -- say a qudit -- from asymptotic infinity and throwing it into the black hole from the right side. In the CFT description, we bring in an external qudit, couple it \emph{locally} (i.e., to an infinitesimally localized region)  to the right side of the thermofield double state and then evolve with the time evolution operator $e^{-itH_R}$ on the right. From the bulk point of view, the particle falls into the black hole. From the boundary CFT point of view, the size of the region over which the information of the qudit is encoded grows with time. One way to see this effect is to ask the following question: what is the size of the smallest subregion $A$ in the right boundary such that the entanglement wedge of $A$ in the bulk geometry contains the in-falling particle in the bulk (see figure \ref{fig:Entanglement_Wedge}). By the entanglement wedge reconstruction paradigm \cite{Czech:2012bh, Headrick:2014cta, Cotler:2017erl, Faulkner:2017vdd, Parrikar:2024zbb}, the information of the infalling particle is then almost entirely encoded in $A$. But as the particle falls deeper and deeper into the bulk, the size of the corresponding region $A$ on the boundary has to be larger and larger, so that the corresponding Ryu-Takayanagi surface can reach deep enough to enclose the bulk particle in its entanglement wedge. Thus, the size of the region over which the information of the qudit has spread out grows with time. In fact, at sufficiently late times, the size of this region grows linearly in time with the velocity given by the butterfly velocity $v_B$, i.e., the velocity which controls the spread of the out-of-time-ordered correlators in the dual CFT \cite{Mezei_Stanford_2017, Mezei_2017, Dong:2022ucb, Chua:2025vig}. 

To see this more quantitatively, consider the near horizon geometry of a general, static, planar black hole:
\beq 
g = -\left[\rho^2+O(\rho^4)\right] \left(\frac{2\pi}{\beta}\right)^2 dt^2 + d\rho^2 + \left(r_H^2+ \frac{2\pi r_H}{\beta}\rho^2 + O(\rho^4) \right)dy^\alpha dy^{\alpha},
\eeq 
where $y^\alpha$ are the transverse coordinates parametrizing the horizon of the black hole, and we have set $\ell_{AdS} = 1$. Note that after a sufficiently long time, the trajectory of the infalling particle will be exponentially close to the black hole horizon, and can be well-approximated by a null geodesic in the near-horizon geometry. We wish to consider the RT surface of a spherically symmetric region on the boundary, so that the surface is just about big enough to enclose the particle in its entanglement wedge. In this discussion, it suffices to look at the part of this RT surface which is very close to the black hole horizon, because once the surface exits the near horizon region, it reaches the asymptotic boundary in a distance of $O(1)$ in the $y$ directions. Thus, up to such $O(1)$ corrections, the main contribution to the operator size comes from the portion of the RT surface very close to the black hole horizon. 
Suppose that we parametrize the surface in this region by $x^+ = \epsilon\,u^+(y^\alpha)$ and $x^-=\epsilon\,u^-(y^{\alpha})$, where $x^+ = \rho e^{\frac{2\pi t}{\beta}}$ and $x^- = -\rho e^{-\frac{2\pi t}{\beta}}$. In fact, since all our RT surfaces will lie on a constant time slice, we will set $u^+ = u(y^\alpha)\,e^{\frac{2\pi t}{\beta}}$, and $u^- = -u(y^\alpha)\,e^{-\frac{2\pi t}{\beta}}$ henceforth, and simply keep track of the radial shape function $u$. The induced metric on this surface is given by
\beq 
h_{\alpha\beta} = \epsilon^2 \frac{\pa u}{\pa y^{\alpha}}\frac{\pa u}{\pa y^{\beta}}+\epsilon^2 \left(r_H^2 + \frac{2\pi r_H}{\beta}u^2 + O(\epsilon^4)\right)\delta_{\alpha\beta}.
\eeq 
Then, to leading order in $\epsilon$, the area of the surface is given by
\beq 
\text{Area} = \int d^{d-1}y^{\alpha}\,\left[r_H^{d-1}+\frac{1}{2}r_H^{d-3}\epsilon^2\left(\delta^{\alpha\beta}\pa_{\alpha}u\pa_{\beta}u+\mu^2 u^2 \right)+\cdots\right],
\eeq 
where $\mu =\frac{2\pi r_H(d-1)}{\beta}$, and $\cdots$ are terms higher order in $\epsilon$. The radial profile $u(y^{\alpha})$ of the RT surface thus satisfies the equation: 
\beq 
-\delta^{\alpha\beta}\partial_{\alpha}\partial_{\beta} u + \mu^2 u=0.
\eeq 
We can simplify the equation by focusing attention on spherically symmetric regions. In this case, $u$ is a function only of the radial coordinate $r= \sqrt{y^{\alpha}y_{\alpha}}$, and we get 
\beq 
r^{2}\pa_r^2 u + (d-2)r\pa_ru - \frac{2\pi r_H(d-1)r^{2}}{\beta}u=0.
\eeq
The solution in $d=2$ is given by
\beq \label{2d}
u(r) = u(0) e^{\mu r},
\eeq 
and in higher dimensions is given by
\beq \label{genD}
u(r) = c\,r^{-\nu}I_{\nu}(\mu r)u(0),
\eeq 
where $\mu = \frac{2\pi r_H}{\beta}$, $\nu = \frac{d-3}{2}$ and $c=\frac{2^{\nu}\Gamma(1+\nu)}{\mu^{\nu}}$.  
Under a change in the boundary time $t$ at which the RT surface is anchored, the bulk particle falls deeper in the bulk, and correspondingly we want the change in the location of the turning point of the RT surface to be given by
\beq \label{tdep}
u(r=0)=u_0\,e^{-\frac{2\pi}{\beta}(t-t_0)}.
\eeq 
From equations \eqref{2d}, \eqref{genD} and \eqref{tdep}
, we see that the shape function of the RT surface is given by
\beq \label{2dt}
u(t,r) \sim  e^{\mu r-\frac{2\pi t}{\beta}},
\eeq 
in $d=2$, while in higher dimensions it is given by
\beq \label{genDt}
u(t,r) \sim r^{-\nu}I_{\nu}(\mu r)e^{-\frac{2\pi t}{\beta}} \sim r^{-\nu-\frac{1}{2}}e^{\mu r-\frac{2\pi t}{\beta}}.
\eeq 
where we have dropped overall constant factors. Note that the shape function of the RT surface at late times and at large $r$ approximately takes the form of a travelling wave solution, with the velocity
\beq 
v_B = \frac{2\pi}{\mu \beta}.
\eeq 
The radius at which the RT surface exits the near horizon region will move outwards precisely with the above velocity. Thus, the information of the in-falling particle spreads out in the CFT in the form of an effective light cone with the velocity $v_B$. Remarkable, the velocity calculated above coincides with the butterfly velocity, i.e., the velocity at which the OTOC spreads. While we have reviewed this calculation in the case of Einstein gravity, this agreement between the two velocities has also been checked in general higher-derivative theories of gravity \cite{Mezei_Stanford_2017, Mezei_2017, Dong:2022ucb, Chua:2025vig}. 

The goal of this paper is to understand the above phenomenon -- namely, the emergence of a sharp light cone for the spread of quantum information, propagating at the butterfly velocity -- from the point of view of the dual quantum mechanical theory (see \cite{Mezei_Stanford_2017, Rampp:2023zmr} for previous work on chaotic spin chains and random circuits). The existence of a sharp entanglement light cone is a necessary condition for the emergence of a dual spacetime satisfying an RT-like formula for entanglement entropy, and thus provides an information theoretic signature for the emergence of bulk locality (see also \cite{Chandrasekaran:2021tkb, Chandrasekaran:2022qmq} for related previous work).

\subsection{Information spreading and quantum error correction}

Let us consider the thermofield double state $|\Omega\rangle$ in a bi-partite quantum mechanical Hilbert space $\mathcal{H}_L \otimes \mathcal{H}_{R}$. Here $L$ and $R$ should be thought of as the two sides of the eternal black hole \cite{Maldacena:2001kr} from the previous section. In this paper, we will focus on the case where both $L$ and $R$ have one spatial dimension, although many of the techniques can be generalized to higher dimensions without trouble. The idea now is to inject some quantum information -- let's say an external qudit with a $d$-dimensional Hilbert space $\mathcal{H}_{a}$ -- into the system $R$, then evolve with some suitable time evolution operator $U_{a,R}(T)$ (which couples the qudit to the rest of the system $R$), and study how the information spreads within $R$. In the context of the gravity calculation from the previous section, the unitary $U_{a,R}(T)$ can be thought of as some coupling between the qudit and the right side, followed by time evolution $e^{-iTH_R}$. 

In order to formulate the question of information spreading, it is helpful here to use the language of quantum error correction, where we think of the qudit being injected as creating a ``code subspace'':
\beq \label{EncState}
|\psi_i\rangle = U_{a,R}(T)\left( |i\rangle_{a}\otimes |\Omega\rangle_{LR}\right).
\eeq 
This seems appropriate because in the gravity calculation of the previous section, the notion of operator spreading came from asking for the smallest region whose entanglement wedge contained the in-falling particle. Without any reference to a bulk geometry (which may not exist in a general quantum system), we will interpret ``being in the entanglement wedge of a subsystem'' in terms of quantum error correction \cite{Almheiri:2014lwa, Dong:2016eik, Harlow:2016vwg} as ``being protected from the erasure of the complement of the subsytem''. There are various ways of characterizing such quantum error correction properties, but for our purposes, the most useful characterization is in terms of the \emph{decoupling principle} \cite{Schumacher:1996dy}: given a code subspace spanned by the (orthonormal) basis states $\{|\psi_i\rangle\}$, we construct the maximally entangled state on the code subspace 
\beq \label{MaxState}
|\Psi\rangle = \frac{1}{\sqrt{d}}\sum_{i}|i\rangle_{\text{ref}}\otimes |\psi_i\rangle_{a,R,L} 
\eeq 
where we have introduced an auxiliary reference system ``$\text{ref}$'' isomorphic to the code subspace with a choice of orthonormal basis vectors $\{|i\rangle_{\text{ref}}\}$ (which we can choose to label with the same index $i$). Let us now partition the system $R$ into two complementary subsystems $B$ and $C$ such that $\mathcal{H}_R = \mathcal{H}_B \otimes \mathcal{H}_C$. The code subspace is protected from erasure of the complement of the subsystem $B$ (i.e., under partial trace over $\mathcal{H}_{L}\otimes \mathcal{H}_C$) if and only if the reduced density matrix for $\Psi$ over $\text{ref}\cup L\cup C$ factorizes:
\beq \label{decoupling}
\rho^{\Psi}_{\text{ref}, L,C} = \rho^{\Psi}_{\text{ref}} \otimes \rho^{\Psi}_{L,C},
\eeq 
or equivalently, the mutual information between the reference system and $L\cup C$ vanishes, $I_{\Psi}(\text{ref}:L\cup C) = 0$. The vanishing of this mutual information guarantees that for any errors that act on the subsystem $L\cup C$, one can always find a recovery channel that exactly reverses the action of the errors on any state supported on the code subspace. Physically, equation \eqref{decoupling} says that no information about the injected qudit has leaked into the subsystem $L\cup C$, and thus the information of the qudit is entirely contained within $a \cup B$. Of course, in a more realistic setting, the mutual information will not be exactly zero but could be very small, and in this case the smallness of the mutual information guarantees the existence of a recovery channel with high fidelity \cite{Schumacher2001}.   

So, we need to compute the following mutual information: 
\beqn\label{MIRC}
I_{\Psi}(\text{ref}:L\cup C)& =& S_{\Psi}(\text{ref})+ S_{\Psi}(L\cup C) - S_{\Psi}(\text{ref}\cup L \cup C)\nonumber \\
&=&  S_{\Psi}(\text{ref})+ S_{\Psi}(\text{ref}\cup a \cup B) - S_{\Psi}(a\cup B)\nonumber \\
&=& 2 S_{\Psi}(\text{ref}) - I_{\Psi}(\text{ref}:a\cup B)\nonumber\\
&=& 2 \log\,d - I_{\Psi}(\text{ref}:a\cup B),
\eeqn 
where in the second line we have used the fact that $\Psi$ is a pure state, and in the last line the fact that the reduced density matrix on the reference system is maximally mixed.  We can take $B$ to be a ball-shaped region of radius $\ell$ around the point at which the initial qudit was injected. For $T \ll \ell$, we expect that the information of the qudit should be contained within $a\cup B$ and so $I_{\Psi}(\text{ref}:a\cup B)$ should be close to its maximal value of $2\log\,d$; correspondingly, the mutual information with the complement should be close to zero. On the other hand, at late times the information should leak out and so the mutual information $I_{\Psi}(\text{ref}:a\cup B)$ should drop to a small value. Equivalently, we could study the mutual information at a fixed value of $T$ and for different values of $\ell$ -- in this case, strong subadditivity implies that $I_{\Psi}(\text{ref}:a\cup B)$ has to be a monotonically increasing function of $\ell$. In general quantum systems, the growth of this mutual information can be gradual, with no sharp features. However, as we discussed in the previous section, in a holographic theory with a local bulk dual, the entanglement wedge reconstruction paradigm implies a \emph{sharp} transition in this mutual information (for well-localized probes) from near zero to the near maximal value of $2\log\,d$ at $\ell \sim v_B t$, where $v_B$ is the butterfly velocity. Our goal now is to look for such a sharp transition in a tractable model, to develop some insight into the underlying mechanism.


\subsection{Setup}

In this paper, we will study the above mutual information in a quantum mechanical system with one spatial dimension. Each system $L$ and $R$ comprises of a one-dimensional chain, labelled by the position index $u \in \mathbb{Z}$. Every node of this chain comprises of $N$ Majorana fermions. For all $u \neq 0$, all the fermions will be treated as being part of $R$ or $L$ respectively. At $u =0$, $N_2 = (N - N_1)$ Majorana fermions will be treated as being part of $R$ or $L$ respectively; the remaining $N_1$ fermions in $R$ at $u=0$ will be treated as being part of the ``external qudit'' $a$ which is being injected into the system, while the remaining $N_1$ fermions at $u =0$ in $L$ will correspond to the reference system ``ref'' introduced above. We will work in the limit $N\to \infty$, $N_1 \to \infty$ with $\lambda = \frac{N_1}{N}$ held fixed, but small. The initial state $\ket{\Omega}$ will be taken to be the maximally entangled state, which is essentially the thermofield double state at infinite temperature. The time evolution operator $U_{a,R}$ will be taken to be a Brownian version of the SYK Hamiltonian, where at every site there is an all-to-all, random, $q$-body coupling term, while neighbouring sites are coupled by a term quadratic in the fermions, but once again with all-to-all, random couplings. The word ``Brownian'' here means that the couplings are all drawn from a Gaussian ensemble, independently at each instance of time. More details will follow in section \ref{sec:BSYK}; for now, the reader may think of the unitary $U_{a,R}$ as being the time ordered exponential of some time-dependent Hamiltonian of all the Majorana fermions in $a\cup R$. Finally, we will take $B$ to consist of the nodes in $R$ between $-\ell \leq u \leq \ell$, including the $N_2$ fermions at the site $u=0$ (where the external qudit was injected), and $C$ will be its complement in $R$. 

\begin{figure}
    \centering
    \includegraphics[height=3cm]{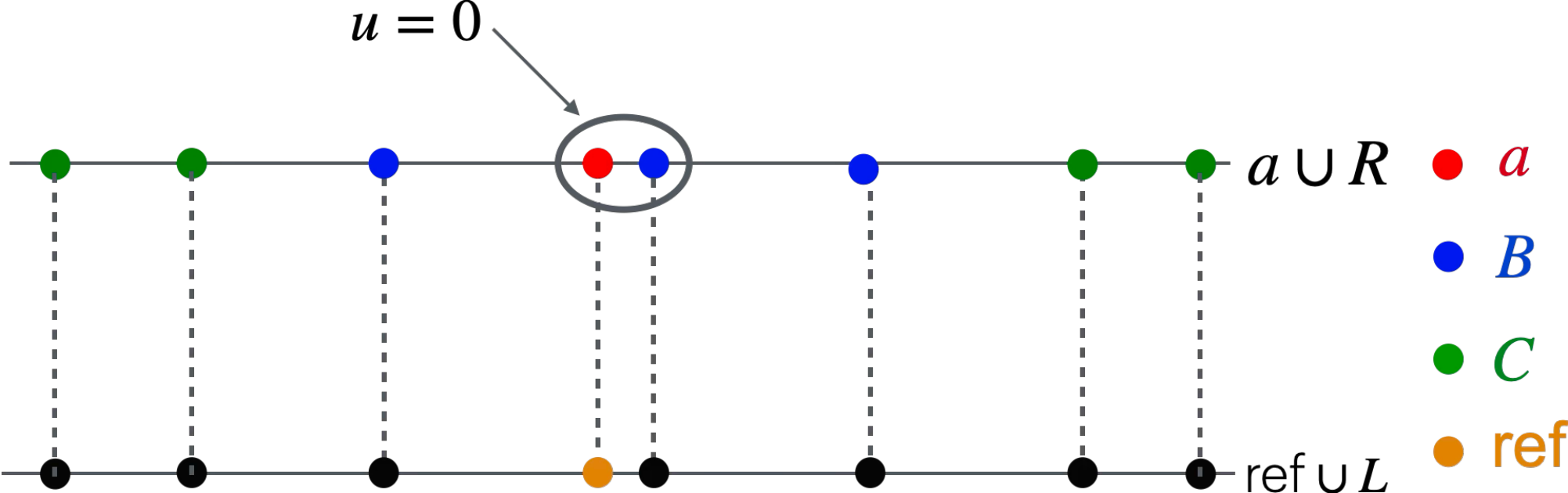}
    \caption{\small{\textbf{Setup:}} The red dot on the chain $R$ denotes the subsystem $a$ which is the ``external qudit" injected at $\sigma = 0$ in the chain. The blue dots on the chain constitute the subsystem B and the rest (green dots) belong to subsystem C.}
    \label{fig:setup}
\end{figure}


As discussed above, in order to test whether the injected qudit is in the ``entanglement wedge'' of the region $a\cup B$, we need to compute the mutual information between $\text{ref}$ and ${a\cup B}$ in the state $\ket{\Psi}$ defined in equation \eqref{MaxState}. In order to compute this mutual information, we need to compute the individual entropies for the subsystems $\text{ref}$, $a\cup B$ and $\text{ref}\cup a \cup B$. Since $U_{a,R}$ is a unitary operator acting on $\cH_a \otimes \cH_{B}\otimes \cH_C$, the reference system is always maximally entangled with the rest of the system. Thus, 
\beq
S_{\Psi}(\re)  = \log d,
\eeq 
independent of the time $T$. 
Moreover, since $\ket{\Omega}_{L,R}$ is a maximally entangled state between $B \cup C$ and its complement, the state $\Psi$ is maximally entangled state between $a \cup B \cup C$ and $\re \cup L$ for any time $T$. Therefore, the reduced state on $a \cup B \cup C$ is maximally mixed. This further implies that the reduced state on $a \cup B$ is maximally mixed, and so
\beq 
S_{\Psi}(a \cup B) = \log{d}+ \log \dim\,\cH_B.
\eeq
Thus, we have:
\begin{equation} \label{MIRC2}
I_{\Psi}(\text{ref}: a\cup B) = 2 \log d + \log \dim \cH_B - S_\Psi (\re \cup a\cup B), 
\end{equation} 
and we are left to evaluate $ S_\Psi (\re \cup a\cup B) $, where
\begin{equation}\label{eq:reduced_density_matrix}
    \rho_{\re \cup a \cup B} = \Tr_{\cH_C \otimes \cH_L} \left( U_{a,R}(T) \ket{\Psi} \bra{\Psi} U^\dagger_{a,R}(T) \right) .
\end{equation}
In general, calculating von Neumann entropy can be a formidable task, but we will be able to do it in the limit $N\to \infty$ and then $\lambda \to 0$, using perturbation theory in $\lambda.$

\subsection{Boundary Conditions for the $n$-th Renyi Entropy}


In order to calculate the above entropy, we need to know the spectrum of the reduced density matrix. One way to access the spectrum is by calculating the moments $\Tr \left(\rho^n_{\re\cup a \cup B}\right)$ of the density matrix, where $n$ is a positive integer. $\Tr \left(\rho^n_{\re\cup a \cup B}\right)$ is a non-trivial function of the time $T$, and it can be computed using a Schwinger-Keldysh path integral, where each instance of the time evolution operator $U_{a,R}$ in the above formula is given by a Lorentzian path integral along a forward time contour, while each instance of $U^{\dagger}_{a,R}$ is given by a Lorentzian path integral along a backward time contour. Thus, the full path integral for the $n$-th Renyi Entropy is performed on a contour that consists of `$n$' forward time evolution contours and `$n$' backward time evolution contours, as shown in figure \ref{fig:contour} (for $n = 2$). In our case, these path integrals are in the Brownian SYK model, and since the time evolution only acts on $a\cup B\cup C$, we need only keep the corresponding fermions $(\psi_a, \psi_B, \psi_C)$ in the path integral. This path integral takes the schematic form: 
\begin{equation}
 \Tr (\rho_{\text{ref}\cup a\cup B}^n) =   \int  \prod_{\omega \in(a,B,C)} \prod_{j=1}^{2 n} \D \psi_\omega^{j}  \,  \exp \left[i \sum_{k = 1}^{n} \int_0^T \d t \left(  \mathcal{L}(t,\psi_\omega^{2k-1}(t)) - \mathcal{L}(t,{\psi}_\omega^{2k}(t) \right) \right],
\end{equation}
where $\psi_\omega^{2k-1}$ and $\psi_\omega^{2k}$ are fermions running along the forward and backward time directions respectively, and the Lagrangian $\mathcal{L}$ is given by: \begin{align}
    \mathcal{L}(t,\psi_\omega(t))= i \sum_{\omega\in(a,B,C)} \frac{\psi_\omega (t)\partial_t \psi_\omega(t)}{2} - H(t,\psi_\omega(t)) ,
\end{align}
with $H$ being the time-dependent Hamiltonian. In addition, we also have to specify boundary conditions on the fermions at $t=0$ and $t=T$. 

\subsubsection*{Boundary conditions at $t=T$} 
The boundary conditions at $t=T$ are dictated by the nature of partial traces in $\Tr \left(\rho^n_{\re\cup a \cup B}\right)$: 
\begin{align*}
\nt   \psi^{2 k }_\omega(T)& = \psi^{2k + 1}_\omega(T), \quad \text{for  } k < n, \quad  \quad \psi^{1}_\omega(T) = -\psi^{2n}_\omega(T)\;\;\quad \;\; \omega \in (a, B) \\
   \psi^{2k -1}_C (T) &= \psi^{2k}_C(T) \quad \text{for} \quad k \leq n .
\end{align*}
The boundary conditions in the first line follow from cyclic contraction of indices in $a \cup B$, while the boundary condition in the second line corresponds to tracing out $C$. The above path integral with forward and backward time-evolution contours can be re-interpreted as a path integral with only forward time-evolution contours if we make the transformation ${\psi}_\omega^{2j}  \rightarrow  - i {\psi}_\omega^{2j}$.  In this interpretation, the boundary conditions $t = T$ define a final state $\bra{\Gamma_f}$ in the Hilbert space $(\cH \otimes \cH^*)^{\otimes n}$, where $\cH = \cH_a \otimes \cH_B \otimes \cH_C$. In this language, the final state satisfies 
\begin{align*}\label{eq:final_a}
\nt  \bra{\Gamma_f} \psi^{2k}_\omega& =  i\bra{\Gamma_f} \psi^{2k +1}_\omega, \quad \text{for   } k < n, \,\; \bra{\Gamma_f}\psi^{1}_\omega =  i\bra{\Gamma_f}\psi^{2n}_1\;\;\quad \;\;\omega \in (a, B) \\
  \bra{\Gamma_f} \psi^{2k-1}_C  &=  i\bra{\Gamma_f} {\psi}^{2k}_C,\;\; \quad \text{ for }  \; k \leq n.\end{align*}

\subsubsection*{Boundary conditions at $t=0$} 
\begin{figure}
    \centering
    \includegraphics[width =\textwidth]{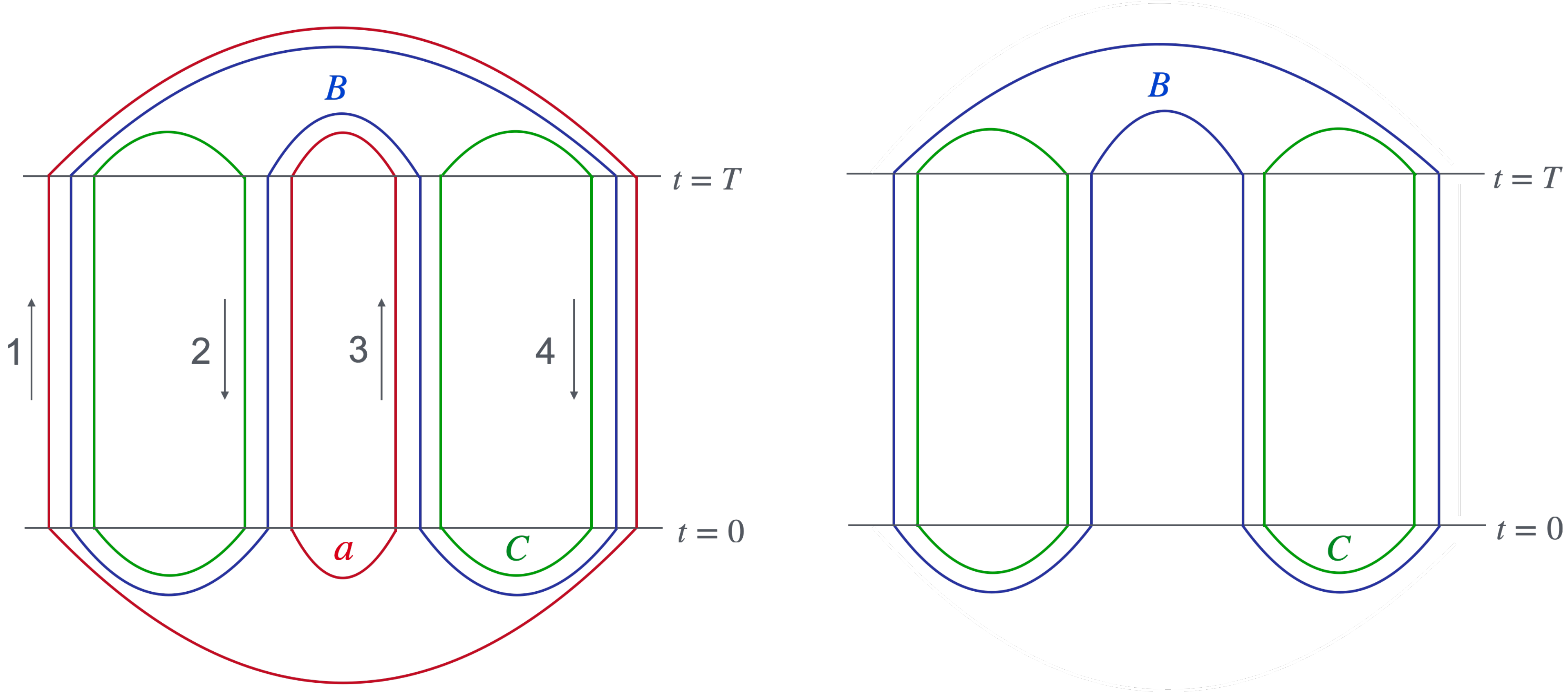}
    \caption{\small{\textbf{Left:} Path integral contour for fermions in subsystems $a$ (red), $B$(blue), and $C$ (green).} \textbf{Right:} Path integral contour ignoring the contribution from the subsystem $a$.}
    \label{fig:contour}
\end{figure}

To fix the initial boundary conditions, first note that $B$ and  $C$ are maximally entangled with corresponding subsystems $\bar{B}$ and $\bar{C}$ in $L$. Since $\bar{B}$ and $\bar{C}$ have trivial time evolution, tracing them out corresponds to inserting a maximally mixed state as the initial state for $B$ and $C$ between contours $2k-1$ and $2k$ for $k \leq n$ (see figure \ref{fig:contour}). Taking into account the re-interpretation of backward time contour as a forward time contour with the change of variables $\psi \to -i\psi$, the above boundary conditions corresponds to an initial state $\ket{\Gamma_i}$ in the Hilbert space $(\cH \otimes \cH*)^{\otimes n}$ that satisfies
\begin{equation}\label{eq:init_BC}
  \nt \psi_\omega^{2k-1} \ket{\Gamma_i}  = -i \psi_\omega^{2 k }\ket{\Gamma_i}, \quad \text{for} \quad k \leq n,  \quad \omega \in (B,C).
\end{equation}
On the other hand, while $\re$ evolves trivially in time, its indices in the density matrix $\rho_{\re\cup a\cup B}$ are contracted in a cyclic manner. This again corresponds to the insertion of maximally mixed states for $a$ at $t=0$, but now, these are inserted between contour number  one and four, and between contour number two and three.  Thus, the initial conditions for $a$ are given by:
\begin{align*}\label{eq:init_a}
 \nt       \psi_a^{2k} \ket{\Gamma_i}  &=  -i \psi_a^{2k+1} \ket{\Gamma_i}, \quad \text{for} \quad k < n, \quad \psi_a^{1} \ket{\Gamma_i} =  - i \psi_a^{2n} \ket{\Gamma}_i.  
\end{align*}

It is instructive to consider the path-integral for $n = 2$  before we consider arbitrary $n$. 
For $n=2$, while the above boundary conditions are stated in terms of the fermion operators, we will see below that the large-$N$, disorder averaged path integral organizes itself in terms of the following collective variables \cite{Stanford:2021bhl}: 
\begin{equation} 
\nt x_\omega  = -2i \, \psi^1_\omega \psi_\omega^2 , \quad   y_\omega =  2 \psi^1_\omega \psi^3_\omega, \quad z_\omega=  -2i\, \psi^1_\omega \psi^4_\omega ,  \quad \omega \in \{a,B,C\}.  
  \end{equation}
We can summarize the initial and final boundary conditions for our path integral in terms of these collective variables as follows:
\begin{enumerate}
    \item For the fermions in $a$: \begin{align*}\label{eq:bc_a}
 \nt       z_a(0) =  1, \quad x_a(0) = -y_a(0),\\ 
        z_a(T) = 1,\quad x_a(T) = y_a(T).
    \end{align*} 
    \item  For the fermions in $B$: \begin{align*}\label{eq:bc_B}
\nt        x_B(0) =  1, \quad z_B(0) = y_B(0),\\ 
        z_B(T) = 1,\quad x_B(T) = y_B(T).
    \end{align*} 
    \item For the fermions in $C$: \begin{align*}\label{eq:bc_C}
     \nt   x_C(0) =  1, \quad z_C(0) = y_C(0),\\ 
        x_C(T) = 1,\quad z_C(T) = -y_C(T).
    \end{align*} 
\end{enumerate}
In the next section, we turn to the explicit computation of the Lorentzian path integral using the above boundary conditions, first for the $n = 2$ case, and then for general $n$. 

\section{Mutual Information in Brownian SYK chain}\label{sec:BSYK}

The Brownian SYK chain is a theory consisiting of a one-dimensional lattice of SYK nodes, with each node comprising of $N$ Majorana fermions. These fermions interact with an on-site, all-to-all, $q$-body term with time-dependent random couplings, together with a nearest-neighbor, all-to-all, $\tilde{q}$-body ``hopping'' term, also with time-dependent, random couplings\footnote{We can also consider entanglement dynamics in a Brownian SYK chain with conserved charges. See \cite{Zhang:2023vpm} for entanglement growth in the presence of a conserved $U(1)$ charge.}:  
\begin{align*}\label{eq:Hamiltonian_saddle}
\nt   H(t) &= \sum_{u}\left(  i^{q/2} \sum_{I_1 \dots I_q} J^u_{I_1 \dots I_q}  (t)\psi_{I_1,u} \dots \psi_{I_q,u} \right. + \\ & \qquad  \left.\qquad   i^{\qt/2} \sum_{I_1 \dots I_{\qt/2};J_1 \dots J_{\qt/2} }  J^{u,u+1}_{I_1 \dots I_{\qt/2}; J_1 \dots J_{\qt/2}} (t) \,\psi_{I_1,u} \dots \psi_{I_{\qt/2},u} \, \psi_{J_1,u+1} \dots \psi_{J_{\qt/2},u+1}  \right),
\end{align*}
where the label $u$ denotes the lattice site, while $I,J,\cdots$ are the fermion flavor indices; we have suppressed the contour indices in the above formula. The couplings in the Hamiltonian are chosen randomly and independently at each instance of time and for every lattice site from a Gaussian distribution with zero mean and the following variances:
\begin{align*}\label{eq:variance}
\nt    \langle  J^{u}_{I_1 \dots I_q}(t) J^{u}_{I_1 \dots I_q}(t') \rangle = \frac{J (q-1)!}{N^{q-1}}\delta(t-t') ,\\  \langle J^{u,u+1}_{I_1 \dots I_{\qt/2} J_1 \dots J_{\qt/2}}(t) J^{u,u+1}_{I_1 \dots I_{\qt/2} J_1 \dots J_{\qt/2}} (t') \rangle = \frac{\jt \left(\frac{\qt}{2}!\right)^2}{\qt N^{\qt -1}} \delta(t-t').
\end{align*}
In the following analysis, we will set $\qt =  2$.

Since the couplings in the Brownian SYK theory are sampled from a random distribution, we can only hope to calculate the averaged moments of the density matrix in equation \eqref{eq:reduced_density_matrix}.  Therefore, the quantity we must consider is $\mathbb{E}\left[\text{Tr}\,\rho^n_{\text{ref}\cup a\cup B}\right]$, 
where $\mathbb{E}[\cdots]$ denotes ensemble average over the couplings. 
In the standard SYK model, the averaging over couplings is done in a time-independent manner, and this results in an effective action which is non-local in time. The advantage in the Brownian SYK model is that the averaging happens independently at each instance of time, and as a result we get an action in the path integral which is local in time: 
\beq 
\mathbb{E}\left[ \Tr \rho_{\text{ref}\cup a\cup B}^n \right] =  \frac{1}{\sqrt{2}^{\,nNL}}\int \prod_{I,j,u,t}d\psi^j_{I,u}(t) \; e^{-S},
\eeq
wher
\beqn 
S&=& \int_0^T  \d t \sum_{u} \left\{  \sum_{I=1}^N \sum_{j=1}^{2n} \frac{\psi^j_{I,u}(t) \partial_t \psi^j_{I,u}(t)}{2} 
+ \frac{N J}{2 q}  \, \sum_{j,k=1}^{2n} s_j s_k \left(\sum_{I=1}^N \frac{ \psi^j_{I,u}(t) \psi^k _{I,u}(t) }{N} \right)^q \right.\nonumber \\
&+& \left.\frac{N \jt}{4}  \sum_{j,k=1}^{2n}  \left( \sum_{I=1}^N \frac{\psi_{I,u}^j(t) \psi_{I,u}^k(t)  }{N} \right) \left( \sum_{J=1}^N  \frac{\psi_{J,u+1}^j (t) \psi_{J,u+1}^k(t)  }{N} \right)\right\},
\eeqn
where we have now restored the contour indices $j,k$ etc., and 
\begin{equation}
    s_j = \begin{cases}
        1 & \text{if} \quad j \in \{1,3\} \\
        -i^q & \text{if} \quad j \in \{2,4\}.    \end{cases}
\end{equation}
Let us first focus on the second and third terms in the action above. Since the action is local in time and fermions enter the action in a flavour-summed form, we can easily calculate the terms in the action where fermions on the same contour appear together using the operator identity $(\psi^j_{I,u})^2 = \frac{1}{2} $.  Such ``contour-diagonal'' terms give the following contribution to the action:
\begin{align*}
\nt  S_{\text{diag.}} &= -  \sum_j \int  \d t \left[ \frac{N J}{2 q} \left( 
\sum_{i=1}^N\frac{ \psi_{I,u}^j \psi_{I,u}^j}{N} \right)^q   + \frac{N \jt}{4}\left(\sum_{I=1}^N
 \frac{ \psi_{I,u}^j \psi_{I,u}^j}{N}  \right) \left( \sum_{J=1}^N\frac{ \psi_{J,u+1}^j \psi_{J,u+1}^j}{N}  \right)\right] \\&= \sum_{u} 2n \left(\frac{N JT}{2^{q+1} q }+ \frac{N \jt T}{16} \right) .
\end{align*}
The contribution to the action coming from fermion pairs from different contours is more non-trivial. Together with the kinetic term, this is given by:
\begin{align*}
\nt    S_{\text{rest}} =\int  \d t\sum_{I=1}^N\frac{\psi_{I,u}^j \partial_t \psi_{I,u}^j}{2} + &\frac{N J}{2 q} \int \d t  \, \sum_{j \neq k} s_j s_k \left( \sum_{I=1}^N\frac{ \psi_{I,u}^j \psi_{I,u}^k}{N}  \right)^q\\& +\frac{N \jt}{4} \int \d t  \,\sum_{j\neq k}  \left( \sum_{I=1}^N\frac{ \psi_{I,u}^j \psi_{I,u}^k}{N} \right)\left(\sum_{J=1}^N\frac{ \psi_{J,u+1}^j \psi_{J,u+1}^k}{N}  \right).
\end{align*}
At this point, we must make explicit the contributions to the action from fermions belonging to the various subsystems $a, B $ and $C$. Recall that these subsystems are defined as follows:\begin{align*}
\nt    a &= \{\psi_{I,0}^j | I \in \{1,2,\cdots,N_1\}\}, \quad \text{where}\;\;  \frac{N_1}{N} = \lambda,  \\
    B &=   \{\psi_{I,0}^j | I \in \{N_1+1,\cdots,N\}\}\cup \{\psi^j_{I,u}| I \in \{1,2,\cdots N\}, \; u \in [-\ell,\ell ]\setminus \{0\} \} ,  \\
    C &=\{\psi_{I,u}^j | \forall I,\, u \notin [-\ell,\ell]\}. 
\end{align*}
In words, $a$ is the external qudit which we inject into the system at $T=0$ at $u=0$. The qudit is taken to comprise of the first $N_1$ fermions at $u=0$, where $N_1 = \lambda N$, with $\lambda$ held fixed in the large $N$ limit. The subsystem $B$ comprises the rest of the $(N-N_1)$ fermions at $u=0$ together with the other nodes between $-\ell \leq u \leq \ell$. Finally, the subsystem $C$ comprises the rest of the nodes in $R$. Making the dependence on the $N_1$ external fermions at $u=0$ manifest, we get:
\begin{align*}
\nt  - \frac{S_{\text{rest}}}{N} =&-\int  \d t \sum_{I,j , u }\frac{\psi_{I,u}^j \partial_t \psi_{I,u}^j}{2} 
 -   \sum_{u \neq 0}\frac{J}{2 q} \int \d t  \, \sum_{j \neq k} s_j s_k \left( \sum_{I=1}^N\frac{ \psi_{I,u}^j \psi_{I,u}^k}{N}  \right)^q 
 \\&   -   \frac{J}{2 q} \int \d t  \, \sum_{j \neq k} s_j s_k  \left(\lambda  \sum_{I=1}^{N_1} \frac{ \psi_{I,0}^j \psi_{I,0}^k}{N_1} + (1-\lambda)  \sum_{I=N_1+1}^{N} \frac{ \psi_{I,0}^j \psi_{I,0}^k}{N-N_1} \right)^q 
 \\& -\sum_{u \notin \{-1,0\}}\frac{ \jt}{4} \int \d t  \,\sum_{j\neq k}   \left( \sum_{I=1}^N\frac{ \psi_{I,u}^j \psi_{I,u}^k}{N} \right) \left(\sum_{I=1}^N\frac{ \psi_{I,u+1}^j \psi_{I,u+1}^k}{N}  \right)\\& -\frac{\jt}{4} \int \d t  \,\sum_{j\neq k} \left( \lambda  \sum_{I=1}^{N_1} \frac{ \psi_{I,0}^j \psi_{I,0}^k}{N_1} + (1-\lambda)  \sum_{I=N_1+1}^{N} \frac{ \psi_{I,0}^j \psi_{I,0}^k}{N- N_1}\right) \left(\sum_{I=1}^N\frac{ \psi_{I,1}^j \psi_{I,1}^k}{N}  \right)
 \\& -\frac{\jt}{4} \int \d t  \,\sum_{j\neq k} \left( \lambda  \sum_{I=1}^{N_1} \frac{ \psi_{I,0}^j \psi_{I,0}^k}{N_1} + (1-\lambda)  \sum_{I=N_1+1}^{N} \frac{ \psi_{I,0}^j \psi_{I,0}^k}{N- N_1}\right) \left(\sum_{I=1}^N\frac{ \psi_{I,-1}^j \psi_{I,-1}^k}{N}  \right).
\end{align*}
To compute the path integral, we introduce the collective variables $g^{j,k}_{\omega,u}$ -- one for each subsystem and each pair of contours (where $\omega \in \{a, B, C\}$ is the subsystem label and $j,k$ are the contour labels) and the corresponding Lagrange multipliers $\sigma^{j,k}_{\omega,u}$ which impose the following constraints:
\begin{align*}
\nt    g^{j,k}_{ a,0} &= \sum_{I \leq N_1 }  \frac{\psi_{I,0} ^{j} \psi_{I,0}^{k}}{N_1 } , \\
    g^{j,k}_{B,0} &= \sum_{I = N_1+1 }^N \frac{\psi_{I,0} ^{j} \psi_{I,0}^{k}}{N - N_1 }, \\ 
    g^{j,k}_{ B,u \neq 0} &= \sum_{I \leq N } \frac{\psi_{I,u} ^{j} \psi_{I,u}^{k}}{N }, \quad  \cdots \;\;(u \in [-l,l]), \\
    g^{j,k}_{C,u} &= \sum_{I \leq N} \frac{\psi_{I,u} ^{k} \psi_{I,u}^{k}}{N }, \quad \cdots \;\; (u \notin [-l,l]).
\end{align*}
After integrating out the fermions, the effective action in terms of the $(g,\sigma)$ fields is given by: 
\begin{align*}\label{eq:g_sigma_action}
\nt    -S_{\text{eff}} =  & \frac{1}{2} \log \det (\partial_t - \sigma) - \frac{N}{2} \int \d t \sum_{j,k} \left(\lambda   \sigma^{j,k}_{a,0}\,  g^{j,k}_{a,0}  + (1 - \lambda) \sigma^{j,k}_{B,0}  \, g^{j,k}_{B, 0} \right)  \\&  - \frac{N}{2} \int \d t \sum_{u \neq 0} \sigma^{j,k}_{\omega(u),u} \, g^{j,k}_{\omega(u),u}   - \frac{N J}{2 q} \int \d t  \, \sum_{j \neq k} s_j s_k  \;  \sum_{u \neq 0  } \left(  g^{j,k}_{\omega(u), u}  \right)^q  \\&   -   \frac{N J}{2 q} \int \d t  \, \sum_{j \neq k} s_j s_k  \left(\lambda  g^{j,k }_{a,0} + (1-\lambda)  g^{j,k }_{B,0}\right)^q
\\& -\sum_{u \notin \{-1,0\}}\frac{N \jt}{4} \int \d t  \,\sum_{j\neq k}   \, g^{j,k}_{\omega(u),u } \;  g^{j,k}_{\omega(u+1),u+1}  \\& -\frac{N \jt}{4} \int \d t  \,\sum_{j\neq k}   \left( \lambda  g^{j,k }_{a,0} + (1-\lambda)  g^{j,k }_{B,0}\right)  \left(g^{j,k}_{\omega(1), 1} + g^{j,k}_{\omega(-1), -1}\right) ,
\end{align*}
where
\begin{align*}
\nt    \omega(u) &=  B  \quad \text{if} \quad  u\in [-\ell,\ell] \\ &= C  \quad \text{elsewhere}.
\end{align*}

\subsection{The $g, \sigma$ equations} 
In the large $N$ limit (keeping $\lambda$ fixed), the path integral can be evaluated in the saddle point approximation. The classical equations of motion for the $(g, \sigma)$ variables are as follows:
\begin{align*}\label{eq:g_sigma}
 \nt   \partial_t g^{i,j}_{\omega(u), u} &= \sum_k \left( \sigma^{i,k}_{\omega(u),u} g^{k,j}_{\omega(u),u} - g^{i,k}_{\omega(u),u} \sigma^{k,j}_{\omega(u),u} \right) = [\sigma_{\omega(u),u},g_{\omega(u),u}]^{i,j}, \\
    \sigma^{i,j}_{\omega(u),u} &\equiv  \sigma^{i,j}_{u} = - J s_i s_j \,  (g^{i,j}_{u})^{q-1} -  \frac{\jt}{2}\, \left( g^{i,j}_{u-1} + g^{i,j}_{u+1} \right),
\end{align*}
where,\begin{align*}
\nt    g^{i,j}_{0} &= \lambda g^{i,j}_{a, 0 } + (1-\lambda) g^{i,j}_{B, 0 }, \\  
    g^{i,j}_{u\neq 0} &= g^{i,j}_{\omega(u), u},
\end{align*}
are the total correlation functions.

\subsection{Saddle point analysis and background solutions}
\subsubsection*{Warm up: $n=2$}
The saddle point equations can be further simplified for $n = 2$ by noting that the boundary conditions and the time evolution preserves the following equations \cite{Stanford:2021bhl}: 
\begin{align*}\label{eq:symmetry}
\nt    g^{12} = g^{34}, \quad g^{14} = g^{23}, \quad g^{13} = - g^{24},
\end{align*}
where we have only displayed the contour labels. This implies that in contour space, instead of having a $g$-variable for each pair of contours, we have a total of only three variables, which we can choose to be:
\begin{equation}\label{eq:def_xyz}
    x_{\omega, u} \equiv -2i g^{1,2}_{\omega,u} \quad    y_{\omega, u} \equiv 2 g^{1,3}_{\omega,u} \quad  z_{\omega, u} \equiv -2i g^{1,4}_{\omega,u}.
\end{equation}It is convenient to also define the total variables:
\begin{equation}
        x_{ u} = -2i g^{1,2}_{u} \quad    y_{ u} = 2 g^{1,3}_{u} \quad  z_{ u} = -2i g^{1,4}_{u}.\end{equation}
        In terms of these variables, the equations of motion take the following nicer form: 
\begin{align*}
\nt    \partial_t x_{\omega,u} &= \frac{J}{2^{q-2}} \left(z_{\omega,u}  y^{q-1}_u  -  y_{\omega,u}  z^{q-1}_u\right) + \jt \left(z_{\omega,u } \partial_u^2 y_u  - y_{\omega,u}  \partial_u^2 z_u \right), \\  \partial_t y_{\omega, u} &= \frac{J}{2^{q-2}} \left(  z_{\omega,u}x^{q-1}_u  -  x_{\omega,u}z^{q-1}_u  \right) + \jt \left( z_{\omega,u} \partial_u^2 x_u - x_{\omega,u } \partial_u^2 z_u  \right), \\
    \partial_t z_{\omega,u} &= \frac{J}{2^{q-2}} \left(y_{\omega, u} x^{q-1}_u  -x_{\omega,u} y^{q-1}_u \right) + \jt \left( y_{\omega,u} \partial_u^2 x_u - x_{\omega,u} \partial_u^2 y_u \right),
\end{align*}where we have defined the discrete Laplacian on the chain as:
\begin{equation}
 \nt   \partial^2_u f_u \equiv \frac{1}{2} \left( f_{u+1} + f_{u-1} - 2\,  f_u \right).
\end{equation}
In terms of the total variables $(x_u,y_u,z_u)$, these equations become
\begin{align*}\label{eq:diff_xyz}
\nt    \partial_t x_{u} &= \frac{J}{2^{q-2}} \left(z_{u}  y^{q-1}_u  -  y_{u}  z^{q-1}_u\right) + \jt \left(z_{u } \partial_u^2 y_u  - y_{u}  \partial_u^2 z_u \right), \\  \partial_t y_{u} &= \frac{J}{2^{q-2}} \left(  z_{u}x^{q-1}_u  -  x_{u}z^{q-1}_u  \right) + \jt \left( z_{u} \partial_u^2 x_u - x_{u } \partial_u^2 z_u  \right), \\
    \partial_t z_{u} &= \frac{J}{2^{q-2}} \left(y_{ u} x^{q-1}_u  -x_{u} y^{q-1}_u \right) + \jt \left( y_{u} \partial_u^2 x_u - x_{u} \partial_u^2 y_u \right),
\end{align*}
It is worth mentioning that these equations can be cast in the form of Hamilton's equations. Although the above equations are written in terms of three variables, it is easy to check that the quantity 
$$f(x_u,y_u,z_u)=(x_u^2- y_u^2 + z_u^2)$$ is a constant of motion. The Hamiltonian is given by 
\beq \label{eq:effective_hamiltonian}
H = \sum_u \tilde{J}\left[(\pa_u x_u)^2- (\pa_u y_u)^2+(\pa_u z_u)^2\right]+\sum_u \frac{J}{2^{q-2}}\left[(x_u)^q- (y_u)^q+( z_u)^q\right].
\eeq 
We will sometimes think of this system in the continuum limit, which is defined by taking $\sigma = \epsilon\,u$, $\tilde{J} = \frac{1}{\epsilon}\,\tilde{J}_c$ and $J = \epsilon\,J_c$, and sending $\epsilon \to 0$, holding $\tilde{J}_c$ and $J_c$ fixed. In this limit, the Hamiltonian becomes:
\beq 
H = \int d\sigma\, \tilde{J}_c\left[(\pa_\sigma x)^2- (\pa_\sigma y)^2+(\pa_\sigma z)^2\right]+\frac{J_{\text{c}}}{2^{q-2}} \int d\sigma\,\left[x^q- y^q+z^q\right].
\eeq

\subsubsection*{Background Solution for $n=2$}

We will solve the equations perturbatively in the fraction $\lambda$.  At $O(\lambda^0)$, we can ignore $a$. We notice that $B$ and $C$ satisfy the same boundary conditions at $t = 0$. In fact, the initial state of $B \cup C$ is maximally mixed. Therefore, at the leading order in $\lambda$ the initial state does not evolve in time under a unitary time evolution. In terms of the dynamical variables, this means that the initial boundary conditions are satisfied at all times. 
\begin{align*}
\nt      z_C(0) = y_C(0) &\implies z_C(t) = y_C(t) \\
             z_B(0) = y_B(0) &\implies z_B(t) = y_B(t) \\
             x_C(0) = x_B(0) = 1 &\implies x_C(t) = x_B(t) = 1.
 \end{align*}
We can use the above equations to derive explicit boundary conditions. Combining the above boundary conditions with those in equations (\ref{eq:bc_a} - \ref{eq:bc_C}), we have
\begin{align*}
\nt    z_C(T) = -y_C(T), \;  z_C(t) = y_C(t) &\implies z_C(T) = y_C(T) = 0 \\ x_B(T) = y_B(T), \;   z_B(t) = y_B(t),\;  x_B(t) = 1\;  &\implies x_B(T) = y_B(T) = z_B(T) = 1 \\ 
x_C(t) = 1 &\implies x_C(T) = 1 \,.
\end{align*}To summarize, we can solve the differential equations at $O(\lambda^0)$ by setting 
\begin{equation}\label{eq:max_mixed_state}
    x_u(t) = 1, \quad y_u(t) = z_u(t)  
\end{equation}everywhere.  Due to this relation the family of three differential equations reduces to a single first order differential equation: 
\begin{align*}\label{eq:bg_fkpp}
  \nt     \partial_t  y_{u} &= \frac{J}{2^{q-2}} \left( y_{u} - y^{q-1}_u \right) - \jt\,  \partial_u^2 y_{u}, 
\end{align*}with a step function boundary condition at $t=T$:
\begin{align*}\nt \label{eq:bc_fkpp} y_C(T) = 0, \quad y_B(T) = 1.\end{align*}Notice that although the two-point functions $y(t)$ and $z(t)$ depend on $T$ through the above boundary conditions, the total action is independent of $T$ because the initial state, which is maximally mixed, does not evolve with $T$. 

\begin{figure}[t]
    \centering
    \includegraphics[width=0.99\linewidth]{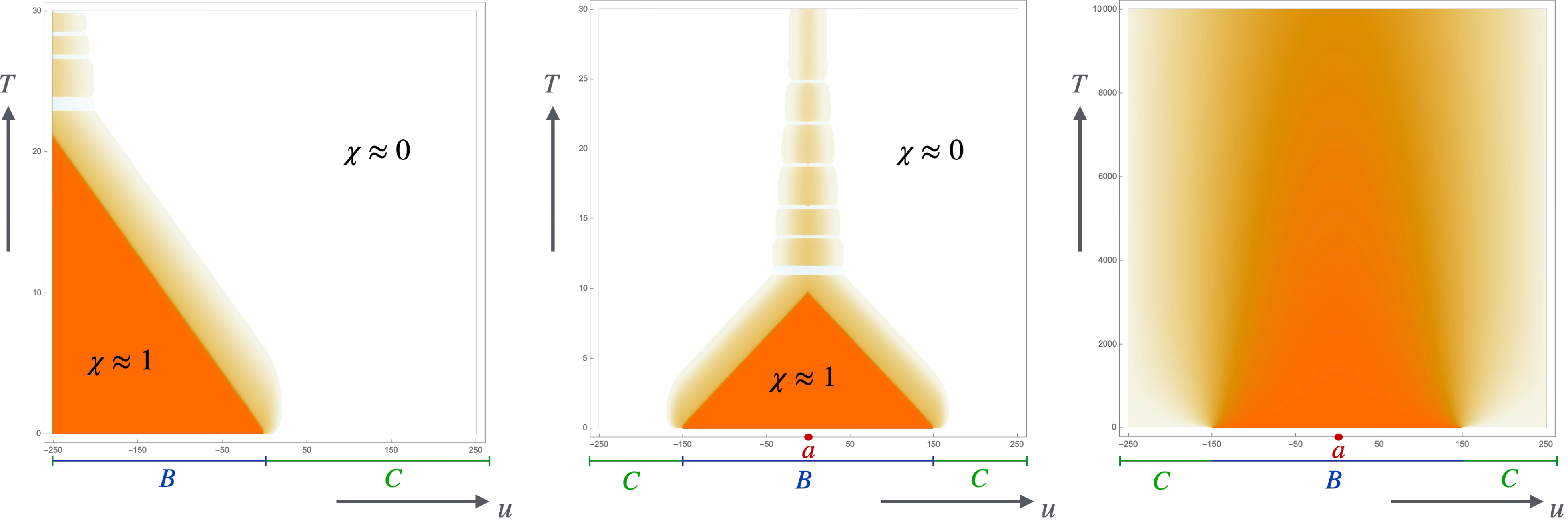}
    \caption{\small{\textbf{Left:} A solution of the FKPP equation  (\ref{eq:bg_fkpp_inverted})} for $\frac{J}{2^{q-2}} = 10$ and $\jt = 1$. The solution is a traveling wave that propagates towards the unstable region with velocity $v_B$. \textbf{Middle:} Solution to the FKPP equation when $B$ is bounded.  \textbf{Right:} Solution of the diffusion equation ($J = 0$).}
    \label{fig:fkpp}
\end{figure}

The coefficient of the diffusion term in the above differential equation is negative. To get the equation in the conventional form, we replace the $y_u(t)$ with $\chi_u(T-t)$. Then $\chi_u(t)$ satisfies the following equation 
\begin{align*} \nt
     \partial_t  \chi_{u} &= -\frac{J}{2^{q-2}} \left( \chi_{u} - \chi^{q-1}_u \right) + \jt\,  \partial_u^2 \chi_{u}, 
\end{align*}
with the following boundary conditions at $T- t = 0$:
\begin{align*}
    \nt  \chi_C(0) = 0, \quad \chi_B(0) = 1.
\end{align*}We will now explain some qualitative features of the solution. For this purpose, it is convenient to take the continuum limit of the differential equation where it takes the following form:
\begin{equation}\label{eq:bg_fkpp_inverted} \partial_t  \chi(t,\sigma) = -\frac{J_c}{2^{q-2}} \left( \chi - \chi^{q-1} \right) + \jt_c\,  \partial_\sigma^2 \chi, 
\end{equation}This equation is an example of a more general class of equations known as the Fisher-Kolmogorov-Petrovksii-Piskunov (FKPP) equation \cite{Fisher, KPP} (see appendix \ref{sec:FKPP} for a brief discussion).  The non-linear term in the differential equation has a stable fixed point at $\chi = 0$ and an unstable fixed point at $\chi =1$. Note that when $J=0$, this equation reduces to the ordinary diffusion equation, and the solution has no sharp features (see the right panel of figure \ref{fig:fkpp}). Remarkably, when $J \neq 0$ and for step-function boundary conditions, it was shown in \cite{Fisher, KPP} that $\chi_u(t)$ approaches a domain wall/traveling wave solution as $ t$ increases (see the left panel of figure \ref{fig:fkpp}). To find the velocity of the traveling wave, it is sufficient to analyze the equation near the unstable fixed point $\chi =1.$ The equation takes the form:
\begin{equation}\label{eq:approx_fkpp}
      \nt     \partial_t  \chi \approx - \frac{J_c (q-2)} {2^{q-2}} \left( 1 - \chi_u\right) + \jt_c\,  \partial_\sigma^2 \chi \,.
\end{equation}The solution of the linearized equation with the boundary conditions in equation \eqref{eq:bc_fkpp} has a front that moves with the velocity $v_B = \sqrt{\frac{\tilde{J}_cJ_c (q-2) }{2^{q-4}}}$, towards the domain with the unstable fixed point i.e. region $B$. The spatial width of the front scales as $\sqrt{\frac{\tilde{J}_c}{J_c}}$. 

When $\chi(t,\sigma)$ approaches a traveling wave solution, we can determine its asymptotic behavior away from the front by using the traveling wave ansatz. Let us assume that region $B$ lies to the left of region $C$. Then $\chi(t,\sigma) \rightarrow \chi(\sigma + v_B t)$ and it satisfies the following equation:\begin{align*}
    \nt  v_B \partial_\sigma \chi = -\frac{J_c}{2^{q-2}} (\chi - \chi^{q-1}) + \jt_c \partial_\sigma^2 \chi
\end{align*}Away from the front, $\chi(t,\sigma)$ is near one of the fixed points. From the boundary conditions, it follows that $\chi(t,\sigma)$ changes from the fixed point $\chi = 0$ in region $C$ to the fixed point $\chi = 1$ in region $B$. In both regions, we can linearize the above equation to find the asymptotic behavior. We find
\begin{align*}\nt
    \chi(\sigma) \sim \begin{cases}
     \alpha_0\,   e^{-\lambda_0 \sigma} \quad \text{for} \quad \sigma \rightarrow \infty \\ 
      \, 1- \alpha_1 e^{\lambda_1 \sigma} \quad\; \text{for} \quad \sigma \rightarrow -\infty\,,
    \end{cases} 
\end{align*}
where \begin{align*}\nt
    \lambda_{0} = \frac{v_B}{2 \jt_c}\left(\sqrt{\frac{q-1}{q-2}} -1\right), \quad \lambda_1 =\frac{v_B}{2\jt_c}\,,
\end{align*}
and $\alpha_{0}, \alpha_1$ are positive numbers. 

In figure \ref{fig:fkpp}, we have shown the solution of equation \eqref{eq:bg_fkpp_inverted} for the case when $B$ is a bounded region on the lattice.  In this case, the solution contains two fronts which originate at the boundaries of region $B$ and move towards each other with velocity $v_B$. The unstable region with $\chi \approx 1$ is bounded between the two fronts. As the fronts approach each other, the unstable region shrinks in size until it completely vanishes at the time when the two fronts meet.

\subsection*{Physical origin of the domain wall solution}
It is worth understanding qualitatively why the FKPP equation has domain wall solutions. The FKPP equation contains two terms (i) an onsite term with an unstable fixed point at  $\chi = 1$ and a stable fixed point at $\chi =0$, and (ii) a diffusive term. Consider, for instance, the boundary condition, $\chi_u(0) = \theta(u)$. At $t = 0$, diffusion lowers the value of $\chi$ at $u=1$ because $\dot{\chi}_1(t)|_{t=0} = -\jt/2$. As $\chi_1(t)$ decreases due to diffusion, the onsite term at $u=1$ becomes important and drags the value of $\chi_1(t)$ at $u=1$ to the stable fixed point $\chi_1=0$. On the other hand, $\chi_0(t)$ always remains close to 0, since it is a stable fixed point. Once $(1 - \chi_1(t)) \sim O(1)$, the diffusive term acts on $\chi_2(t)$, and drags it to the stable fixed point, and so on. This qualitatively explains the existence of a domain wall solutions to the FKPP equation. 

As we will explain later, the physical ingredients which go into the above explanation follow from general properties of operator growth in chaotic quantum systems. Very briefly, imagine a typical operator supported only in the complement region $C$, with no support inside $a \cup B$. Consider the averaged size of such an operator, defined as 
\begin{equation} \phi_u = \frac{1}{N}\langle K_u\rangle, \end{equation}
where $K_u$ is the average number of Majorana fermions that constitute the operator at the site $u$. For a typical operator supported in $C$, $\phi_u \sim 1/2$ at points in $C$, and $\phi_u\sim 0$ at points inside $a\cup B$. As we will show in the next section, the function $\chi_u(t)$ is simply related to $\phi_u$ by the formula:
\beq \label{eq:yvu}
\chi_u(t) = 1-2\phi_u(t).
\eeq 
From this point of view, the domain wall solutions we found above emerge from an underlying light-cone structure in operator growth. Indeed, the unstable fixed point $\chi_u=1$ maps to $\phi_u=0$, which corresponds to the identity operator and the fermion parity operator, while the stable fixed point $\chi_u=0$ maps to $\phi_u=1/2$, which corresponds to the size $\frac{N}{2}$ operators, which are entropically favored. This correspondence makes the physical origins of the domain wall solutions clear -- under time evolution, an operator that is entirely supported in $C$ develops some small non-trivial support at the nearest neighbor sites in $B$ (owing to the presence of kinetic terms in the Hamiltonian), at which point the local chaotic dynamics takes over and drags the operator into the entropically favored sector of size $N/2$ operators -- the time scale for this to happen is of course controlled by the on-site coupling constant. 

\subsubsection*{Background Solution for general $n$}

\begin{figure}
    \centering
    \includegraphics[width=\linewidth]{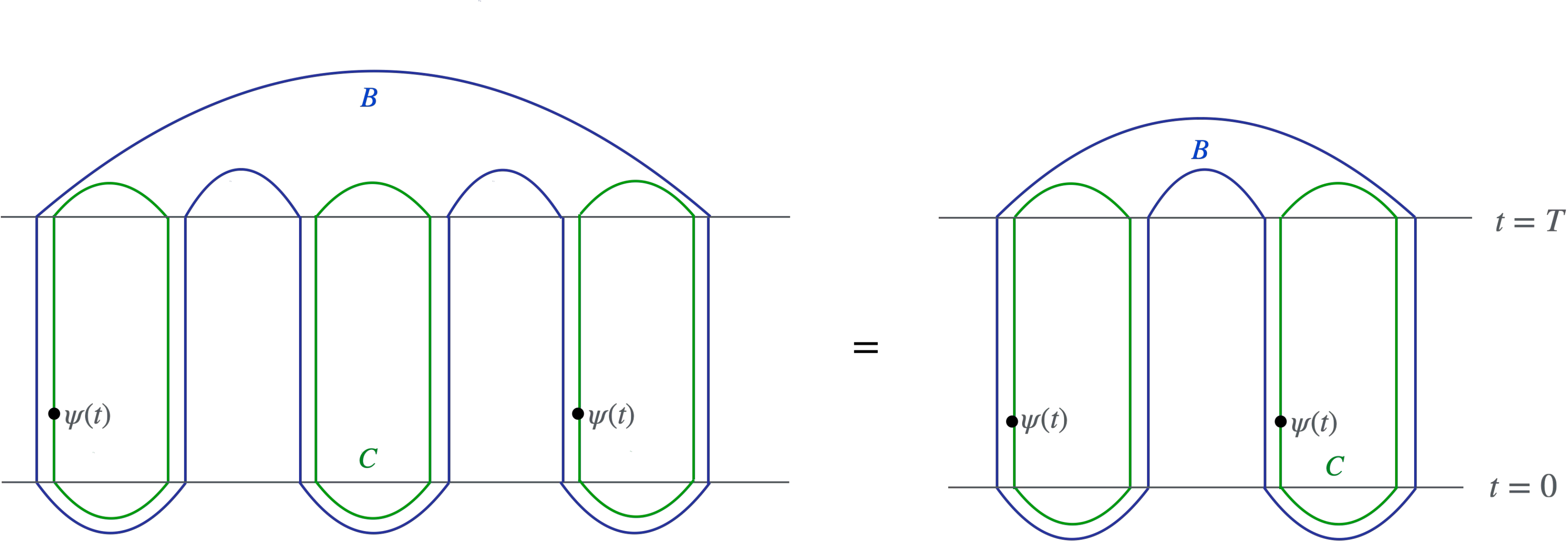}
    \caption{\small{The path integral for $n = 3$ on the left has three legs but since the fermions are inserted only on contours 1 and 5, we can pull the second leg up to $t = T$ without changing the value of the path integral. Therefore the right contour gives the same answer as the left contour.}}
    \label{fig:bkgd_n}
\end{figure}
To compute the $n$th moment $S^{(n)}(\re \cup a \cup B)$, we need to solve the saddle point equations \eqref{eq:g_sigma} for general $n$. For $n > 2$, at first sight it seems hard to solve these equations. However, using the fact that $B\cup C$ is in a maximally mixed state at $t = 0$, we can easily find the background solution for arbitrary $n > 2$ without solving the equations explicitly. To find $g^{ij}(t)$, we must evaluate the path-integral with fermion insertions on contours `$i$' and`$j$' at time $t$.  Since a maximally mixed state evolves trivially in time, any pair of contours $(2k-1,2k)$ with no fermion insertion can be pulled up to $t = T$ without changing the path integral (see figure \ref{fig:bkgd_n}). After we have pulled up all such contours, we get a fermion 2-point function on a contour corresponding to $n =2$. Moreover, the initial boundary conditions in equation (\ref{eq:init_BC}) are satisfied at all times (due to trivial evolution of the maximally mixed state). Using these boundary conditions and the known solution for $n =2$, we get (suppressing the spatial indices)
\begin{align*} \nt 
    g^{2i-1,2j-1} &= -i g^{2i-1, 2j} = - g^{2i,2j} = -i g^{2i, 2j -1} = \frac{y}{2},  \quad \\ 
    g^{2i-1,2i} &= \frac{i}{2} \quad \text{for} \quad 1 \leq i < j \leq  n.
\end{align*}
As a $2n \times 2n$ matrix, $g(t)$ takes the following form:
\begin{equation}
    g(t) = -\frac{1}{2} \sigma_y \otimes I  + \frac{y(t)}{2}\sigma^+ \otimes M, \quad  \sigma^+ = \sigma_z + i\sigma_x, \quad  
    M_{i,j} = \begin{cases}
        &\text{sgn}(j -i) \quad  \text{if} \quad i \neq j \\
        & 0 \quad \text{otherwise}
        \end{cases} \,.
\end{equation}It is straightforward to interpret the two contributions to $g(t)$ in the above sum. The first term is the equal time two-point function between fermions inserted on the forward and backward contour of the same density matrix. The second term is the equal time two-point function between fermions inserted on contours of different density matrices. $M$ is an anti-symmetric matrix related to the ordering of the $n$ copies of the density matrix in the $n$-th R\'enyi entropy.

From the equation for $\sigma(t)$ in (\ref{eq:g_sigma}), we get:
\begin{align*}
    \sigma^{2i-1,2j-1} &= -i \sigma^{2i-1, 2j} = - \sigma^{2i,2j} = -i \sigma^{2i, 2j -1} = \sigma^{1,3},  \quad \\ 
    \sigma^{2i-1,2i} &= \frac{-i}{2} \jeff \quad \text{for} \quad 1 \leq i < j \leq  n, \qquad J_\text{eff} =  \left(\frac{J}{2^{q-2}} + \jt\right)\,.
\end{align*}
In matrix representation, we can write $\sigma(t)$ as:
\begin{equation}
    \sigma(t) = \frac{\jeff}{2} \sigma_{y} \otimes I + \sigma^{1,3}(t) \sigma^+ \otimes M.
\end{equation}
To summarize, the background saddle point solution for general $n$ can be obtained once again in terms of one function $y_u(t)$, which satisfies the same FKPP equation with step-function boundary conditions at $t=T$. For completeness, we re-write the equation of motion of $y(t)$:
\begin{equation}
    \frac{\d }{\d t} y(t) = \jeff\,  y(t) + 2 \sigma^{1,3}(t)
\end{equation}

\subsection{Saddle point action at $O(\lambda)$}

Let us consider the total action upto $O(\lambda)$.  For general $n$, this is given in (\ref{eq:g_sigma_action}). At $O(\lambda^0)$, only $B\cup C$ contributes to the action. Since the state of $B\cup C$ is time-independent at this order, their contribution to the action is independent of $T$. To evaluate the action at $O(\lambda)$, we vary the action w.r.t. $T$ and use the Hamilton-Jacobi equation. 
\begin{equation}
     \frac{\d S}{\d T} = -H.
\end{equation}
At $O(\lambda)$, the non-zero contribution to the Hamiltonian comes from the difference $ g^{i,j}_{a,0} - g^{i,j}_{B,0}$ at site $u = 0$ where $a$ was injected. Therefore, we have:
\begin{align*}\label{eq:h_variation}
    \frac{-2 H}{N \lambda } &= - \sum_{i,j} (g^{i,j}_{a} - g^{i,j}_{B,0})\sigma^{i,j}_{B,0}  = \Tr(\sigma (g_{a} - g_{B,0})) \\
    & =  \frac{\jeff}{2} \Tr( \sigma_y \otimes I\, (g_{a} - g_{B,0})) + \sigma^{1,3} \Tr( \sigma^+ \otimes M (g_{a} - g_{B,0})) \\ &= \frac{\jeff}{2} \Tr( \sigma_y \otimes I\, g_{a}) + \frac{\jeff}{2} \Tr(\frac{1}{2}\sigma_y^2 \otimes I )+ \sigma^{1,3} \Tr( \sigma^+ \otimes M g_{a}) \\  &= \frac{\jeff}{2}  \left( \Tr( \sigma_y \otimes I\, g_{a}) + n \right) + \sigma^{1,3} \Tr( \sigma^+ \otimes M g_{a}) 
\end{align*}
We will find it convenient to define the following variables: 
\begin{equation}
    u_k(t) = \Tr(\sigma_y \otimes M^{2k} g_a(t)), \quad  v_k(t) = \Tr(\sigma^+ \otimes M^{2k+1} g_a(t)) 
\end{equation}As we show below equation \eqref{eq:pert_g_sigma}, the  time evolution of $g_a(t)$ in these variables has a simple description.
In terms of these variables, the Hamiltonian is
\begin{equation}
    H = -\frac{N \lambda}{2} \left(\frac{\jeff}{2} (u_0(0) +n) + \sigma^{1,3}(0) v_0(0)\right) + O(\lambda^2)  \,.
\end{equation}The above expression of the Hamiltonian contains an explicit factor of $\lambda$. Therefore, it is sufficient to solve the saddle point equations in $a$ at $O(\lambda^0)$ to determine the Hamiltonian at $O(\lambda)$. Note that the saddle point solution in $B\cup C$ also changes at $O(\lambda)$ but it contributes to the Hamiltonian at $O(\lambda^2)$. Hence, we will not compute those corrections in this paper.

\subsubsection*{Boundary Conditions and saddle point equations for $a$}
The boundary conditions for $u_k, v_k$ at $t = 0$ and $t= T$ follow from (\ref{eq:init_a})
and (\ref{eq:final_a}).  In terms of $g^{i,j}_{a}$, the boundary conditions are:
\begin{align*}
&g_{a}^{2k, 2k+1}(t) = g_{a}^{1,2n}(t) = \frac{i}{2}  &&\hspace{-1.2cm}\text{for} \quad 1\leq k < n, \quad t = 0, T \\ 
&g_{a}^{j, 2k}(0) + i g_{a}^{j, 2k+1}(0) = 0,  \quad &&g_{a}^{j, 2k}(T) - i g_{a}^{j, 2k+1}(T) = 0 \\
&g_{a}^{1, j}(0) + i g_{a}^{2n, j}(0) = 0,  \quad &&g_{a}^{1, j}(T) - i g_{a}^{2n, j}(T) = 0 
\end{align*}
In appendix \ref{matrix_bc}, we show that they imply the following boundary conditions on $u_k,v_k$:\begin{align*} \nt \label{eq:u_v_bdy} 
v_{k-1}(0) &- u_{k}(0) =  u_k(T) - v_k(T) =  \alpha_k
\end{align*}
where \begin{equation} \alpha_k = - \Tr(M^{2k}) \end{equation}

\begin{figure}
    \centering
\includegraphics[width=0.49\linewidth]{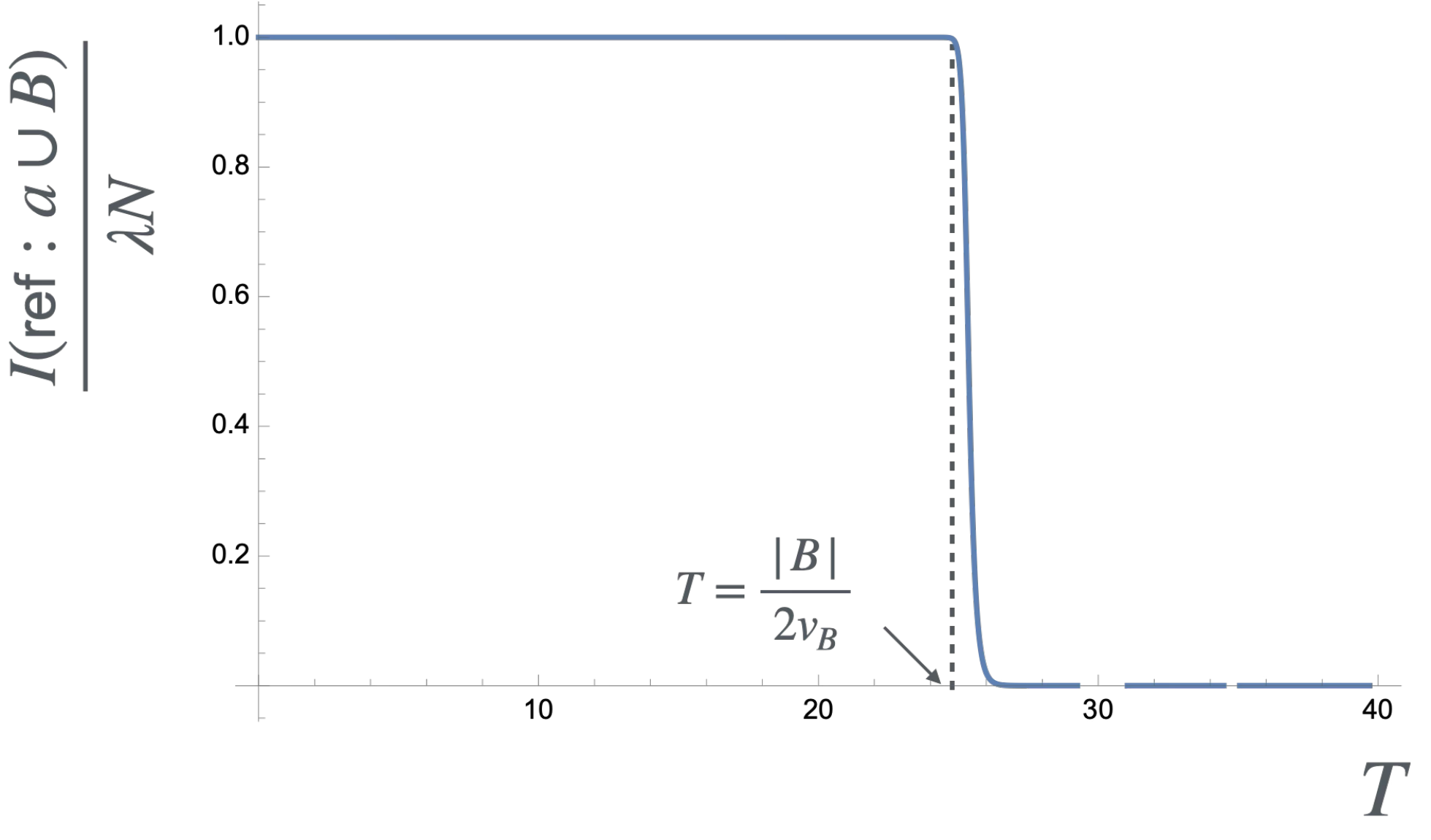} 
\includegraphics[width=0.49\linewidth]{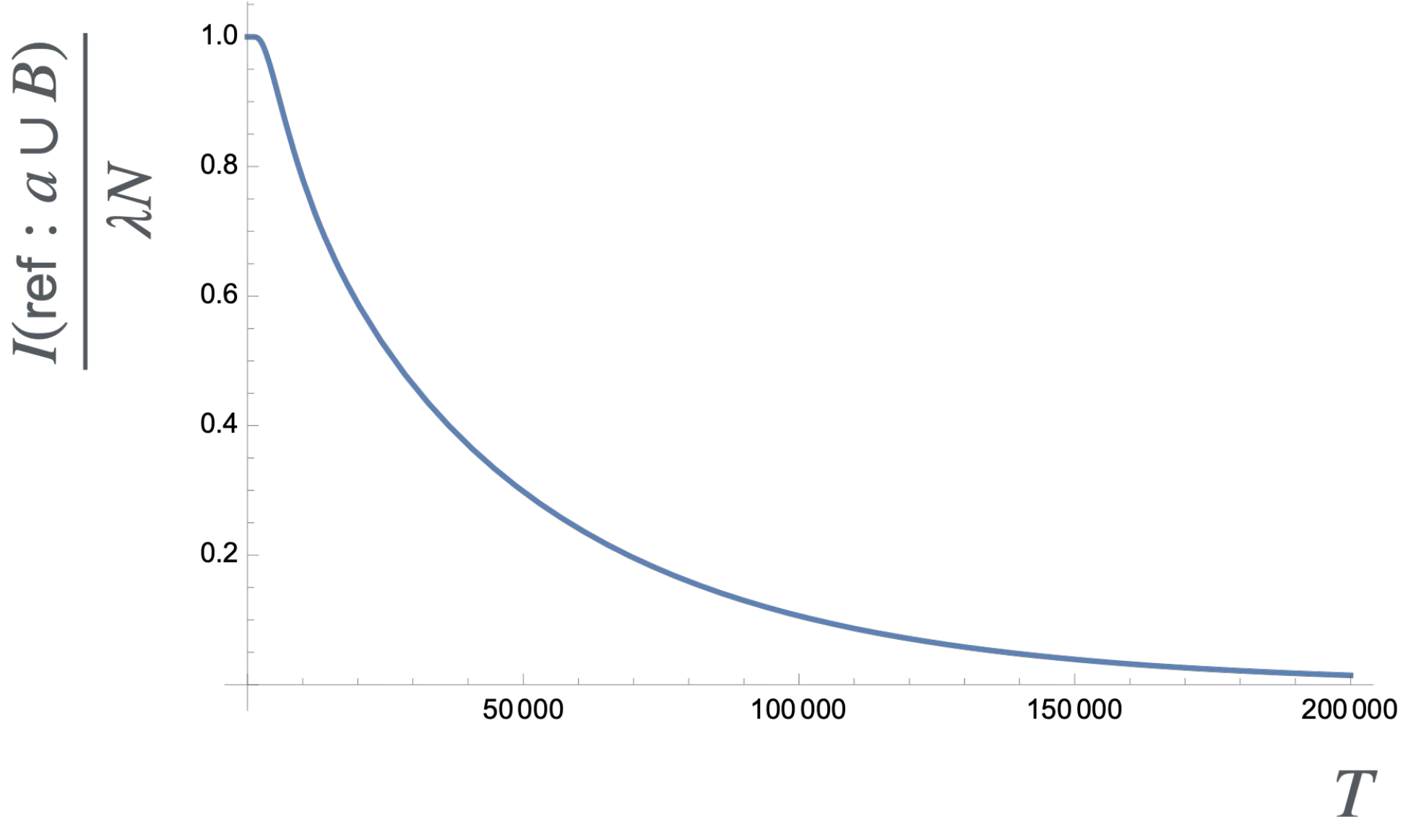}
    \caption{Mutual Information $I(\re: a \cup B)$ for the example shown in Figure \ref{fig:fkpp}. \textbf{Left:} For $\frac{J}{2^{q-2}} = 10$ the mutual information sharply goes to 0 at $T = \frac{|B|}{2 v_B}$. \textbf{Right:} $ J =0$ the mutual information saturates to a small non-zero value at large $T$.    }
    \label{fig:mutual_information}
\end{figure}

\begin{figure}
    \centering
\includegraphics[width=0.6\linewidth]{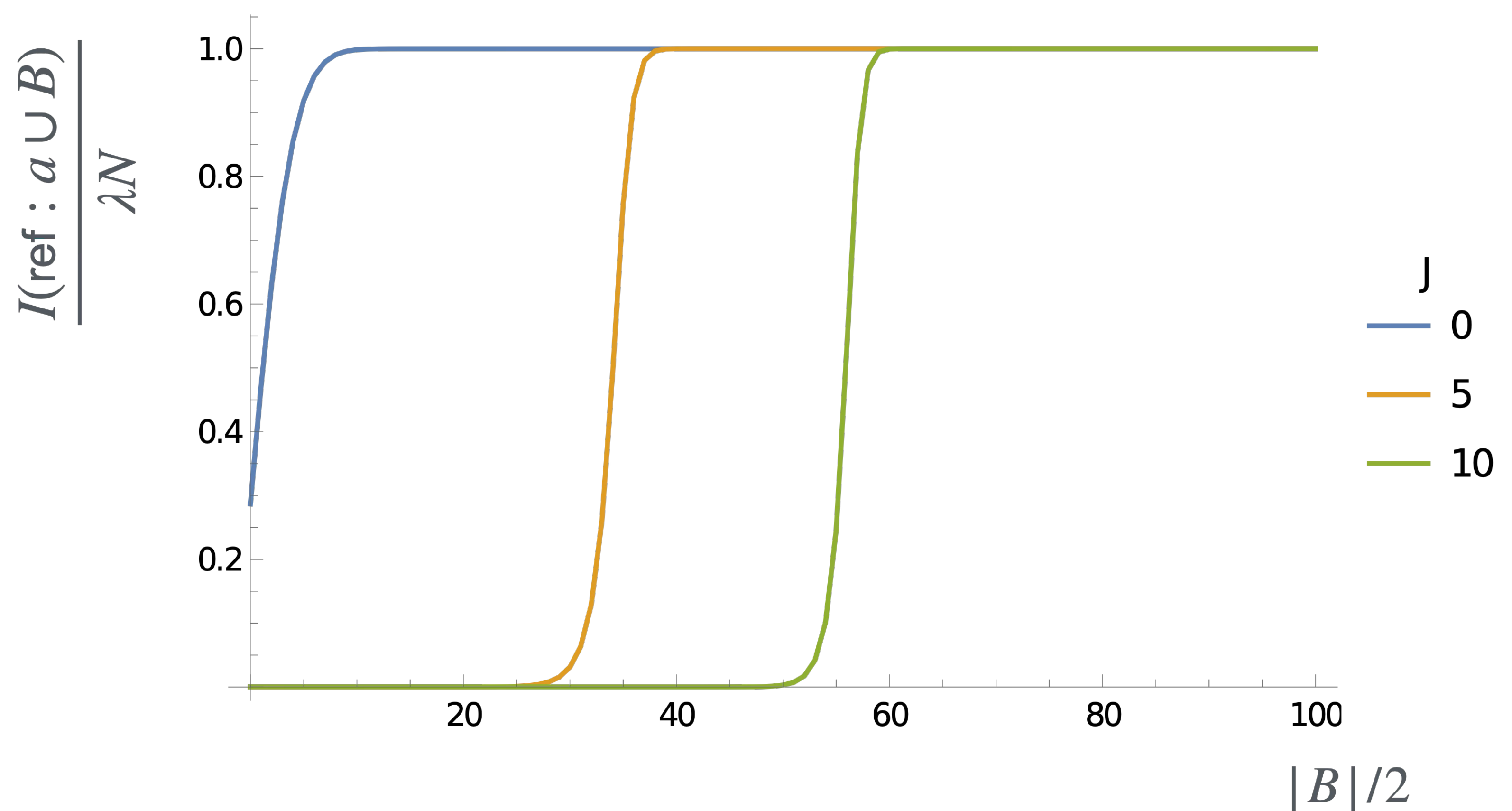}
    \caption{$I(\re: a \cup B)$ as a function of the $|B|$ at a fixed time $T = 10$. Notice that for $J=0$ (diffusive case), the mutual information starts with $O(1)$ value while it is almost zero when $J$ is non-zero.}
\label{fig:mutual_information_B_dep}
\end{figure}

To evaluate the Hamiltonian, we must solve the differential equations for $g_a(t)$. $g_a(t)$ satisfies the following differential equation:\begin{align*} \label{eq:pert_g_sigma}
    \nt \frac{\d}{\d t} g_a(t) &= [\sigma(t),g_a(t)] \\ 
  &= \frac{\jeff}{2} [\sigma_{y} \otimes I, g_a(t) ] +\sigma^{1,3}(t)  [\sigma^+ \otimes M, g_a(t)]\,.
\end{align*}This is a first order linear differential equation analogous to the Heisenberg's equation of operator evolution by $\sigma(t)$. It is convenient to solve the equation in a basis where the action of $\sigma(t)$ on $g_a(t)$ is block diagonal.  This basis is generated by $u_k(t)$ and $v_k(t)$: \begin{align*}\nt
\frac{\d}{\d t} u_k(t) &= \Tr(\sigma_y \otimes M^{2k} [\sigma(t), g_a(t)])  \\ &= \sigma^{1,3}(t)\Tr(\sigma_y \otimes M^{2k} [\sigma^+ \otimes M, g_a(t)])  \\ &= \sigma^{1,3}(t)\Tr([\sigma_y \otimes M^{2k}, \sigma^+ \otimes M], g_a(t))  \\&=  \,2\sigma^{1,3}(t)v_k(t)  \\ \nt
\frac{\d}{\d t} v_k(t) &= \Tr(\sigma^+ \otimes M^{2k+1} [\sigma(t), g_a(t)]) \\ &= \frac{\jeff}{2} \Tr(\sigma^+ \otimes M [\sigma_y \otimes I, g_a(t)]) \\&= - \jeff \, v_k(t) \\ \implies v_k(t) &= e^{-\jeff t}v_k(0)
\end{align*}Thus we get a compact representation for the equation of motion of $g_a(t)$ when expressed in terms of $u_k(t)$ and $v_k(t)$:
\begin{align*} \nt
  \frac{\d}{\d t}  \begin{pmatrix}
        u_k(t) \\ v_k(t) 
    \end{pmatrix} = \begin{pmatrix}
        0 & 2 \sigma^{1,3} \\ 0 & -\jeff 
    \end{pmatrix} \begin{pmatrix} u_k{(t)} \\ v_k(t)\end{pmatrix}
\end{align*} It is straightforward to solve the above equations. Inserting the solution of $v_k(t)$ to the equation of $u_k(t)$, we get:\begin{align*}\nt \label{eq:u_v relation}
    u_k(t) - u_k(0)  &=  v_k(0)  \int_0^t \d t \, 2 \sigma^{1,3}(t) e^{-\jeff t} \\ &= v_k(0) \left( y_0(t) e^{-\jeff t} - y_0(0) \right)
\end{align*}
In particular, the saddle point equations imply:
\begin{align*}
    u_k(T) - u_k(0) &= v_k(0) \left( y_0(T) e^{-\jeff T} - y_0(0) \right) = v_k(0) \left(e^{-\jeff T} - y_0(0) \right)  \\ v_k(T) &= e^{-\jeff  T} v_k(0)
\end{align*}
We can combine the two equations in (\ref{eq:u_v_bdy}) and the above equations to get the following recursion relation for $v_k(0)$:
\begin{align*}
  u_{k}(T) =   v_k(0) e^{-\jeff T}  + \alpha_k &= v_{k-1}(0) - \alpha_k + ( e^{-\jeff T} - y_0(0)) v_{k}(0) \\
    \implies v_k(0) &= \frac{v_{k-1}(0)- 2 \alpha_k}{y_0(0)} 
 \end{align*}Solving the recursion relation, we find:  \begin{equation}\label{eq:sol_rec}
     v_m(0) = \frac{v_0(0)}{y_0^m} - 2 \sum_{r = 0}^{m-1} \frac{\alpha_{m-r}}{y_0^{r+1}}
 \end{equation}
Since $M$ is an $n\times n$ matrix, it satisfies a characteristic equation. This implies a linear relation among $v_k(0)$. $M$ has the following characteristic equation: \begin{align*} \nt 
 M^{ n } + \sum_{m = 1}^{l} \beta_{l-m} M^{n- 2m} &= 0,\quad l = \lfloor n/2 \rfloor, 
\quad  \beta_{l -m } = \binom{ n }{2 m }
\end{align*}
This implies the following relation among the variables $v_k$: 
\begin{equation}
   \frac{v_0(0)}{y^l} - 2 \sum_{r = 0}^{l-1} \frac{\alpha_{l-r}}{y^{r+1}} =v_l(0) = -\sum_{m = 0}^{l-1} \beta_m v_m(0) 
\end{equation}
Inserting the relation (\ref{eq:sol_rec}) to the above equation, we can solve for $v_0(0)$:
\begin{align*} \nt
   \qquad  v_0 \left( \sum_{m= 0}^{l-1} \frac{\beta_m}{y_0^m} + \frac{1}{y_0^l}\right) &= 2 \left( \sum_{m =1}^{l-1} \beta_m \sum_{r = 0}^{m-1} \frac{\alpha_{m-r}}{ y_0^{r+1}} + \sum_{m =0}^{l-1} \frac{\alpha_{l-r}}{y_0^{r+1}}\right) \\ \implies 
    v_0(0)  &= 2 \frac{\left( \alpha_1 + \sum_{r = 0}^{l-2} \left( \alpha_{l-r}  + \sum_{m = r+1}^{l-1} \beta_m \alpha_{m-r}\right) y_0^{l-r-1}\right)}{\left( 1 + \sum_{m = 0}^{l-1} \beta_m y_0^{l-m}\right)} \\ & = 4\frac{\d}{\d y_0} \log \left( 1 +\sum_{m = 0}^{l-1} \beta_m y_0^{l-m} \right)
\end{align*}
In the third step, we have used the following relation satisfied by $\alpha_k$ : 
\begin{equation}\label{eq:matrix_rel}
 \sum_{m= r+1}^{l} \binom{n}{2(l - m)} \alpha_{m-r} =   2 (l-r) \binom{n}{2(l - r)}
\end{equation}
A partial proof of this relation for even $n$ is given in appendix (\ref{app:matrix_rel}). We do not have a complete proof of the relation, but we have verified it for a few different values of $n$ in the notebook \cite{TraceIdentity}.
To evaluate the Hamiltonian at $O(\lambda)$, we need $u_0(0)$:
\begin{align*}\nt
     u_0(0)  &= u_0(T) - \left( e^{-\jeff T} - y_0(0)\right)\, v_0(0) \\
     &= ( v_0(T) + \alpha_0) - (e^{-\jeff T} - y_0(0)) v_0(0) \\ 
     &=  ( v_0(0)e^{-\jeff T} + \alpha_0) - (e^{-\jeff T} - y_0(0)) v_0(0)\\ 
     &= y_0(0) v_0(0) + \alpha_0 \\ &= y_0(0) v_0(0) - n 
 \end{align*}
 Now, we can evaluate the Hamiltonian:  \begin{align*} \nt
     H
     &= - \frac{N\lambda}{2}\left( \frac{\jeff}{2} \left( u_0(0) + n\right) + \sigma^{1,3} v_0(0) \right) \\ 
     &= -\frac{N \lambda v_0(0)}{2} \left( \frac{\jeff}{2} y_0(0)  + \sigma^{1,3}\right)  \\ &= \frac{ N \lambda v_0(0)}{4}\times \frac{-\d }{\d t} y_0(0)\\ &=  N \lambda \,  \partial_{y_0(0)} \log(1 + \sum_{m = 0}^{l-1} \beta_m y_0^{l -m}(0)) \times \frac{\d }{\d T} y_0(0) \\&  =  N \lambda\,  \frac{\d }{\d T} \log(1 + \sum_{m = 0}^{l-1} \beta_m y_0^{l -m}(0)) 
 \end{align*}The total action is:\begin{equation}
 - S = - (n-1)  \left( \log \dim \cH_{B}  + N \lambda \log 2  \right) +   N \lambda \log(1 + \sum_{m = 0}^{l-1} \beta_m y^{l -m}(0)) 
\end{equation}We can re-write the expression in the logarithm as: 
\begin{align*}\nt
    1 + \sum_{m =0}^{l-1} \beta_m y^{l-m}(0)& = 2^{n-1} \left( \left( \frac{1 + \sqrt{y_0(0)}}{2} \right)^n + \left(  \frac{ 1 - \sqrt{y_0(0)}}{2} \right)^n\right) \\&= 2^{n-1} \left( \left( \frac{1 + \sqrt{\chi_0(T)}}{2} \right)^n + \left(  \frac{ 1 - \sqrt{\chi_0(T)}}{2} \right)^n\right)
\end{align*}
We have the final expression for the total action:
\begin{align*}\nt
    -S = -(n-1) \log \dim \cH_B \, + \lambda  N \log\left( \left( \frac{1 + \sqrt{\chi_0(T)}}{2} \right)^n + \left(  \frac{ 1 - \sqrt{\chi_0(T)}}{2} \right)^n\right)
\end{align*} 
Thus, the $n$th R\'enyi entropy is given by:
\begin{equation}
    S_\Psi^{(n)}(\re \cup a \cup B) = \log \dim \cH_B - \frac{\lambda N}{n-1}  \log \left( \left( \frac{1 + \sqrt{\chi_0(T)}}{2} \right)^n + \left(  \frac{ 1 - \sqrt{\chi_0(T)}}{2} \right)^n\right).
\end{equation}
From the R\'enyi entropies, we can obtain the entanglement entropy by taking the $n\to 1$ limit, and we get:
\begin{equation}\label{EF}
       S_\Psi(\re \cup a \cup B) = \log \dim \cH_B +  \lambda N H\left(\frac{1 + \sqrt{\chi_0(T)}}{2}\right),
\end{equation}
where, $H(p)$ is the Shannon entropy:
\begin{equation}
    H(p) = - p \log p - (1-p) \log (1- p),
\end{equation}
and recall that $\chi_u(T)$ is the solution to the FKPP equation \eqref{eq:bg_fkpp_inverted} evaluated at $t=0$ and $u=0$ (i.e., the time and place at which the external qudit is injected) with step function boundary conditions at $t=T$. Going back to equations \eqref{MIRC} and \eqref{MIRC2}, we can now evaluate the mutual information:
\begin{eqnarray}
    I_{\Psi}(\text{ref}:a\cup B) &=& 2\log d - \lambda N H\left(\frac{1 + \sqrt{\chi_0(T)}}{2}\right),\nonumber\\
    I_{\Psi}(\text{ref}:L\cup C) &=& \lambda N H\left(\frac{1 + \sqrt{\chi_0(T)}}{2}\right).
\end{eqnarray}
The physical interpretation of these formulas is particularly simple and pleasing at strong coupling (large $J$): in this case, the solution to the FKPP equation has a sharp emergent light cone structure (see figure \ref{fig:fkpp}) where $\chi \sim 1$ inside the light-cone and $\chi\sim 0$ outside the light-cone. When the site at which the external qudit is injected lies within this light-cone, it does not contribute to the entropy of $\text{ref} \cup a \cup B$; intuitively, this is because all the information of $a$ is contained inside $a \cup B$ in this case, and so $\text{ref}\cup a \cup B$ together constitute a pure state. Correspondingly, the mutual information $I_{\Psi}(\text{ref}:L\cup C)$ vanishes, and the qudit is protected against the erasure of $L\cup C$.  On the other hand, when the injected qudit lies outside the light-cone, then it contributes $\log(d)$ to the entropy of $\text{ref} \cup a \cup B$; in this case, most of the information of the qudit has leaked out to $C$ and it is no longer reconstructible inside $a \cup B$. It is tempting to make an analogy between the entropy formula equation \eqref{EF} and the Ryu-Takayanagi formula -- the $\log \text{dim}\,\mathcal{H}_B$ term is analogous to the classical ``area'' term in the RT formula, while the contribution from the qudit (depending on whether it lies inside or outside the light-cone) is analogous to the FLM bulk correction, with the emergent FKPP light-cone playing the role of the entanglement wedge.\footnote{A note of caution: the analogy with FLM should perhaps be interpreted with care, because in our context the qudit contribution also scales with $N$ and comes from a saddle point calculation. We thank Mark Mezei for pointing this out.} Indeed, this structure for the entropy was expected from the quantum error correction picture of \cite{Harlow:2016vwg}, but the interesting point is the emergence of a sharp light-cone at strong coupling within which the information of the external qudit spreads and is protected against errors in the complement (see figures \ref{fig:mutual_information} and \ref{fig:mutual_information_B_dep}). 


We end this section with a few remarks:
\begin{enumerate}
\item As noted above, at strong coupling the mutual information $I_{\Psi}(\text{ref}:a\cup B)$ drops sharply once the time interval $T$ becomes ever so slightly larger than the critical value of $ \ell/v_B$ and all the information leaks into the complement (see figure \ref{fig:mutual_information}). This behavior is reminiscent of a holographic theory, where the sharp drop in the mutual information happens when the bulk probe particle crosses the RT surface and leaves the entanglement wedge of $a \cup B$. It is also somewhat analogous to what happens in the Hayden-Preskill setup \cite{Hayden:2007cs} (see also \cite{Chandrasekaran:2021tkb, Chandrasekaran:2022qmq}). 

\item Our derivation of entanglement entropy as a function of the variable $\chi_u(T-t)$ is valid for a general Brownian SYK Hamiltonian. The derivation only depends on the fact that the effective action obtained by averaging over the couplings splits as a sum over functions of the fermion bi-linear variables $g^{i,j}$. 

\item Although we derived the entanglement entropy assuming that subsystem $a$ contains a small fraction $\lambda$ of Majorana fermions at site $u = 0$, our derivation can be extended to subsystems containing small fraction of fermions from multiple sites. If $a$ contains $\lambda_u N$ Majorana fermions from site $u$, the single site result easily extends to the following result. 
\begin{align*}\label{EF2}\nt
        S_\Psi^{(n)}(\re \cup a \cup B) &= \log \dim \cH_B - \sum_{u }\frac{ \lambda_u N}{n-1}  \log \left( \left( \frac{1 + \sqrt{\chi_u(T)}}{2} \right)^n + \left(  \frac{ 1 - \sqrt{\chi_u(T)}}{2} \right)^n\right) ,
\end{align*}as long as 
\begin{equation}
   \sum_u{\lambda_u} = \lambda \ll 1 .
\end{equation}
This is due to the fact that the R\'enyi entropy due to fermions is additive at $O(\lambda)$. 

\item In this paper, we have only considered the $O(\lambda)$ contribution to the mutual information $I_{\Psi}(\text{ref}:L\cup C)$. At $O(\lambda^2)$, the background solution in $B\cup C$ receives back-reaction from $a$. Such back-reaction will affect the growth of the mutual information at early times. However, for $T \gg 1/J$,  we expect that the information contained in $a$ spreads on the chain with the butterfly velocity $v_B$. This is because any operator in $a$ spreads on the lattice with the butterfly velocity at late enought times.

\end{enumerate}

\section{Operator growth and entropy inequalities}\label{sec:OGEE}
In this section, we will reinterpret the FKPP equation for $y_u(t)$ as an operator growth equation (see \cite{Xu_2019, PhysRevB.107.014201} for previous work on the connection between FKPP and operator growth), and show that the same velocity $v_B$ controls the spread of entanglement and operator growth in the Brownian SYK chain. We will also check that the formula we derived for the entanglement entropy of $S_{\Psi}(\text{ref}\cup a \cup B)$ satisfies subadditivity and strong-subadditivity, and discuss consequences for operator growth.
\subsection{Operator growth}
\begin{figure}
    \centering
    \includegraphics[width=0.45\linewidth]{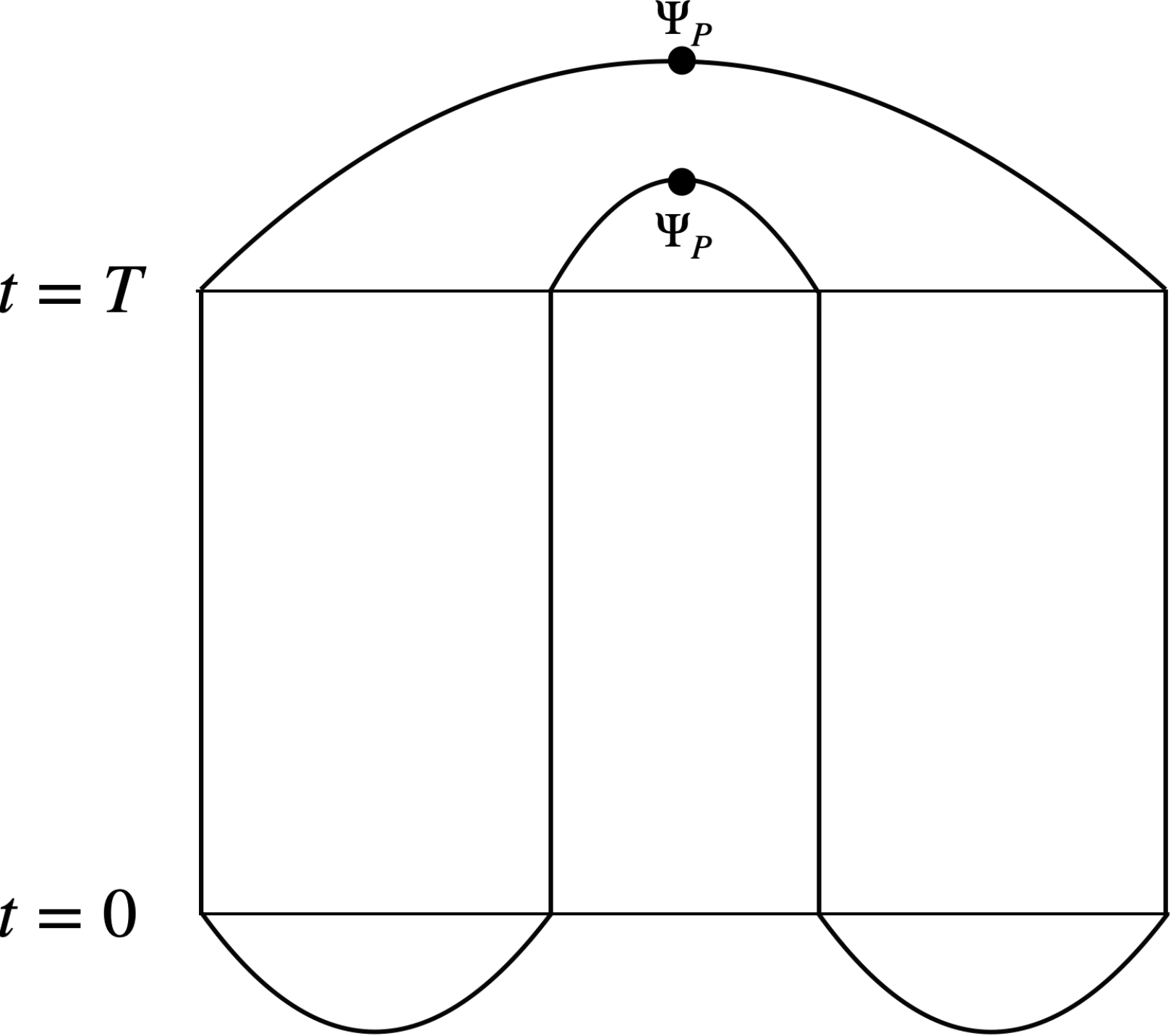}
    \caption{Path Integral for Operator Growth}
    \label{fig:operator_growth}
\end{figure}

Consider a path integral representation of the following trivial quantity:
\begin{equation}\label{eq:path_integral_operator}
1 =\langle \Tr (\Psi_P^2) \rangle     =  \langle \Tr ( \Psi_P U(T)   U^\dagger(T) \Psi_P  U(T) U^\dagger(T)  \rangle 
\end{equation}
where $\Psi_P$ is some product of Majorana fermions belonging to $a \cup B \cup C $ and the angular brackets denote the average over random couplings. A path integral representation of the quantity on the right hand side involves four contours. The corresponding contour is shown in figure (\ref{fig:operator_growth}).  From the contour representation, it is clear that the initial state $\ket{\Gamma_i}$ satisfies the initial conditions in equation (\ref{eq:init_BC})
\begin{equation}
   \psi_\omega^{1} \ket{\Gamma_i}  = -i \psi_\omega^{2} \ket{\Gamma_i}, \;\;\psi_\omega^{3} \ket{\Gamma_i}  = -i \psi_\omega^{4} \ket{\Gamma_i}, \;\; \cdots \;\;  \omega \in (a, B,C).
\end{equation}
 In terms of the $g^{i,j}$ variables defined in the previous section, this would mean that the initial condition has the same symmetries as in equation (\ref{eq:symmetry}): \begin{align*}\label{eq:symmetry_rep}
\nt    g^{12} = g^{34}, \quad g^{14} = g^{23}, \quad g^{13} = - g^{24}.
\end{align*}
Since the averaged dynamics preserves this symmetry, the action of the above path integral can again be written in terms of the $x,y,z$ variables defined in equation (\ref{eq:def_xyz}) and these variables satisfy the differential equations in (\ref{eq:diff_xyz}). Due to the symmetry (\ref{eq:symmetry_rep}), we can set $x = 1$ and $y = z$ everywhere. This leads us to the same differential equation as (\ref{eq:bg_fkpp}):
\begin{equation}
    \partial_t y_u = \frac{J}{2^{q-2}} ( y_u - y_u^{q-1}) - \jt \partial_u^2 y_u.
\end{equation} 
For the path integral in (\ref{eq:path_integral_operator}), $\langle y_u(T-t) \rangle$ is related to the size of $\Psi_P$ at site $u$ when it is evolved backwards by time $t$.
\begin{align*}\nt \label{eq:op_interpretation}
\langle y_u(T-t) \rangle     &=  \frac{1}{N} \sum_{i = 1}^N \langle \Tr \left( \Psi_P U(t) \psi_{i,u} U^\dagger(t)  \Psi_P U(t) \psi_{i,u}  U^\dagger(t)  \right)\rangle  \\ &=\frac{1}{N} \sum_{i = 1}^N \langle \Tr \left(U^\dagger(t)  \Psi_P U(t) \psi_{i,u} U^\dagger(t)  \Psi_P U(t) \psi_{i,u}    \right)\rangle .
\end{align*}
Since the set of all possible products of Majorana fermions form a basis of operators acting on $\cH$, we can consider expanding $U^\dagger(t) \Psi_P U(t)$ in that basis. 
\begin{equation}
    U^\dagger(t) \Psi_P U(t) = \sum_K c_K(t) \frac{\Psi_K}{\sqrt{\dim \cH}},
\end{equation}
where $K$ runs over all possible products of Majorana Fermions. Inserting this equation in the relation (\ref{eq:op_interpretation}), we obtain:
\begin{align*}\nt\label{op_interpretation}
\langle y_u(T- t) \rangle    &=  \sum_{K} \langle |c_K(t)|^2 \rangle  \frac{1}{N} \sum_{i = 1}^N \frac{\Tr \left(\Psi_K \psi_{i,u} \Psi_K \psi_{i,u} \right)}{\dim \cH}  \\ &= (-1)^{|P|}   \sum_{K} \langle |c_K(t)|^2 \rangle  \left( 1 - \frac{2 |K_u|}{N} \right) \\&= (-1)^{|P|} \left(1 -2\phi_u(t)\right),
\end{align*}
where we have defined 
\begin{equation}
    \phi_u(t) = \frac{\langle |P_u|(t)\rangle}{N} = \frac{1}{N}\sum_{K} \langle |c_K(t)|^2 \rangle |K_u|,
\end{equation}
as the average size of fermions at position $u$ that contribute to $\Psi_P(t)$. From the equation of $\langle y_u(t) \rangle$, we get:
\begin{equation}
    \partial_t \phi_u(t) = \frac{J}{2^{q-1}} (1 - 2 \phi_u(t)) \left( 1 - (1 - 2\phi_u(t))^{q-2}\right) +  \jt \partial_u^2 \phi_u(t).
\end{equation}
In particular for $q = 4$, the equation is
\begin{equation}\label{eq:operator_growth}
    \partial_t \phi_u(t) = J \phi_u(t) (\frac{1}{2} - \phi_u(t) ) (1 - \phi_u(t)) + \jt \partial_u^2 \phi_u(t).
\end{equation}
It is easy to interpret the origin of the fixed points: $\phi_u = 0$ corresponds to the identity operator whose time evolution is trivial, while $\phi_u = 1$  corresponds to the product of all fermions at all the sites (which is the parity operator).  Since the Hamiltonian is bosonic, the parity operator is conserved. The fixed point $\phi_u = \frac{1}{2}$ appears because at large $N$,  operators of size $m_u = N/2$ are most favorable entropically. 

If $a \cup B$ contains all fermions in the interval $[x_L,x_R]$, then using equations \eqref{EF} and \eqref{EF2}, we can express $S_\Psi(\re \cup a \cup B)$ and the mutual information $I_{\Psi}(\text{ref}: L\cup C)$ as a function of the operator size variable: 
\begin{equation}\label{eq:mutinf_opgrowth}
       S_\Psi(\re \cup a \cup B) = \log \dim \cH_B + \sum_{u} N\lambda_u  h\left(\phi_u(T,x_L,x_R)\right),
\end{equation}
\begin{equation} \label{eq:mutinf_opgrowth2}
I_{\Psi}(\text{ref}: L\cup C) = \sum_{u } N \lambda_u  h\left(\phi_u(T,x_L,x_R)\right),
\end{equation}
where
\begin{equation}
    h(\phi) = H\left( \frac{1 + \sqrt{1 - 2 \phi}}{2}\right).
\end{equation}
Here, $\phi_u(T,x_L,x_R)$ is the size of an operator at site $ u $ at time $t = T$, that evolves in time following the equation (\ref{eq:operator_growth}) and the initial condition:
\begin{equation}
    \phi_u(0) = \begin{cases}
    0 \text{\,\,\,\, if } u \in [x_R,x_R]\\ \frac{1}{2}  \text{ \, if  } u \notin [x_L,x_R] 
       \end{cases}
\end{equation}
The initial boundary condition corresponds to an operator that has size $N/2$ at every site outside the interval $[x_L,x_R]$ but has void in $[x_L,x_R]$. Equations \eqref{eq:mutinf_opgrowth} and \eqref{eq:mutinf_opgrowth2} are quite remarkable in that they express entanglement entropy and mutual information in terms of operator size. In words, when operators of local size $N/2$ supported in $L\cup C$ and a void in $a\cup B$ grow to a size proportional to $N/2$ in $a$, the mutual information between the reference system and $L\cup C$ grows to its maximal value. Since operator growth, on general grounds, can be argued to have an emergent light-cone structure at strong coupling (see the discussion around equation \eqref{eq:yvu} for the argument), any theory which satisfies some analog of this relationship will have an emergent locality in entanglement spreading. Furthermore, since both the spread of entanglement and operator growth are controlled by the same underlying equation in Brownian SYK chains, we see that the emergent light-cone structure is controlled by the same butterfly velocity in both cases.

\subsection{Entropy inequalities} 
\begin{figure}
    \centering
\includegraphics[width=0.5\linewidth]{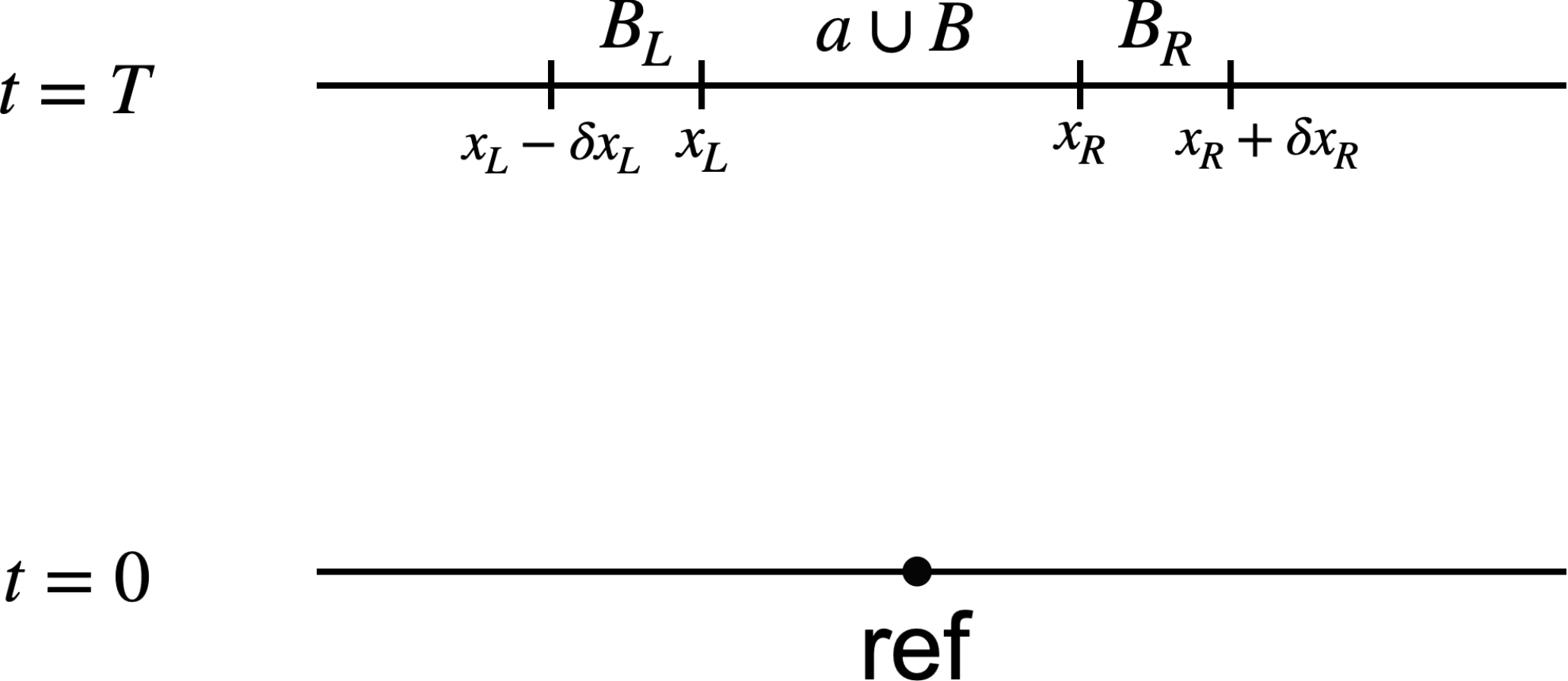}
    \caption{We consider deforming the sub-region $a \cup B$  by adding small sub-regions $B_L$ and $B_R$ of widths $\delta x_L$ and $\delta x_R$ respectively.}
    \label{fig:subadditivity_BSYK}
\end{figure}

Since the entanglement entropy at leading order is completely determined from operator growth, it is interesting to consider the implications of entropy inequalities on operator growth. We consider the sub-additivity and strong sub-additivity of entanglement entropy. Throughout this section, we will denote $a \cup B$ by $\hat{B}$.

Strong subadditivity (i.e., monotonicity of mutual information) implies the following:
\begin{align*}  \nt  & \qquad I_{\Psi}(\re: \hat{B}) \leq  I_{\Psi}(\re: \hat{B} \cup B_L) \\& \implies S(\re \cup \hat{B} \cup B_L) -  S(\re \cup \hat{B}) \leq    S(\hat{B} \cup B_L) - S(\hat{B})  \\ & \implies   h(\phi_0(T, -x_L-\delta x_L, x_R)) - h(\phi_0(T,-x_L,x_R)) \leq 0,
\end{align*}
where $B_L$ and $B_R$ are infinitesimal regions which can be appended to the left and right ends of $B$ respectively (see Figure \ref{fig:subadditivity_BSYK}). It is easy to interpret the above relation:
in the domain $0 \leq \phi \leq \frac{1}{2}$, $h(\phi)$ monotonically increases with $\phi$.  Therefore, the above constraint is equivalent to the following statement. \begin{align*}\label{eq:opgrowth_left} \nt
    \phi_0(T, -x_L - \delta x_L, x_R) \leq \phi_0(T, -x_L, x_R) \implies -\partial_{x_L}  \phi_0(T, -x_L, x_R) \leq 0.
\end{align*}
We can replace $B_L$ with $B_R$ to get a similar relation for the right boundary, namely:
\begin{equation}\label{eq:opgrowth_right}
    \partial_{x_R} \phi_0(T, -x_L, x_R) \leq 0 .
\end{equation}
Here, by $\partial_{L/R}$ we mean the  derivative\footnote{Since we're working on a lattice, we define $\partial_x f(x) = f(x+1) - f(x)$ } wrt. the coordinate of the left/right end-point of the interval. This implies that the operator growth at a site is slower when the initial void is bigger. 
\begin{figure}
    \centering
\includegraphics[width=0.49\linewidth]{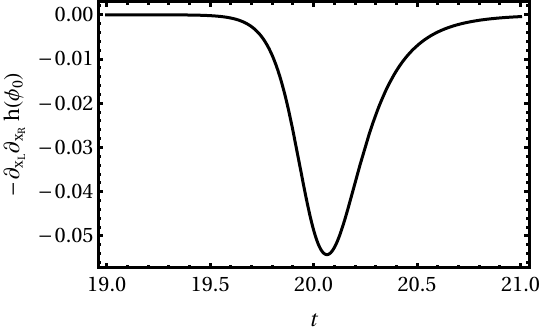}   
    \caption{We show the time dependence of $-\partial_{x_L} \pa_{x_R} h(\phi)$ (left figure) for the $q = 4$, $\tilde q = 2$ Brownian SYK Hamiltonian. We chose a lattice of 500 sites with $x_L = 100$ and $x_R = 400$. The plot is for $\phi$ at site 250.  We find that $-\partial_{x_L} \pa_{x_R} h(\phi) \leq 0$ in agreement with the strong sub-additivity.}
    \label{fig:approxstrongsub}
\end{figure}

We can also apply strong sub-additivity for the following different choice of subregions:
\begin{equation}
I_{\Psi}(B_L:\text{ref}\cup \hat{B})\leq I_{\Psi}(B_L:\text{ref}\cup \hat{B}\cup B_R).
\end{equation}
Expanding this out in terms of entropies gives:
\begin{align*} \nt
      &S(\re \cup \hat{B} \cup B_L \cup B_R)  + S(\re \cup \hat{B}) \\ &\hspace{3cm}\leq  S(\re \cup \hat{B} \cup B_L) + S(\re \cup \hat{B} \cup B_R), \\  \implies&   h(\phi_{0}(T, x_L-\delta x_L,x_R+\delta x_R)) + h(\phi_{0}(T, x_L ,x_R))  \\ &\hspace{3 cm}\leq h(\phi_{0}(T, x_L-\delta x_L, x_R)) + h(\phi_{0}(T, x_L, x_R + \delta x_R)),  \\ \implies   &h(\phi_{0}(T, x_L-\delta x_l,x_R+\delta x_R)) - h(\phi_{0}(T, x_L, x_R +\delta x_R)) \\ &\hspace{3cm} -   h(\phi_{0}(T, x_L-\delta x_L, x_R)) - h(\phi_{0}(T, x_L,x_R))   \leq 0, \\\implies& -\partial_{x_L} \partial_{x_R} h(\phi_0(T,-x_L,x_R)) \leq 0. \nt \label{strong_subadditivity}
\end{align*} 
A crude reason that this inequality works goes as follows: let us assume that $\phi_0(T,x_L,x_R)$ varies slowly as a function of $x_L$ and $x_R$. Then we use chain rule to expand the double derivative on the LHS of the inequality (To keep notation simple, we will suppress the dependence of $\phi_0$ on $T,x_L,x_R$):
\begin{align*}\nt \label{eq:approx_strong_sub}
    -\partial_{x_L} \partial_{x_R} h(\phi_0) \approx -  h'(\phi_0)\partial_{x_L} \partial_{x_R} \phi_0 - \partial_{x_L} \phi_0 \partial_{x_R} \phi_0 h''(\phi_0).
\end{align*}
Since $h(\phi)$ is a monotonically increasing concave function of $\phi$ for $0\leq\phi \leq1/2$, $h''(\phi) \leq 0$ and $h'(\phi) \geq 0$. Moreover, it follows from equations (\ref{eq:opgrowth_right}, \ref{eq:opgrowth_left}) that the product $ -\partial_{x_R} \phi_0 \partial_{x_L} \phi_0 \geq 0$. Therefore, if  $- \partial_{x_L} \partial_{x_R} \phi_0 \leq 0$ then the inequality (\ref{strong_subadditivity}) follows trivially. Let us assume that $\phi_u(t)$ satisfies an equation of the following form:
\begin{equation}
    \partial_t \phi_u(t) =  f(\phi_u) + \partial^2_u \phi_u(t),
\end{equation}
where $f(\phi)$ is some concave function such that $f(\phi) \geq 0,\;  f(0) = f(\frac{1}{2}) = 0$. 

Define $\phi_u$, $\phi_u^L$, $\phi_u^R$ and $\phi_u^{L+R}$ as solutions of above differential equation with the initial conditions such that the boundaries of the voids exist at coordinates $(x_L,x_R)$,  $(x_L-1, x_R)$, $(x_L, x_R+1)$ and $(x_L-1, x_R+1)$ respectively. The time evolution of $-\partial_L \partial_R \phi_u(t)$ is determined from the following differential equation:
\begin{align*}
 \partial_t \left( -\partial_L \partial_R \phi_u \right) &=    \partial_t \left( \phi_u + \phi^{L+R}_{u} - \phi^{L}_{u} - \phi^{R}_{u}\right) \\&= \left(  f(\phi_u) + f(\phi_u^{L+R}) - f(\phi_u^{L}) - f(\phi_u^R)\right)  + \partial_u^2 \left(  -\partial_{x_L} \partial_{x_R} \phi_u \right) \\ &\approx f''(\phi_u) \left(-\partial_{x_L} \phi \partial_{x_R} \phi_u\right) + f'(\phi_u) \left( -\partial_{x_L} \partial_{x_R} \phi_u\right) + \partial_u^2(-\partial_{x_L} \partial_{x_R} \phi_u).
\end{align*}

At $t = 0$, $\partial_{x_{L}} \partial_{x_R} \phi =0$. Since $f''(\phi) \leq 0$, the first term on the RHS is always non-positive i.e. it can only make $-\partial_{x_L} \partial_{x_R} \phi$ negative. Moreover, the next two terms can not change its sign. Therefore, $-\partial_{x_L} \partial_{x_R} \phi \leq 0$. This explains the inequality (\ref{strong_subadditivity}). In figure (\ref{fig:approxstrongsub}), we plot $-\partial_{x_L} \partial_{x_R} h (\phi_0)$ for the Brownian SYK Hamiltonian in (\ref{eq:Hamiltonian_saddle}). We see that our formula indeed satisfies strong sub-additivity.

\section{Discussion}

The spread of quantum information in holographic theories has a sharp, emergent light-cone structure. This is a direct manifestation of the emergence of a local bulk spacetime satisfying an RT-like formula for entanglement entropy of boundary subregions. Furthermore, the velocity at which this emergent light-cone spreads in holographic theories precisely coincides with the butterfly velocity at which OTOCs propagate, a signature of the equivalence principle in the bulk. In the hope of developing some insight into these characteristic features of bulk locality, in this work we studied the spatial spread of quantum information pertaining to an external qudit $a$ injected at a point $p$ in a Brownian SYK chain at infinite temperature in the large $N$ limit. As a measure of the spread of information, we calculated the time evolution of the mutual information between a reference system (initially maximally entangled with the qudit $a$) and an interval of length $2 \ell$ centered at $p$. By taking $a$ to consist of $\lambda N$ fermions (with $\lambda$ fixed to be some infinitesimally small number as $N\to \infty$) and perturbatively expanding in $\lambda$, we were able to explicitly compute this mutual information via analytic continuation from the corresponding R\'enyi entropies. At large $N$, the mutual information is controlled by a non-linear generalization of the diffusion equation, called the FKPP equation. At strong coupling, this equation admits sharp domain wall solutions, which leads to the emergence of a sharp light-cone structure for entanglement spreading, quite analogous to the emergence of bulk locality in holographic conformal field theories. By relating this entanglement entropy calculation to operator growth, we argued that the velocity underlying the spread of quantum information is precisely the same as the butterfly velocity which controls operator growth and the spread of OTOCs in this model. This too agrees with expectations from holographic theories. 

\subsection{Towards a general argument for sharp entanglement spreading}

In section \ref{sec:OGEE}, we related the entanglement entropy to operator spreading in Brownian SYK chain; this gave us an intuitive way of thinking about the emergence of a sharp light-cone in entanglement spreading in the strong coupling limit. In this section, we would like to sketch out a general argument relating the second R\'enyi entropy to OTOCs with typical operators, along the lines of \cite{Hosur_2016}. 

\begin{figure}
    \centering
    \includegraphics[width=1\linewidth]{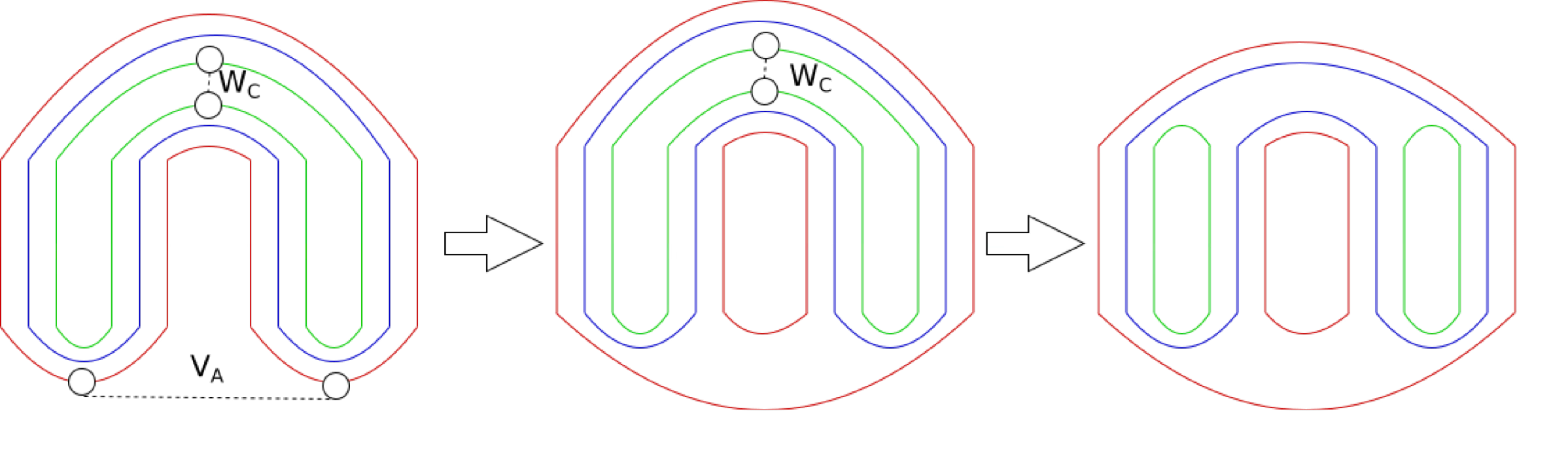}
    \caption{Deforming OTOC contour to second Renyi entropy contour via operator averaging over a and C subspaces.}
    \label{fig:Deforming OTOC contour to second Renyi entropy}
\end{figure}

Consider a maximally mixed state on two copies of the Hilbert space $L$ and $R$, and consider an OTOC of the form 
\begin{equation} \label{OTOC}
\langle V_a(0) W_C(T) V_a(0) W_C(T)\rangle,
\end{equation}
where $V_a$ is an operator inserted on the code-subspace site and $W_C$ is an operator inserted on the region $C$, where we are using the same notation for the regions $a, B$ and $C$ as before. This OTOC is a good probe to study how fast the operators inserted in the $a$ region spread into the $C$ region. We will take $W_C$ to be a \emph{locally typical} operator, by which we mean that at every site it is supported within the subspace of typical operators, where ``typical'' is defined in some suitable way (see below). In this case, we can equivalently think of the above OTOC as a measure of how locally-typical operators $W_C$ in the $ C$ region spread outside $C$. Several definitions of locally typical are possible. For instance, starting with some operator $\phi$ with non-trivial support on all of $C$ and such that $\text{Tr}\,\phi=0$, we can construct:
\begin{equation}
    W_C = \mathcal{U}\,\phi\,\mathcal{U}^{\dagger}, \,\,\mathcal{U} = \prod_{u\in C}U_{(u)},
\end{equation}
where each $U_{(u)}$ is a unitary operator on the Hilbert space at site $u$ drawn independently and randomly from the Haar ensemble. If we now compute the Haar average of the OTOC in equation \eqref{OTOC}, the integration over the local Haar random unitaries essentially implements a re-wiring of the contours on the $C$ subregion, so that the new contour is precisely the one used for computations of the second R\'enyi entropy (see figure \ref{fig:Deforming OTOC contour to second Renyi entropy}). Thus, the second R\'enyi entropy is closely related to OTOCs of locally typical operators. In any theory with strongly chaotic dynamics at each site, we expect that locally typical operators spread in a sharp light cone. In any such theory, the above argument suggests that the second R\'enyi entropy will also spread in a sharp, emergent light-cone. It would be good to make this heurisitic argument more precise, but we will not attempt to do so here.  



\subsection{Entanglement membranes}

In \cite{jonay2018} (see also \cite{Zhou:2018myl, Zhou:2019pob, vardhan_moudgalya2024}), it was conjectured that an emergent locality in entanglement spreading is a general property of chaotic quantum systems. These authors wrote down an effective theory for entanglement dynamics in terms of an effective \emph{membrane}. Let us consider a chain of local quantum systems, and let $S_{\ket{\chi}}(x,T)$ denote the entanglement entropy of all sites to the left of the point $x$ in a state $\ket{\chi(T)} = U(T) \ket{\chi(0)}$, where $U(T)$ is some unitary time evolution operator. In \cite{jonay2018}, the authors conjectured that $S_{\ket{\chi}}(x,T)$ satisfies the membrane formula :
\begin{equation}
    S_{\ket{\chi}}(x,T) = \min_y \left\{ s_{eq}\,  T\,\mathcal{E}\left(\frac{x-y}{T} \right) + S_{|\chi\rangle}(y,0)\right\},
\end{equation}
where the function $\mathcal{E} (v)$ is called the membrane tension and $s_{eq}$ is the local entropy density. For systems at infinite temperature, $s_{eq}$ is the logarithm of the local Hilbert space dimension. The authors further gave a simple, alternate way of obtaining the membrane tension: consider the maximally entanglement state $|\Omega\rangle \in \mathcal{H}\otimes \mathcal{H}^*$ on two copies of the chain, and let 
\begin{equation}
    \ket{U(T)} = U(T) |\Omega\rangle,
\end{equation}
where the unitary $U$ only acts on $\mathcal{H}^*$. Let us define the \emph{operator entanglement} $S_{U(T)}(x,y)$ of $U$ as the entanglement entropy of the intervals $[-\infty, x]$ in $\mathcal{H}$ and the interval $[-\infty, y]$ in $\mathcal{H}^*$ in the state $\ket{U(T)}$.\footnote{For a finite system, by $[-\infty,x]$ we mean all points to the left of $x$} Then, it was argued in \cite{jonay2018} that the membrane tension $\mathcal{E}(v)$ is related to the operator entanglement in the limit of large times and interval sizes as:
\begin{equation}\label{eq:membrane_tension} \mathcal{E} (v) =\frac{1}{T} S_{U(T)}(\sigma, \sigma+ vT). \end{equation}
For a homogeneous system, $\sigma$ is arbitrary as long as it is far from boundary. From this identification, we can easily derive the following properties of $\mathcal{E}(v)$ using entropy inequalities (see appendix \ref{sec:EM} for a proof):
\begin{equation}\label{eq:inequalities}
 (i) \quad    \mathcal{E}(v) \geq |v|, \quad (ii) \quad  |\mathcal{E}'(v)| \leq 1 \quad \text{and} \quad (iii) \quad  \mathcal{E}''(v) \geq 0.
\end{equation}
Only velocities which satisfy $|v|\leq v_B$ are needed in the entanglement membrane formula in systems with the effective butterfly light-cone for entanglement spreading. 

We expect that an elegant membrane theory should govern entanglement spreading in Brownian SYK chains as well. Indeed, it would be interesting to see whether the sharp domain wall solutions that we encountered in this work are related to entanglement membranes.\footnote{It is also possible to show sharp entanglement spreading by assuming the entanglement membrane picture \cite{shreya_note}. We thank Shreya Vardhan for explaining this to us.} We will leave a more detailed study to future work.

\subsection{Perturbative and non-perturbative corrections}

Throughout this paper, we assumed a strict $N \rightarrow \infty$ limit, ignoring $\frac{1}{N}$ corrections to the operator growth equation. However, it was shown in \cite{Xu_2019, Stanford:2023npy} that such corrections have two important effects on operator growth: (i) The butterfly velocity reduces by $O(1/\log^2 N)$ and (ii) there is a diffusive broadening of the wavefront where the diffusion constant scales as $1/\log^3 N$. When $ T \sim \log^3 N$, the wavefront starts to smoothen out. According to equation (\ref{eq:mutinf_opgrowth}), we expect that due to this diffusive broadening at late times, the width of transition of the mutual information $I_\Psi(\re: a\cup B)$ would grow as the size of $a \cup B$ grows.

There are also non-perturbative corrections coming from other saddles. For instance, although the mutual information $I_\Psi(\re:a \cup B)$ decreases monotonically, it will not saturate to zero in a finite size system. Let $\ket{\psi_1}$ and $\ket{\psi_2}$ be two orthogonal states in the code subspace discussed in equation (\ref{EncState}). If we model $U_{a,R}(T)$ as $T \rightarrow \infty$ as a Haar random unitary, then it's easy to see that their corresponding reduced density matrices on $C \cup L$ have a non-zero overlap of order $\dim \cH_a/\dim \cH_{C}$ \cite{Stanford:2021bhl}. This non-zero overlap implies that $\ket{\psi}_1$ and $\ket{\psi}_2$ can not be exactly recovered from the knowledge of the reduced density matrices on $C\cup L$. 

In the Brownian SYK chain, the effective Hamiltonian that appears in the calculation of $S^{(n)}_\Psi(\re \cup a\cup B)$ has multiple ground states. We are interested in calculating the following quantity:\begin{equation}
    e^{-S^{(n)} (\re \cup a \cup B)} = \bra{\Gamma_f^{(n)}} e^{-H^{(n)}_\eff T}\ket{\Gamma_i^{(n)}}.
\end{equation}

At late times, we can estimate the sub-leading contributions from these ground states. For instance, consider $S^{(2)}_\Psi(\re:a \cup B)$. In this case, the Effective Hamiltonian is given in equation \ref{eq:effective_hamiltonian}.
It has four degenerate ground states ($x_u \ket{\pm} = \pm \ket{\pm}$, $z_u \ket{1} = \ket{1}$ and $z_u \ket{0} = - \ket{0}$). The initial and final states are
\begin{equation}
    \ket{\Gamma_i} = \ket{+}_{a} \otimes \ket{\uparrow}_{B} \otimes \ket{\uparrow}_C, \; \; \quad   \ket{\Gamma_f} = \ket{+}_a \otimes \ket{+}_B \otimes \ket{\uparrow}_C.
\end{equation}
At late times, we get: 
\begin{align*}
    e^{-S^{(2)} (\re \cup B \cup C)} &\approx \langle \Gamma_f | \uparrow\rangle \langle \uparrow|\Gamma_i\rangle +  \langle \Gamma_f | +\rangle \langle +|\Gamma_i\rangle  \\ &=  \frac{1}{ \dim \cH_B}  \left( \frac{1}{\dim^2 \cH_a} +   \frac{1}{\dim^2 \cH_C} \right)  = \frac{1}{\dim \cH_B \dim^2 \cH_a} \left( 1 + \frac{\dim^2 \cH_a}{\dim^2 \cH_C}\right).
\end{align*}
Therefore, the $I^{(2)} \left(\re: a \cup B   \right)$ approaches $\frac{\dim^2 \cH_a}{\dim^2 \cH_C}$ at late times. It would be interesting to find the saddle point corresponding to this correction.

\section*{Acknowledgments}
We thank Kedar Damle, Abhijit Gadde, Gautam Mandal, Shiraz Minwalla, Tridib Sadhu and Shreya Vardhan for helpful discussions. We also thank Mark Mezei and Shreya Vardhan for their insightful comments on an earlier version of the manuscript. We acknowledge support from the Department of
Atomic Energy, Government of India, under project identification number RTI 4002, and from the Infosys Endowment for the study of the Quantum Structure of Spacetime.

\appendix

\section{A brief review of the Fisher-KPP equation}\label{sec:FKPP}

The FKPP equation is
\begin{equation}\label{eq:fkpp}
    \partial_t u =  \frac{1}{2} \partial_x^2 u + f(u)
\end{equation}
$f(u)$ is a continuous function with the following properties for $0 < u < 1$: 
\begin{align*}\label{eq:cond_f}
\nt   &f(0) = f(1) = 0, \quad  f(u) > 0 \\
    &f'(0) = 1, \quad f'(u) \leq 1 
\end{align*}
Kolmogorov, Petrovsky and Piskunov showed that if $u(0,x) = \Theta(-x)$, then \begin{align*}
  \nt     u(t,x) \rightarrow w(x &- m(t)) \text{  as  } t \rightarrow \infty \\ \text{s.t.} \lim_{t \rightarrow \infty } m(t)/t &= \sqrt{2}
\end{align*}
$w(x)$ is a monotonically decreasing function with the following properties: 
\begin{align*}\label{eq:cond_fkpp}
 \nt  0 \leq w(x) \leq 1 \quad &\text{for}\quad  x \in \mathbb{R} \\ \lim_{x \rightarrow -\infty} w(x) = 1, \quad &\text{and}
 \lim_{x \rightarrow -\infty} w(x) = 0. 
\end{align*}Assuming that $u(t,x) \rightarrow w(x - m(t))$ as $t \rightarrow \infty$, we will show that $\dot{m}(t) \rightarrow \sqrt{2}$. 
\begin{equation}
    f(u) = \begin{cases}
        u, & \text{if} \quad  0 \leq u \leq 1/2,\\
        1 - u, & \text{if} \quad 1/2 < u \leq 1. 
    \end{cases}
\end{equation}

\subsection{Traveling Wave Solutions}

Let us consider the traveling wave solutions of the FKPP equation which satisfy (\ref{eq:cond_fkpp}). If $u(t,x+\lambda t) = w_\lambda(x)$, then $\wl(x)$ satisfies the following equation:\begin{equation}
   \lambda \partial_x \wl(x) + \frac{1}{2} \partial_x^2 \wl(x) + f(\wl(x)) = 0 
\end{equation} Assuming that $\wl(x)$ is monotonically decreasing in $x$, it is easy to solve the above equation. Due to the translational invariance of the equation, we can fix the origin by demanding that $\wl(0) = \frac{1}{2}$. Then, equation (\ref{eq:fkpp}) simplifies to the following equation. 
\begin{align*}
   \nt    \lambda\partial_x \wl(x) + \frac{1}{2} \partial_x^2 \wl(x) + (1 - \wl(x)) = 0, \quad \text{if} \quad x < 0, \\
     \lambda   \partial_x \wl(x) + \frac{1}{2} \partial_x^2 \wl(x) + \wl(x) = 0, \quad \text{if} \quad x \geq 0,
\end{align*}with the boundary conditions:\begin{equation}\label{eq:bc}
    \wl(-\infty) = 1, \quad    \wl(\infty) = 0, \quad \text{and} \quad \wl(0) = \frac{1}{2}.
\end{equation}
The solution is: 
\begin{equation}\label{eq:sol}
    \wl(x) =\begin{cases} 1 - \frac{1}{2} \exp(\alpha_0 x)  \quad &\text{if} \quad x<0  \\ 
    a      \exp(\alpha_+ x) + (\frac{1}{2} - a) \exp(\alpha_- x), \quad &\text{if} \quad x\geq0 \\
    \end{cases}   
\end{equation}
where, \begin{align*}
    \alpha_0 = -\lambda + \sqrt{\lambda^2 + 2}, \\
    \alpha_\pm = -\lambda \pm \sqrt{\lambda^2 - 2},
\end{align*}
By matching the left and right derivatives of $\wl(x)$ and $x = 0$, we get $a$:\begin{align*}    a = -\frac{\alpha_0 + \alpha_-}{2(\alpha_+ - \alpha_-)}
\end{align*}
Note that we have not yet imposed the condition $ 0\leq \wl(x) \leq 1$. It turns out that this condition is not satisfied by $\wl(x)$ for $\lambda < \sqrt{2}$. To see this, consider the equation: 
\begin{equation}
    \wl(x) = 0, \quad \text{for} \quad x>0
\end{equation}
Using the second line of equation (\ref{eq:sol}), we find that the roots of this equation satisfy:\begin{align*}
    \exp \left( \alpha_+ - \alpha_-\right)x &= -\frac{(1/2 - a)}{a} \\ &=\frac{\alpha_0 + \alpha_+}{\alpha_0 + \alpha_-} \\&= \frac{-2\lambda + \sqrt{\lambda^2 + 2} + \sqrt{\lambda^2 - 2}}{-2 \lambda + \sqrt{\lambda^2 + 2} - \sqrt{\lambda^2 -2}}
\end{align*}
If $\lambda < \sqrt{2}$,  both the LHS and the RHS of the equation are pure phase. Therefore, there are infinitely many solutions to the equation in the range $x > 0 $. For $\lambda >\sqrt{2}$, the RHS is always smaller than 1 while the LHS is always bigger than 1.  Therefore, for $\lambda < \sqrt{2}$, there exists no traveling wave solution to the FKPP equation that satisfies all the conditions in (\ref{eq:cond_fkpp}). 

Although we analyzed the FKPP equation for a special choice of $f(u)$, the conclusions drawn in the previous paragraph hold for more general functions. The analysis for general $f(u)$ can be found in \cite{BramsonLectures,TongLectures}.  

\subsection{Upper bound on the velocity}\label{sec:upper_bound}

Now, we will show that the velocity of the traveling wave is upper bounded by $\sqrt{2}$ if the initial condition is a step function. This can be shown by considering the following differential equation: 
\begin{equation}
    \partial_t v(t,x) =  \frac{1}{2} \partial_x^2 v(t,x) + v(t,x)
\end{equation}
The integral representation of the solution to this equation is: 
\begin{equation}
    v(t,x) = \int_{-\infty}^{\infty} \d y \,   g(t,y-x) v(0,y) +  \int_0^t \d s \int_{-\infty}^{\infty} \d y \,  g(t-s,y-x)\,  v(t-s,y)
\end{equation}
where \begin{equation}
    g(t,y) = \frac{1}{\sqrt{2 \pi t}} \, e^{-\frac{y^2}{2 t}}
\end{equation}
We can also write an integral representation of the solution to the FKPP equation:
\begin{equation}
    u(t,x) =  \int_{-\infty}^{\infty} \d y \,   g(t,x-y) u(0,y) +  \int_0^t \d s \int_{-\infty}^{\infty} \d y \,  g(t-s,x-y)\,  f(u(t-s,y))
\end{equation}
If $u(0,x) = v(0,x)$, then 
\begin{align*}
    v(t,x) - u(t,x) &= \int_0^t \d s \int_{-\infty}^{\infty} \d y \,  g(t-s,x-y)\,\left[  v(t-s,y) - f(u(t-s,y))\right] \\ & \geq \int_0^t \d s \int_{-\infty}^{\infty} \d y \,  g(t-s,x-y)\,\left[  v(t-s,y) - u(t-s,y)\right] \\ &\geq \int_0^t \d s \int_{-\infty}^{\infty} \d y \,  g(t-s,x-y)\, \inf_{z \in \mathbb{R}}\left[  v(t-s,z) - u(t-s,z)\right]  \\ &=\int_0^t \d s  \inf_{z \in \mathbb{R}}\left[  v(t-s,z) - u(t-s,z)\right] 
\end{align*}
In the second line, we used the inequality $f(u) \leq u$. Taking the infimum w.r.t. $x$ on the LHS, we get:
\begin{equation}
\inf_{z \in \mathbb{R}}\left[  v(t,z) - u(t,z)\right]  \geq \int_0^t \d s \inf_{z \in \mathbb{R}}\left[  v(t-s,z) - u(t-s,z)\right] 
\end{equation}
If we define $\sigma(t) =  \int_0^t \d s \inf_{z \in \mathbb{R}}\left[  v(s,z) - u(s,z)\right] $, then the above equation can be rewritten as:
\begin{align*}
\nt   & \sigma'(t) \geq \sigma(t) \\
   \end{align*}This inequality implies that 
   \begin{align*} 
   \nt
& \frac{\d}{\d t} (e^{-t} \sigma(t)) \geq 0  \\ 
   \implies & e^{-t} \sigma(t) > \sigma(0) = 0, \quad \text{if} \; t >0  \\ \implies &\sigma(t) \geq 0  \\\implies & \sigma'(t) \geq 0 \\ \implies& \inf_z \left[v(t,z) - u(t,z) \right] \geq 0
\end{align*}Therefore, \begin{equation}
    v(t,x) \geq u(t,x),\quad 
\end{equation}Now we can solve the differential equation for $v(t,x)$ and use the above inequality to get a bound on $u(t,x)$
\begin{equation}
 u(t,x) \leq    v(t,x) = e^t \int_{-\infty}^{\infty}  \frac{ \d y }{\sqrt{2 \pi t}} \, e^\frac{-(x-y)^2}{2 t}  v(0,y)
\end{equation}
If $v(0,y) = u(0,y) = \Theta(-y)$, then \begin{equation}
    u(t,x) \leq  \int_{-\infty}^{0}  \frac{ \d y }{\sqrt{2 \pi t}} \, e^{t - \frac{(x-y)^2}{2 t}}
\end{equation}Setting $x = \sqrt2 t + z$ in the above inequality, we find that
\begin{align*}\nt
    u(t,\sqrt2 t +z) &\leq \int_{-\infty}^{0}  \frac{ \d y }{\sqrt{2 \pi t}} \, e^{\frac{-(z-y)^2 }{2 t} - \sqrt{2} (z-y)} \\ &= \int_{-\infty}^{z}  \frac{ \d y }{\sqrt{2 \pi t}} \, e^{\frac{-y^2 }{2 t} - \sqrt{2} y} \\
&\leq \frac{e^{-\sqrt{2} z}}{2 \sqrt{\pi t}} 
\end{align*}
We see that \begin{equation}\label{eq:conv}
    u(t,\sqrt{2} t + z) \rightarrow 0, \quad \text{if} \quad z = O(1)
\end{equation} This means that at large $t$, $u(t,x) \rightarrow 0$ in any neighborhood of $O(1)$ size around $x = \sqrt{2} t$. 
Let us assume that $u(t,x)$ approaches a traveling wave solution at late times, i.e. $u(t,z+m(t)) \rightarrow w(z)$ for some $m(t)$. As shown in the previous section, $ \dot{m}(t) \leq \sqrt{2}$, but $ \dot{m} (t)= \sqrt{2}$ is the smallest velocity for which $u(t, m(t) + z) \rightarrow 0$. Therefore, $\dot{m}(t) \rightarrow \sqrt{2}$ as $t \rightarrow \infty$.  In fact, the velocity of the traveling wave approaches $\sqrt{2}$ upto $1/t$ corrections \cite{BramsonLectures}: 
\begin{equation}
    \dot{m}(t) = \sqrt{2} - \frac{3}{2t} + \dots
\end{equation}

\subsection{FKPP on a chain}

In section 3, we found a discrete version of the FKPP equation on the chain. The general analysis of FKPP on chain is more difficult than the continuum version. Nevertheless, we can estimate the velocity dependence of the traveling wave on the lattice. Consider the equation:
\begin{equation}\label{eq:disc_fkpp}
    \partial_t u(t,x) = \frac{1}{2} \left(u(t,x+1) + u(t,x-1) -2u(t,x)\right)  +  J f(u(t,x)), \quad x \in \mathbb{Z}
\end{equation}
where $f(u)$ follows the conditions given in equation (\ref{eq:cond_f}). Following the arguments of section (\ref{sec:upper_bound}), but now in the discrete case, we get a bound on $u(t,x)$: 
\begin{equation}
u(t,x) \leq v(t,x)     
\end{equation}where $v(t,x)$ satisfies the following differential equation: \begin{equation}
    \partial_t v(t,x) = \frac{1}{2} \left(v(t,x+1) + v(t,x-1) - 2v(t,x)\right)  +  J \,v(t,x)
 \end{equation}
 and $v(0,x) = u(0,x)$. $v(t,x)$ has the following integral representation:
 \begin{equation}
     v(t,x) = e^{Jt}  \int_{-\pi}^\pi \frac{\d k}{2 \pi} \sum_{x'\in \mathbb{Z}} \, e^{i k (x-x')-2\sin^2(k/2)t} v(0,x)
 \end{equation}
If $v(0,x) = \Theta(-x)$, then \begin{align*}
     v(t,x) &= e^{Jt}  \int_{-\pi}^\pi \frac{\d k}{2 \pi} \sum_{x'\leq  0} \, e^{i k (x-x')-2\sin^2(k/2)t} \\    &= e^{Jt}  \int_{-\pi}^\pi \frac{\d k}{2 \pi}  \frac{ e^{i k x-2\sin^2(k/2)t}}{1 - e^{-\eps + ik}}
 \end{align*}
 \begin{figure}[t]
    \centering
    \includegraphics[width=0.6\linewidth]{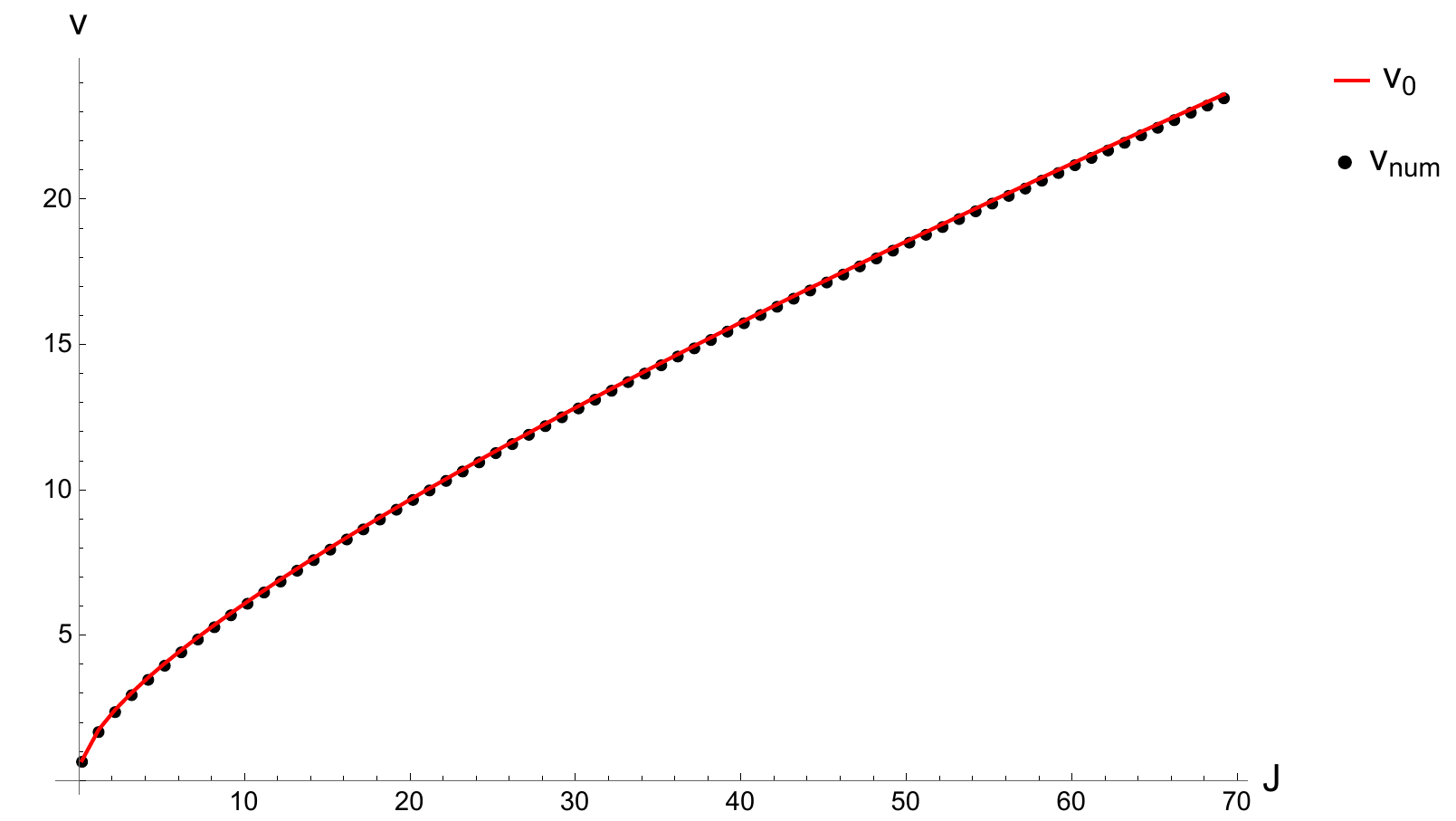}
    \caption{Comparison between $v_0$ (red curve) and the velocity of the traveling wave $\text{v}_\text{num}$ (black dots) obtained numerically. }
    \label{fig:fkpp_lattice}
\end{figure}
Evaluating $v(t,x)$ at $x = v_0t$ for some $\lambda$ in  $t \rightarrow \infty$ limit, we can use saddle point approximation to estimate $v(t,x)$:  \begin{equation}
     v(t,\lambda t) = e^{Jt} \int_{-\pi}^{\pi} \frac{\d k}{2 \pi} \frac{ e^{i( k \lambda + 2i\sin^2(k/2))t}}{1 - e^{-\eps + ik}} \sim  e^{Jt} \frac{ e^{i( lv_0 + 2i\sin^2(l/2))t}}{1 - e^{-\eps + il}} 
 \end{equation}
where $l$ satisfies the saddle point equation:
\begin{align*}
     \lambda + 2 i \sin(l) = 0  \\ \implies   l = i \sinh^{-1}\frac{\lambda}{2}
\end{align*}\begin{equation}
    v(t, \lambda t) \sim  \# \, e^{\left(J  - v_0 \sinh^{-1} \frac{v_0}{2} - \frac{1 - \sqrt{1 + v_0^2/4}}{2} \right)t} 
\end{equation}
To estimate the velocity of the traveling wave, we demand that the $t$ dependence in the exponent vanishes.

This happens when\begin{equation}
 \quad  J =  v_0 \sinh^{-1} \frac{v_0}{2} + \frac{1 - \sqrt{1 + v_0^2/4}}{2} 
\end{equation}
Figure (\ref{fig:fkpp_lattice}) shows a comparison of $v_0$ with the traveling wave velocity obtained by numerically solving equation (\ref{eq:disc_fkpp}). 

\section{Boundary Conditions and Some Properties of $M$}\label{matrix_bc}

The variables $g_a^{i,j}$ satisfy the following boundary conditions: 
\begin{align*}\nt \label{eq:bc_app}
&g_{a}^{2k, 2k+1}(t) = g_{a}^{1,2n}(t) = \frac{i}{2}  &&\hspace{-0.7cm}\text{for} \quad 1\leq k < n, \quad t = 0, T \\ 
&g_{a}^{j, 2k}(0) + i g_{a}^{j, 2k+1}(0) = 0,  \quad &&g_{a}^{j, 2k}(T) - i g_{a}^{j, 2k+1}(T) = 0 \\
&g_{a}^{1, j}(0) + i g_{a}^{2n, j}(0) = 0,  \quad &&g_{a}^{1, j}(T) - i g_{a}^{2n, j}(T) = 0 
\end{align*}
Let's consider $g_a^{i,j}(0)$. For illustration, let's set $n = 3$:
\begin{align*}\nt
g_a(0) = \begin{pmatrix}
0 & -i g^{1,3}_a &  g_a^{1,3} & -ig_a^{1,5} & g^{1,5}_a & \frac{i}{2} \\
ig^{1,3}_a & 0 & \frac{i}{2} & -g^{3,5}_a & - ig^{3,5}_a & - g^{1,3}\\
-g^{1,3}_a  &  - \frac{i}{2} & 0  & - i g_a^{3,5}  & g^{3,5} &  -ig^{1,3} \\
ig_a^{1,5}& g_a^{3,5} & ig_{a}^{3,5} & 0 & \frac{i}{2} & -g^{1,5}_a\\  -g^{1,5}_a&i g_a^{3,5} & -g^{3,5}_a&-\frac{i}{2} &0 & -ig^{1,5}\\ -\frac{i}{2}&g^{1,3}_a &ig_a^{1,3} & g_a^{1,5}& ig_a^{1,5}& 0 
\end{pmatrix} 
\end{align*}
Consider a $2n \times 2n$ matrix $A$ defined as follows:
\begin{equation}
A =
\begin{pmatrix}
0  & 1  & 0  & 0  & \cdots & 0  & 0  \\
0  & 0  & 1  & 0  & \cdots & 0  & 0  \\
0  & 0  & 0  & 1  & \cdots & 0  & 0  \\
\vdots & \vdots & \vdots & \vdots & \ddots & \vdots & \vdots \\
0  & 0  & 0  & 0  & \cdots & 0  & 1  \\
-1  & 0  & 0  & 0  & \cdots & 0  & 0  
\end{pmatrix}
\end{equation}
By conjugating $g_a(0)$ with $A_{6 \times 6}$, we get
\begin{equation}
    \tilde{g}_a = A^T g_a A = 
 \begin{pmatrix}
0 & \frac{i}{2} & -g_a^{1,3} & -ig_a^{1,3} & -g_a^{1,5} & -ig_a^{1,5} \\
-\frac{i}{2} & 0 & -ig_a^{1,3} & g_a^{1,3} & -ig_a^{1,5} & g_a^{1,5} \\
g^{1,3} & ig^{1,3}_a & 0 & \frac{i}{2} & -g^{3,5}_a & - ig^{3,5}_a \\ig^{1,3}& 
-g^{1,3}_a  &  - \frac{i}{2} & 0  & - i g_a^{3,5}  & g^{3,5} \\ g^{1,5}_a&
ig_a^{1,5}& g_a^{3,5} & ig_{a}^{3,5} & 0 & \frac{i}{2} & \\ ig^{1,5}&
-g^{1,5}_a&i g_a^{3,5} & -g^{3,5}_a&-\frac{i}{2} &0 
\end{pmatrix} 
\end{equation}
$\tilde g$ has a compact representation as:
\begin{equation}
    g_a = A \tilde{g} A^T = -A \left(  \frac{\sigma_y}{2} \otimes I + (\sigma_z + i \sigma_x) \otimes g'_a\right) A^T 
\end{equation}
where $g'$ is a $n \times n$ matrix. For $n = 3$:
\begin{equation}
    g' = \begin{pmatrix}
        0&  g_a^{1,3} & g_a^{1,5} \\ -g_a^{1,3}& 0 & g^{1,5}_a \\ -g^{1,5}_a & -g_a^{1,5} & 0 
    \end{pmatrix}
\end{equation}
The first boundary condition in  equation (\ref{eq:u_v_bdy}) corresponds to evaluating $v_{k-1}(0) - u_k(0)$. 
\begin{align*}\label{eq:bc_diff}\nt 
  v_{k-1}(0) - u_k(0) &= \Tr  \left( (\sigma^+ \otimes M^{2k-1} - \sigma_y \otimes M^{2k} )g_a(0) \right)  \\ 
  &= - \Tr  \left( (\sigma^+ \otimes M^{2k-1} - \sigma_y \otimes M^{2k} ) A\, \frac{\sigma_y}{2}\otimes I \, A^T\right)   \\&\hspace{2cm}-  \Tr  \left( (\sigma^+ \otimes M^{2k-1} - \sigma_x \otimes M^{2k} ) A\, \sigma^+ \otimes g'_a (0)\, A^T\right)
\end{align*}
We can show that the second term is always zero. Note that $A$ can be written as:
\begin{align*}\nt 
    A = (\sigma_x + i \sigma_y) \otimes I  + (\sigma_x - i \sigma_y) \otimes A' \\ 
    A^T  = (\sigma_x - i \sigma_y) \otimes I + (\sigma_x + i \sigma_y) \otimes A'^{\, T}
\end{align*}
where $A'$ is the $n \times n$ matrix with similar entries as $A$. Using the above representation of $A$, we can show the following 
\begin{align*} \nt
    \Tr_2 (A^T \sigma^+ \otimes M^{2k-1} A \,\sigma^{+}) &= - 4 (M^{2k - 1} + A' \, M^{2k-1} + M^{2k-1}A'{\, ^T} + A' M^{2k-1} A'^{,T}), \\
     \Tr_2 (A^T \sigma_y \otimes M^{2k} A \,\sigma^{+}) &= 4 (A' M^{2k} - M^{2k} A'^{\, T})
\end{align*}
where $\Tr_2$ denotes trace over the Pauli matrices. 

Taking the difference between the two matrices, we get: 
\begin{align*} \nt 
       & \Tr_2 (A^T \sigma^+ \otimes M^{2k-1} A \,\sigma^{+}) - \Tr_2 (A^T \sigma_y \otimes M^{2k} A \,\sigma^{+}) \\ = & - 4 ( M^{2k-1} + A' M^{2k-1}A'^{\,T}  + A' M^{2k-1} + M^{2k-1} A'^{\,T} + A' M^{2k} - M^{2k} A'^{\, T})  
\end{align*}
We'll show that the above sum is zero. We'll use the following property of $M^r$:
    \begin{equation} M = A' M A'^{\, T} \implies  M^r = A' M^r A'^{\, T} \,  \quad  \forall r \end{equation}
We can simplify the above sum of matrices:
\begin{align*} \nt 
 &   M^{2k-1} + A' M^{2k-1}A'^{\,T}  + A' M^{2k-1} + M^{2k-1} A'^{\,T} + A' M^{2k} - M^{2k} A'^{\, T} \\=& \, 2 M^{2k-1} + (A' + A'^{\, T}) M^{2k-1} + (A'  - A'^{\, T}) M^{2k} \\= & \left[2 + A' + A'^{\, T}  + (A' - A'^{\, T} )M \right]M^{2k-1}  
\end{align*} 
One can easily check that the sum of matrices in the square brackets vanishes. Therefore, the sum is zero. We can further simplify the boundary condition as follows:
\begin{align*}
    v_{k-1}(0) - u_k(0) &= -\Tr \left( (\sigma^+ \otimes M^{2k -1} - \sigma_y \otimes M^{2k}) A \frac{\sigma_y}{2} \otimes I A^T\right ) \\&= n( M^{2k-1}_{1,2} -  M^{2k}_{1,2}) \\& = n \left( M^{2k-1}_{1,2} + \sum_{3}^n M^{2k-1}_{1,3} \right) \\ &= - n M^{2k}_{1,1} \\& = - \Tr \left( M^{2k}\right) 
\end{align*}
Similarly, we can show that
$u_k(T) - v_k(T) = - \Tr(M^{2k})$. 

\section{Partial proof of the trace relation}\label{app:matrix_rel}

In this appendix, we will give a partial proof of the relation (\ref{eq:matrix_rel}):
\begin{equation}
 \sum_{r = m+1}^{l} \binom{n}{2(l-r)} \alpha_{r - m} = 2 (l-m) \binom{n}{2(l-m)}, \quad l = \lfloor n/2 \rfloor.
\end{equation}
Assume $n$ is even. By replacing $ l - m \rightarrow m $, we can rewrite the above relation as follows:
\begin{equation}
    \sum_{r =0}^{m-1} \binom{n}{2r} \alpha_{ m -r } = 2 m \binom{n}{2m}  
\end{equation}
Using the definition of $\alpha_r = - \Tr( M^{2r})$,
we get the following relation:
\begin{equation}
 f_m \equiv   \sum_{r =0}^{m} \binom{n}{2r} \Tr (M^{2m - 2r}) + 2m \binom{n}{2m} = 0
\end{equation}
We will prove that 
\begin{equation}
    f_m + f_{l-m} = 0 
\end{equation}
Consider the characteristic equation of $M$:
\begin{equation}
  \sum_{n = 0}^{l}  M^{n - 2m}  \binom{n}{2m} = 0 
\end{equation}
Multiplying the above characteristic equation by $M^{-n}$, we note that $M$ and $M^{-1}$ satisfy the same characteristic equation. Therefore, $M$ and $M^{-1}$ have the same set of eigenvalues. In particular, this implies that $\Tr(M^{k}) = \Tr (M^{-k})$. Now multiply the characteristic equation by $M^{-2m}$ and take its trace. We get,
\begin{align*}
   & \binom{n}{2m} + \sum_{r = 0}^{l-m-1} \binom{n}{2r} \Tr(M^{2 (l-m)- 2r}) + \sum_{r = 0}^{m-1} \binom{n}{2r} \Tr(M^{2m  -2r})  = 0 \\ 
\implies& (2l - 2m) \binom{n}{2m}+ \sum_{r = 0}^{l-m-1} \binom{n}{2r} \Tr(M^{2 (l-m)- 2r}) + 2m \binom{n}{2m} + \sum_{r = 0}^{m-1} \binom{n}{2r} \Tr(M^{2m  -2r}) = 0 \\ \implies& f_{l-m} + f_m = 0 
\end{align*}

\section{Properties of membrane tension from entanglement inequalities}\label{sec:EM}
\begin{figure}[t]
    \centering
    \includegraphics[width=.3\textwidth]{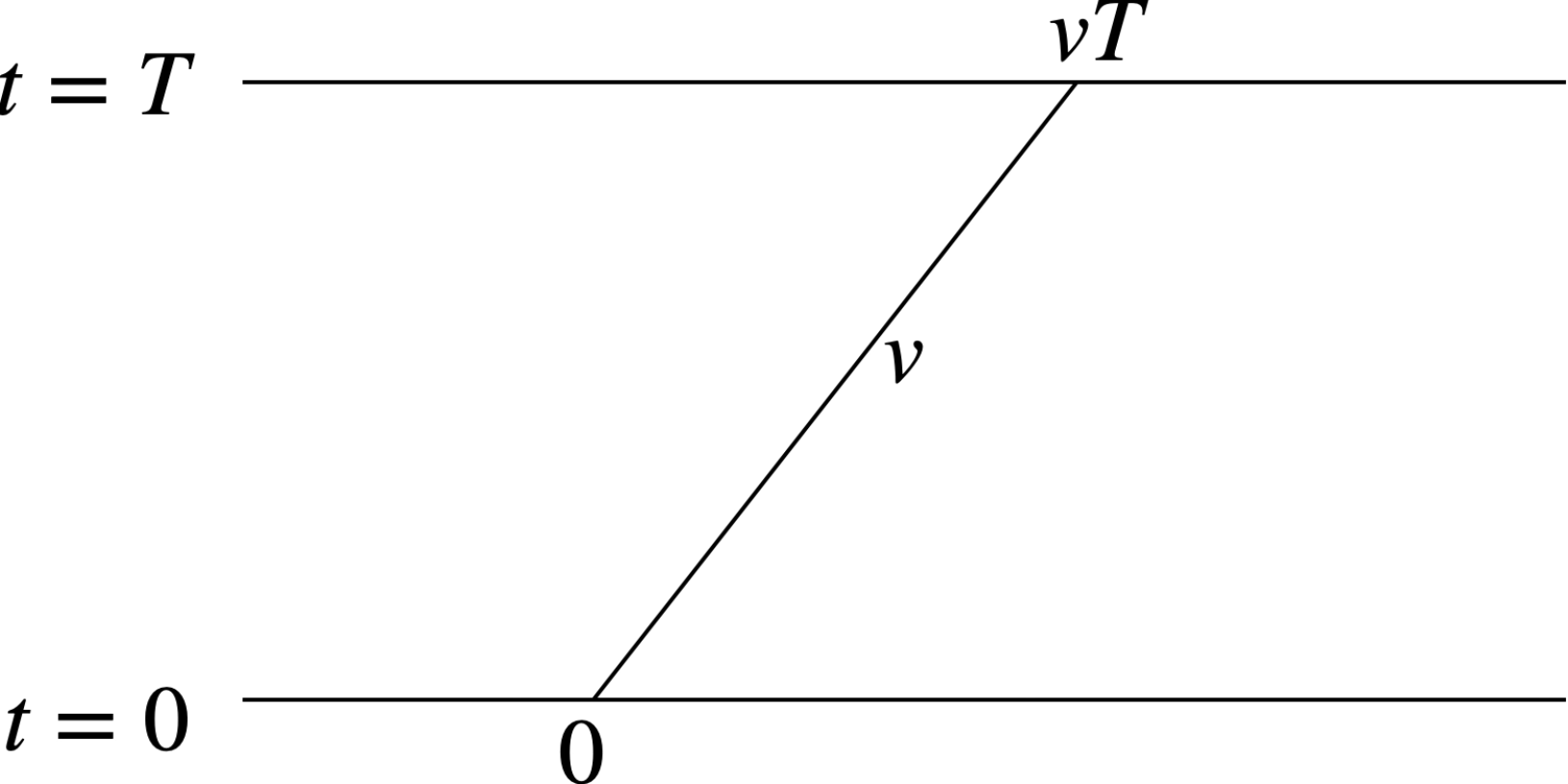}
    \includegraphics[width=.3\textwidth]{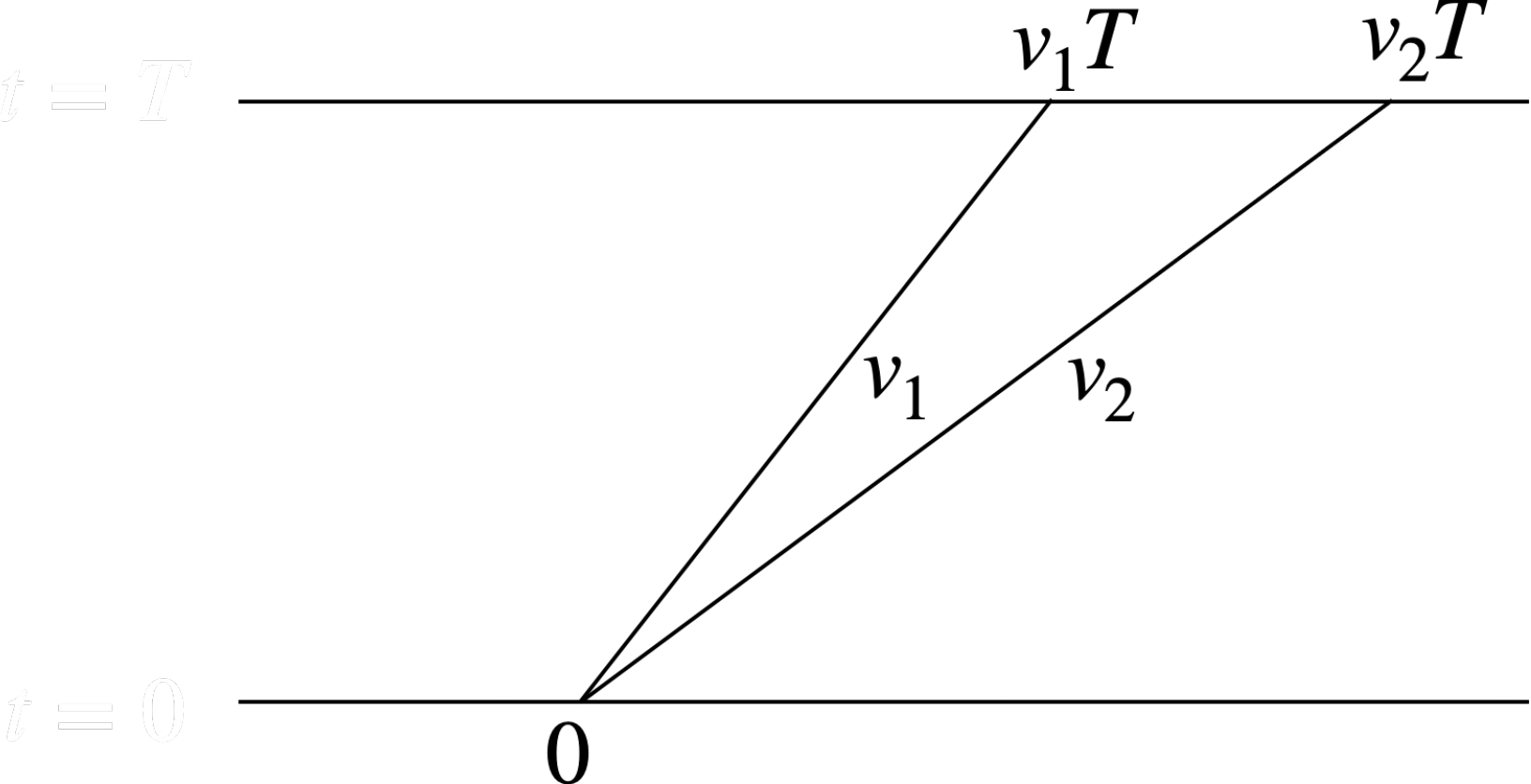}
     \includegraphics[width=.3\textwidth]{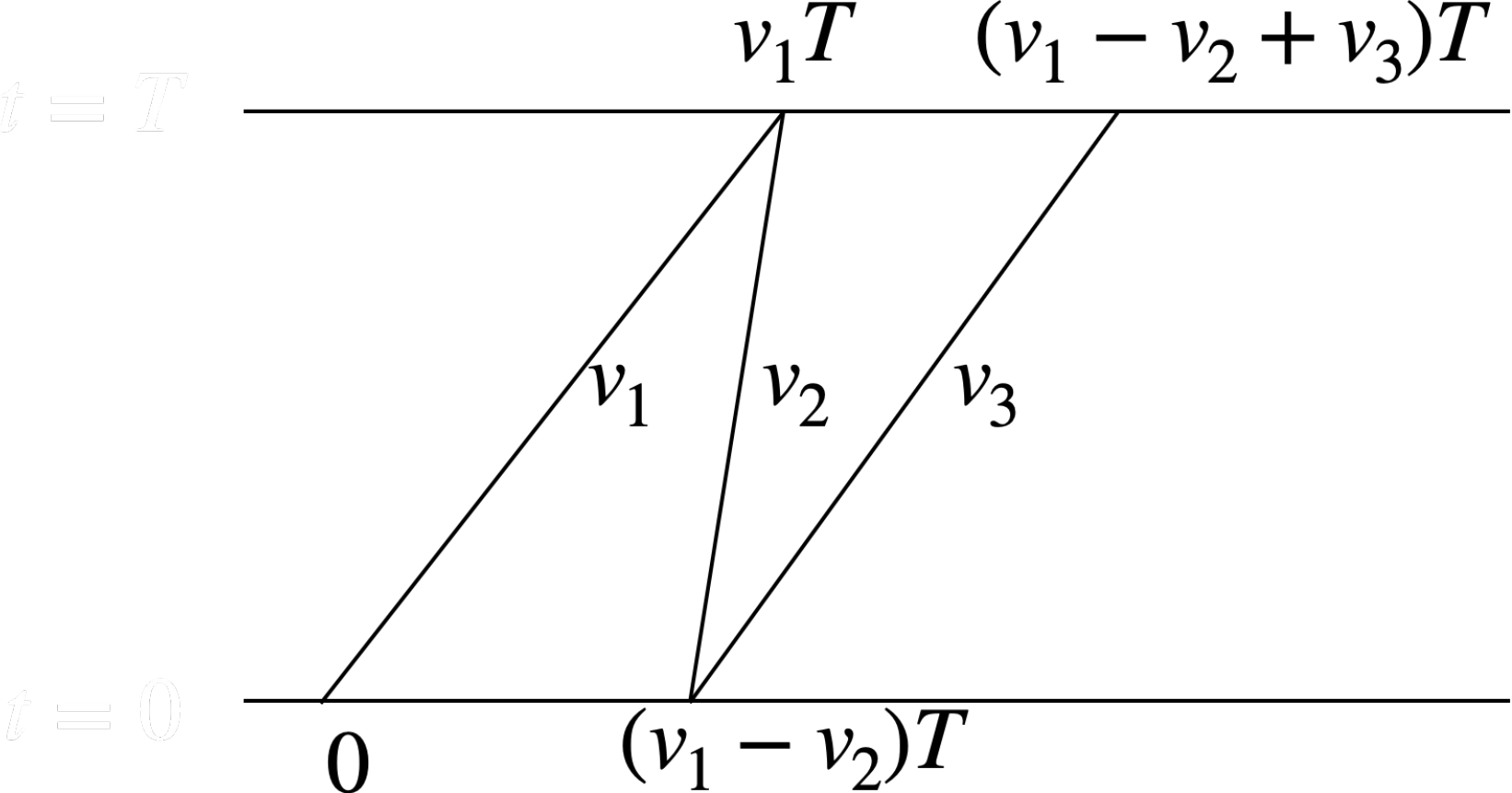}
    \caption{Constraints on Membrane Tension}
    \label{fig:membrane_tension}
\end{figure}
In this section, we will prove the following three inequalities that the membrane tension $\mathcal{E}(v)$ must satisfy:
\begin{equation}\label{eq:inequalities}
 (i) \quad    \mathcal{E}(v) \geq |v|, \quad (ii) \quad  |\mathcal{E}'(v)| \leq 1 \quad \text{and} \quad (iii) \quad  \mathcal{E}''(v) \geq 0.
\end{equation}
 We will use the definition of membrane tension in equation (\ref{eq:membrane_tension}). 
 
Let $S_{eq}([x,y])$ be the entropy of the interval $[x,y]$ in the maximally mixed state. It follows that $S_{eq}([x,y]) = s_{eq}|x-y|$, where $s_{eq}$ is the local Hilbert space dimension. 
For illustration of the various sub-regions used in the proof, see figure (\ref{fig:membrane_tension}).

Proof $(i):$ The inequality follows from sub-additivity of entanglement entropy. 
\begin{align*} \nt \mathcal{E}(v) &= \frac{1}{ s_{eq}T} S_{U(T)}(0, v T) \\ &\geq \frac{|S_{eq}( [ -\infty, 0])  - S_{eq}(-\infty, vT)|}{s_{eq} T} = \frac{s_{eq} | v T |}{ s_{eq}T}  = |v| 
\end{align*}

\text{Proof} (ii):
\begin{align*} \qquad  \nt |\mathcal{E}(v_1) - \mathcal{E}(v_2)| &= \frac{ |S_{U(T)}(0, v_1 T) -  S_{U(T)}(0,  v_2 T)|}{ s_{\text{eq}}T}   \\  &\leq  \frac{S_{eq}([v_1 T,  v_2 T])}{s_{\text{eq} T}}  =  | v_1 - v_2| 
\end{align*}

Proof $(iii):$ 
\begin{align*}  
    S_{U(T)}(0 , v_1 T) +  S_{U(T)}((v_1 - v_2)T, &(v_3 + v_1 - v_2) T) \\ &\leq S_{U(T)}(0, (v_3 + v_1 - v_2) T) + S_{U(T)}((v_1 - v_2 )T, v_1 T) \\  \implies \mathcal{E}(v_1) + \mathcal{E}(v_3) &\leq \mathcal{E}(v_1 + v_3 - v_2) + \mathcal{E}(v_2) \\
    \implies \mathcal{E}(v_1)-  \mathcal{E}(v_2) &\leq \mathcal{E}(v_3 + v_1 - v_2) - \mathcal{E}(v_3) \\ \implies \mathcal{E}'(v_2) (v_1 - v_2) &\leq \mathcal{E}' (v_3) (v_1 - v_2)  \\ \implies \mathcal{E}'(v_2) &\leq \mathcal{E}'(v_3) \quad ( \text{Since } v_1 - v_2 \geq 0)  \\ \implies  0&\leq  \mathcal{E}''(v_3) \quad ( \text{Since } v_2 \leq v_3) \nt 
\end{align*}

\section*{} 
\bibliographystyle{JHEP}
\bibliography{refs}

@article{Ryu:2006bv,
    author = "Ryu, Shinsei and Takayanagi, Tadashi",
    title = "{Holographic derivation of entanglement entropy from AdS/CFT}",
    eprint = "hep-th/0603001",
    archivePrefix = "arXiv",
    reportNumber = "NSF-KITP-06-11",
    doi = "10.1103/PhysRevLett.96.181602",
    journal = "Phys. Rev. Lett.",
    volume = "96",
    pages = "181602",
    year = "2006"
}

@article{Maldacena:2015iua,
    author = "Maldacena, Juan and Simmons-Duffin, David and Zhiboedov, Alexander",
    title = "{Looking for a bulk point}",
    eprint = "1509.03612",
    archivePrefix = "arXiv",
    primaryClass = "hep-th",
    doi = "10.1007/JHEP01(2017)013",
    journal = "JHEP",
    volume = "01",
    pages = "013",
    year = "2017"
}

@article{Caron-Huot:2021enk,
    author = "Caron-Huot, Simon and Mazac, Dalimil and Rastelli, Leonardo and Simmons-Duffin, David",
    title = "{AdS bulk locality from sharp CFT bounds}",
    eprint = "2106.10274",
    archivePrefix = "arXiv",
    primaryClass = "hep-th",
    doi = "10.1007/JHEP11(2021)164",
    journal = "JHEP",
    volume = "11",
    pages = "164",
    year = "2021"
}

@article{Heemskerk:2009pn,
    author = "Heemskerk, Idse and Penedones, Joao and Polchinski, Joseph and Sully, James",
    title = "{Holography from Conformal Field Theory}",
    eprint = "0907.0151",
    archivePrefix = "arXiv",
    primaryClass = "hep-th",
    reportNumber = "NSF-KITP-09-110",
    doi = "10.1088/1126-6708/2009/10/079",
    journal = "JHEP",
    volume = "10",
    pages = "079",
    year = "2009"
}

@article{Hubeny:2007xt,
    author = "Hubeny, Veronika E. and Rangamani, Mukund and Takayanagi, Tadashi",
    title = "{A Covariant holographic entanglement entropy proposal}",
    eprint = "0705.0016",
    archivePrefix = "arXiv",
    primaryClass = "hep-th",
    reportNumber = "DCPT-07-13, KUNS-2069",
    doi = "10.1088/1126-6708/2007/07/062",
    journal = "JHEP",
    volume = "07",
    pages = "062",
    year = "2007"
}

@article{Dong:2022ucb,
    author = "Dong, Xi and Wang, Diandian and Weng, Wayne W. and Wu, Chih-Hung",
    title = "{A tale of two butterflies: an exact equivalence in higher-derivative gravity}",
    eprint = "2203.06189",
    archivePrefix = "arXiv",
    primaryClass = "hep-th",
    doi = "10.1007/JHEP10(2022)009",
    journal = "JHEP",
    volume = "10",
    pages = "009",
    year = "2022"
}

@article{Chua:2025vig,
    author = "Chua, Wan Zhen and Hartman, Thomas and Weng, Wayne W.",
    title = "{Replica manifolds, pole skipping, and the butterfly effect}",
    eprint = "2504.08139",
    archivePrefix = "arXiv",
    primaryClass = "hep-th",
    month = "4",
    year = "2025"
}

@article{Sachdev_1993,
   title={Gapless spin-fluid ground state in a random quantum Heisenberg magnet},
   volume={70},
   ISSN={0031-9007},
   url={http://dx.doi.org/10.1103/PhysRevLett.70.3339},
   DOI={10.1103/physrevlett.70.3339},
   number={21},
   journal={Physical Review Letters},
   publisher={American Physical Society (APS)},
   author={Sachdev, Subir and Ye, Jinwu},
   year={1993},
   month=may, pages={3339–3342} }

@article{Sunderhauf:2019djv,
    author = {S\"underhauf, Christoph and Piroli, Lorenzo and Qi, Xiao-Liang and Schuch, Norbert and Cirac, J. Ignacio},
    title = "{Quantum chaos in the Brownian SYK model with large finite $N$: OTOCs and tripartite information}",
    eprint = "1908.00775",
    archivePrefix = "arXiv",
    primaryClass = "quant-ph",
    doi = "10.1007/JHEP11(2019)038",
    journal = "JHEP",
    volume = "11",
    pages = "038",
    year = "2019"
}

@article{Milekhin:2023bjv,
    author = "Milekhin, Alexey and Xu, Jiuci",
    title = "{Revisiting Brownian SYK and its possible relations to de Sitter}",
    eprint = "2312.03623",
    archivePrefix = "arXiv",
    primaryClass = "hep-th",
    doi = "10.1007/JHEP10(2024)151",
    journal = "JHEP",
    volume = "10",
    pages = "151",
    year = "2024"
}

@article{Jian:2020krd,
    author = "Jian, Shao-Kai and Swingle, Brian",
    title = "{Note on entropy dynamics in the Brownian SYK model}",
    eprint = "2011.08158",
    archivePrefix = "arXiv",
    primaryClass = "cond-mat.stat-mech",
    doi = "10.1007/JHEP03(2021)042",
    journal = "JHEP",
    volume = "03",
    pages = "042",
    year = "2021"
}

@article{Zhang:2023vpm,
    author = "Zhang, Pengfei",
    title = "{Information scrambling and entanglement dynamics of complex Brownian Sachdev-Ye-Kitaev models}",
    eprint = "2301.03189",
    archivePrefix = "arXiv",
    primaryClass = "cond-mat.str-el",
    doi = "10.1007/JHEP04(2023)105",
    journal = "JHEP",
    volume = "04",
    pages = "105",
    year = "2023"
}

@article{PhysRevB.107.014201,
  title = {Hydrodynamic theory of scrambling in chaotic long-range interacting systems},
  author = {Zhou, Tianci and Guo, Andrew and Xu, Shenglong and Chen, Xiao and Swingle, Brian},
  journal = {Phys. Rev. B},
  volume = {107},
  issue = {1},
  pages = {014201},
  numpages = {16},
  year = {2023},
  month = {Jan},
  publisher = {American Physical Society},
  doi = {10.1103/PhysRevB.107.014201},
  url = {https://link.aps.org/doi/10.1103/PhysRevB.107.014201}
}

@article{Fisher,
author = {FISHER, R. A.},
title = {THE WAVE OF ADVANCE OF ADVANTAGEOUS GENES},
journal = {Annals of Eugenics},
volume = {7},
number = {4},
pages = {355-369},
doi = {https://doi.org/10.1111/j.1469-1809.1937.tb02153.x},
url = {https://onlinelibrary.wiley.com/doi/abs/10.1111/j.1469-1809.1937.tb02153.x},
eprint = {https://onlinelibrary.wiley.com/doi/pdf/10.1111/j.1469-1809.1937.tb02153.x},
year = {1937}
}

@article{KPP,
author = {Kolmogorov, A. and Petrovskii, I. and Piskunov, N.},
title = {A study of the diffusion equation with increase in the amount of substance, and its application to a biological problem},
journal = {Selected Works of A. N. Kolmogorov I (translated by V. M. Volosov from Bull. Moscow Univ., Math. Mech. 1, 1–25, 1937)},
pages = {248-270},
year = {1991}
}

@article{Stanford:2023npy,
    author = "Stanford, Douglas and Vardhan, Shreya and Yao, Shunyu",
    title = "{Scramblon loops}",
    eprint = "2311.12121",
    archivePrefix = "arXiv",
    primaryClass = "hep-th",
    doi = "10.1007/JHEP10(2024)073",
    journal = "JHEP",
    volume = "10",
    pages = "073",
    year = "2024"
}

@article{Jian:2021hve,
    author = "Jian, Shao-Kai and Liu, Chunxiao and Chen, Xiao and Swingle, Brian and Zhang, Pengfei",
    title = "{Measurement-Induced Phase Transition in the Monitored Sachdev-Ye-Kitaev Model}",
    eprint = "2104.08270",
    archivePrefix = "arXiv",
    primaryClass = "cond-mat.str-el",
    doi = "10.1103/PhysRevLett.127.140601",
    journal = "Phys. Rev. Lett.",
    volume = "127",
    number = "14",
    pages = "140601",
    year = "2021"
}

@article{Maldacena:2016upp,
    author = "Maldacena, Juan and Stanford, Douglas and Yang, Zhenbin",
    title = "{Conformal symmetry and its breaking in two dimensional Nearly Anti-de-Sitter space}",
    eprint = "1606.01857",
    archivePrefix = "arXiv",
    primaryClass = "hep-th",
    doi = "10.1093/ptep/ptw124",
    journal = "PTEP",
    volume = "2016",
    number = "12",
    pages = "12C104",
    year = "2016"
}

@article{Headrick:2014cta,
    author = "Headrick, Matthew and Hubeny, Veronika E. and Lawrence, Albion and Rangamani, Mukund",
    title = "{Causality \& holographic entanglement entropy}",
    eprint = "1408.6300",
    archivePrefix = "arXiv",
    primaryClass = "hep-th",
    reportNumber = "DCPT-14-33, BRX-TH-6284",
    doi = "10.1007/JHEP12(2014)162",
    journal = "JHEP",
    volume = "12",
    pages = "162",
    year = "2014"
}

@article{Maldacena:2015waa,
    author = "Maldacena, Juan and Shenker, Stephen H. and Stanford, Douglas",
    title = "{A bound on chaos}",
    eprint = "1503.01409",
    archivePrefix = "arXiv",
    primaryClass = "hep-th",
    doi = "10.1007/JHEP08(2016)106",
    journal = "JHEP",
    volume = "08",
    pages = "106",
    year = "2016"
}

@article{Maldacena:2016hyu,
    author = "Maldacena, Juan and Stanford, Douglas",
    title = "{Remarks on the Sachdev-Ye-Kitaev model}",
    eprint = "1604.07818",
    archivePrefix = "arXiv",
    primaryClass = "hep-th",
    doi = "10.1103/PhysRevD.94.106002",
    journal = "Phys. Rev. D",
    volume = "94",
    number = "10",
    pages = "106002",
    year = "2016"
}

@article{Saad:2018bqo,
    author = "Saad, Phil and Shenker, Stephen H. and Stanford, Douglas",
    title = "{A semiclassical ramp in SYK and in gravity}",
    eprint = "1806.06840",
    archivePrefix = "arXiv",
    primaryClass = "hep-th",
    month = "6",
    year = "2018"
}

@article{Stanford:2021bhl,
    author = "Stanford, Douglas and Yang, Zhenbin and Yao, Shunyu",
    title = "{Subleading Weingartens}",
    eprint = "2107.10252",
    archivePrefix = "arXiv",
    primaryClass = "hep-th",
    month = "7",
    year = "2021"
}

@article{Faulkner:2013ana,
    author = "Faulkner, Thomas and Lewkowycz, Aitor and Maldacena, Juan",
    title = "{Quantum corrections to holographic entanglement entropy}",
    eprint = "1307.2892",
    archivePrefix = "arXiv",
    primaryClass = "hep-th",
    doi = "10.1007/JHEP11(2013)074",
    journal = "JHEP",
    volume = "11",
    pages = "074",
    year = "2013"
}

@article{Almheiri:2014lwa,
    author = "Almheiri, Ahmed and Dong, Xi and Harlow, Daniel",
    title = "{Bulk Locality and Quantum Error Correction in AdS/CFT}",
    eprint = "1411.7041",
    archivePrefix = "arXiv",
    primaryClass = "hep-th",
    reportNumber = "SU-ITP-14-30",
    doi = "10.1007/JHEP04(2015)163",
    journal = "JHEP",
    volume = "04",
    pages = "163",
    year = "2015"
}

@article{Dong:2016eik,
    author = "Dong, Xi and Harlow, Daniel and Wall, Aron C.",
    title = "{Reconstruction of Bulk Operators within the Entanglement Wedge in Gauge-Gravity Duality}",
    eprint = "1601.05416",
    archivePrefix = "arXiv",
    primaryClass = "hep-th",
    reportNumber = "NSF-KITP-16-005",
    doi = "10.1103/PhysRevLett.117.021601",
    journal = "Phys. Rev. Lett.",
    volume = "117",
    number = "2",
    pages = "021601",
    year = "2016"
}

@article{Harlow:2016vwg,
    author = "Harlow, Daniel",
    title = "{The Ryu\textendash{}Takayanagi Formula from Quantum Error Correction}",
    eprint = "1607.03901",
    archivePrefix = "arXiv",
    primaryClass = "hep-th",
    doi = "10.1007/s00220-017-2904-z",
    journal = "Commun. Math. Phys.",
    volume = "354",
    number = "3",
    pages = "865--912",
    year = "2017"
}

@article{Schumacher:1996dy,
    author = "Schumacher, Benjamin and Nielsen, M. A.",
    title = "{Quantum data processing and error correction}",
    eprint = "quant-ph/9604022",
    archivePrefix = "arXiv",
    doi = "10.1103/PhysRevA.54.2629",
    journal = "Phys. Rev. A",
    volume = "54",
    pages = "2629",
    year = "1996"
}

@ARTICLE{Schumacher2001,
       author = {{Schumacher}, Benjamin and {Westmoreland}, Michael D.},
        title = "{Approximate quantum error correction}",
      journal = {arXiv e-prints},
         year = 2001,
        month = dec,
          eid = {quant-ph/0112106},
        pages = {quant-ph/0112106},
archivePrefix = {arXiv},
       eprint = {quant-ph/0112106},
 primaryClass = {quant-ph},
       adsurl = {https://ui.adsabs.harvard.edu/abs/2001quant.ph.12106S}
}

@article{Xu_2019,
   title={Locality, Quantum Fluctuations, and Scrambling},
   volume={9},
   ISSN={2160-3308},
   url={http://dx.doi.org/10.1103/PhysRevX.9.031048},
   DOI={10.1103/physrevx.9.031048},
   number={3},
   journal={Physical Review X},
   publisher={American Physical Society (APS)},
   author={Xu, Shenglong and Swingle, Brian},
   year={2019},
   month=sep }

@article{Mezei_2017,
   title={On entanglement spreading from holography},
   volume={2017},
   ISSN={1029-8479},
   url={http://dx.doi.org/10.1007/JHEP05(2017)064},
   DOI={10.1007/jhep05(2017)064},
   number={5},
   journal={Journal of High Energy Physics},
   publisher={Springer Science and Business Media LLC},
   author={Mezei, Márk},
   year={2017},
   month=may }

@article{Zhou:2018myl,
    author = "Zhou, Tianci and Nahum, Adam",
    title = "{Emergent statistical mechanics of entanglement in random unitary circuits}",
    eprint = "1804.09737",
    archivePrefix = "arXiv",
    primaryClass = "cond-mat.stat-mech",
    doi = "10.1103/PhysRevB.99.174205",
    journal = "Phys. Rev. B",
    volume = "99",
    number = "17",
    pages = "174205",
    year = "2019"
}

@article{Hayden:2007cs,
    author = "Hayden, Patrick and Preskill, John",
    title = "{Black holes as mirrors: Quantum information in random subsystems}",
    eprint = "0708.4025",
    archivePrefix = "arXiv",
    primaryClass = "hep-th",
    reportNumber = "CALT-68-2659",
    doi = "10.1088/1126-6708/2007/09/120",
    journal = "JHEP",
    volume = "09",
    pages = "120",
    year = "2007"
}

@article{Rampp:2023zmr,
    author = "Rampp, Michael A. and Claeys, Pieter W.",
    title = "{Hayden-Preskill recovery in chaotic and integrable unitary circuit dynamics}",
    eprint = "2312.03838",
    archivePrefix = "arXiv",
    primaryClass = "quant-ph",
    doi = "10.22331/q-2024-08-08-1434",
    journal = "Quantum",
    volume = "8",
    pages = "1434",
    year = "2024"
}

@article{shreya_note,
    author = "Vardhan, Shreya",
    title = "{private communication}"
}

@article{Chandrasekaran:2021tkb,
    author = "Chandrasekaran, Venkatesa and Faulkner, Thomas and Levine, Adam",
    title = "{Scattering strings off quantum extremal surfaces}",
    eprint = "2108.01093",
    archivePrefix = "arXiv",
    primaryClass = "hep-th",
    doi = "10.1007/JHEP08(2022)143",
    journal = "JHEP",
    volume = "08",
    pages = "143",
    year = "2022"
}

@article{Chandrasekaran:2022qmq,
    author = "Chandrasekaran, Venkatesa and Levine, Adam",
    title = "{Quantum error correction in SYK and bulk emergence}",
    eprint = "2203.05058",
    archivePrefix = "arXiv",
    primaryClass = "hep-th",
    doi = "10.1007/JHEP06(2022)039",
    journal = "JHEP",
    volume = "06",
    pages = "039",
    year = "2022"
}

@article{Balasubramanian:2023xdp,
    author = "Balasubramanian, Vijay and Kar, Arjun and Li, Cathy and Parrikar, Onkar and Rajgadia, Harshit",
    title = "{Quantum error correction from complexity in Brownian SYK}",
    eprint = "2301.07108",
    archivePrefix = "arXiv",
    primaryClass = "hep-th",
    doi = "10.1007/JHEP08(2023)071",
    journal = "JHEP",
    volume = "08",
    pages = "071",
    year = "2023"
}

@article{Czech:2012bh,
    author = "Czech, Bartlomiej and Karczmarek, Joanna L. and Nogueira, Fernando and Van Raamsdonk, Mark",
    title = "{The Gravity Dual of a Density Matrix}",
    eprint = "1204.1330",
    archivePrefix = "arXiv",
    primaryClass = "hep-th",
    doi = "10.1088/0264-9381/29/15/155009",
    journal = "Class. Quant. Grav.",
    volume = "29",
    pages = "155009",
    year = "2012"
}

@article{Cotler:2017erl,
    author = "Cotler, Jordan and Hayden, Patrick and Penington, Geoffrey and Salton, Grant and Swingle, Brian and Walter, Michael",
    title = "{Entanglement Wedge Reconstruction via Universal Recovery Channels}",
    eprint = "1704.05839",
    archivePrefix = "arXiv",
    primaryClass = "hep-th",
    doi = "10.1103/PhysRevX.9.031011",
    journal = "Phys. Rev. X",
    volume = "9",
    number = "3",
    pages = "031011",
    year = "2019"
}

@article{Maldacena:2001kr,
    author = "Maldacena, Juan Martin",
    title = "{Eternal black holes in anti-de Sitter}",
    eprint = "hep-th/0106112",
    archivePrefix = "arXiv",
    reportNumber = "NSF-ITP-01-59",
    doi = "10.1088/1126-6708/2003/04/021",
    journal = "JHEP",
    volume = "04",
    pages = "021",
    year = "2003"
}

@article{Faulkner:2017vdd,
    author = "Faulkner, Thomas and Lewkowycz, Aitor",
    title = "{Bulk locality from modular flow}",
    eprint = "1704.05464",
    archivePrefix = "arXiv",
    primaryClass = "hep-th",
    doi = "10.1007/JHEP07(2017)151",
    journal = "JHEP",
    volume = "07",
    pages = "151",
    year = "2017"
}

@article{Parrikar:2024zbb,
    author = "Parrikar, Onkar and Rajgadia, Harshit and Singh, Vivek and Sorce, Jonathan",
    title = "{Relational bulk reconstruction from modular flow}",
    eprint = "2403.02377",
    archivePrefix = "arXiv",
    primaryClass = "hep-th",
    reportNumber = "MIT-CTP/5688",
    doi = "10.1007/JHEP07(2024)138",
    journal = "JHEP",
    volume = "07",
    pages = "138",
    year = "2024"
}

@article{Zhou:2019pob,
    author = "Zhou, Tianci and Nahum, Adam",
    title = "{Entanglement Membrane in Chaotic Many-Body Systems}",
    eprint = "1912.12311",
    archivePrefix = "arXiv",
    primaryClass = "cond-mat.str-el",
    doi = "10.1103/PhysRevX.10.031066",
    journal = "Phys. Rev. X",
    volume = "10",
    number = "3",
    pages = "031066",
    year = "2020"
}

@misc{jonay2018,
      title={Coarse-grained dynamics of operator and state entanglement}, 
      author={Cheryne Jonay and David A. Huse and Adam Nahum},
      year={2018},
      eprint={1803.00089},
      archivePrefix={arXiv},
      primaryClass={cond-mat.stat-mech},
      url={https://arxiv.org/abs/1803.00089}, 
}

@misc{TongLectures,
  author = "Tong, David",
  title = "{Lectures on Mathematical Biology}",
  howpublished = {\url{https://www.damtp.cam.ac.uk/user/tong/mathbio.html}},
}

@misc{BramsonLectures,
  author = "Bramson, Maury",
  title = "{Kolmogorov Non-Linear Diffusion Equation}",
  howpublished = {\url{https://hdl.handle.net/11299/151582}},
}

@misc{TraceIdentity,
author = {},
title = {Mathematica File For Trace Identity},
  howpublished = {\url{https://github.com/harshit-rajgadia/Trace-Identity.git}},
}

@misc{Kitaev2015v1,
    author = "Kitaev, Alexei",
    title = "{A simple model of quantum holography (part 1)}",
    howpublished = {\url{https://online.kitp.ucsb.edu/online/entangled15/kitaev/}},
    note = {Accessed: 2022-01-06}
}

@misc{Kitaev2015v2,
    author = "Kitaev, Alexei",
    title = "{A simple model of quantum holography (part 2)}",
    howpublished = {\url{https://online.kitp.ucsb.edu/online/entangled15/kitaev2/}},
    note = {Accessed: 2022-01-06}
}

@misc{vardhan_moudgalya2024,
      title={Entanglement dynamics from universal low-lying modes}, 
      author={Shreya Vardhan and Sanjay Moudgalya},
      year={2024},
      eprint={2407.16763},
      archivePrefix={arXiv},
      primaryClass={cond-mat.stat-mech},
      url={https://arxiv.org/abs/2407.16763}, 
}

@article{Mezei_Membrane_Theory_Holography,
  title = {Membrane theory of entanglement dynamics from holography},
  author = {Mezei, M\'ark},
  journal = {Phys. Rev. D},
  volume = {98},
  issue = {10},
  pages = {106025},
  numpages = {9},
  year = {2018},
  month = {Nov},
  publisher = {American Physical Society},
  doi = {10.1103/PhysRevD.98.106025},
  url = {https://link.aps.org/doi/10.1103/PhysRevD.98.106025}
}

@article{Mezei_Stanford_2017,
   title={On entanglement spreading in chaotic systems},
   volume={2017},
   ISSN={1029-8479},
   url={http://dx.doi.org/10.1007/JHEP05(2017)065},
   DOI={10.1007/jhep05(2017)065},
   number={5},
   journal={Journal of High Energy Physics},
   publisher={Springer Science and Business Media LLC},
   author={Mezei, Márk and Stanford, Douglas},
   year={2017},
   month=may }

@article{Engelhardt:2014gca,
    author = "Engelhardt, Netta and Wall, Aron C.",
    title = "{Quantum Extremal Surfaces: Holographic Entanglement Entropy beyond the Classical Regime}",
    eprint = "1408.3203",
    archivePrefix = "arXiv",
    primaryClass = "hep-th",
    doi = "10.1007/JHEP01(2015)073",
    journal = "JHEP",
    volume = "01",
    pages = "073",
    year = "2015"
}

@article{Hosur_2016,
   title={Chaos in quantum channels},
   volume={2016},
   ISSN={1029-8479},
   url={http://dx.doi.org/10.1007/JHEP02(2016)004},
   DOI={10.1007/jhep02(2016)004},
   number={2},
   journal={Journal of High Energy Physics},
   publisher={Springer Science and Business Media LLC},
   author={Hosur, Pavan and Qi, Xiao-Liang and Roberts, Daniel A. and Yoshida, Beni},
   year={2016},
   month=feb }

\end{document}